\documentclass[pdftex,twocolumn,epjc3]{svjour3}          %
\pdfoutput=1
\RequirePackage[T1]{fontenc}
\usepackage{tocloft}

\RequirePackage{graphicx}
\RequirePackage{mathptmx}      %
\RequirePackage{flushend}
\RequirePackage{amsmath}
\RequirePackage[british]{babel} 
\RequirePackage{bm}              
\RequirePackage{lineno}    
\RequirePackage[latin9]{inputenc} 
\RequirePackage{seqsplit}
\RequirePackage{xcolor}
\RequirePackage{textcomp}
\RequirePackage{amssymb}
\RequirePackage{booktabs}
\RequirePackage{xspace}

\usepackage{widetext}

\usepackage[backend=bibtex,style=numeric,sorting=none,doi=false]{biblatex}
\newcommand{\abmp} {ABMP16\xspace}
\newcommand{\nnpdf} {NNPDF3.1\xspace}
\newcommand{\chisq}{\ensuremath{\chi^2}\xspace}
\newcommand{\ndf}{dof\xspace}

\newcommand{\xfitter} {\textsc{xFitter}\xspace}

\newcommand{\lhapdf} {{\textsc{lhapdf}}\xspace}
\newcommand{\xbj}{\ensuremath{x_{\text{Bj}}}\xspace}
\newcommand{\fonll} {\hbox{FONLL-B}\xspace}
\newcommand{\ffns} {\hbox{FFNS~A}\xspace}
\newcommand{\ffnsb} {\hbox{FFNS~B}\xspace}
\newcommand{\ffthreea} {{\hbox{HERAPDF2.0} FF3A}\xspace}
\newcommand{\ffthreeb} {{\hbox{HERAPDF2.0} FF3B}\xspace}

\newcommand\new[1]{{\color{blue} #1}}
\renewcommand\new[1]{{ #1}}

\journalname{DESY Report 19-107}

\RequirePackage[colorlinks,citecolor=blue,urlcolor=blue,linkcolor=blue]{hyperref}
\begin{document}
\sloppy

\renewcommand\rightmark{Probing the strange content of the proton with charm production in charged current at LHeC} 
\renewcommand\leftmark{\xfitter Developers' team:}

\makeatletter %
\def\makeheadbox{{%
\hbox to0pt{\vbox{\baselineskip=10dd\hrule\hbox
to\hsize{\vrule\kern3pt\vbox{\kern3pt
\hbox{\bfseries\@journalname }
\kern3pt}\hfil\kern3pt\vrule}\hrule}%
\hss}}}
\makeatother %

\title{Probing the strange content of the proton with charm production in charged current at LHeC}

\author{\xfitter Developers' team:
     Hamed~Abdolmaleki\thanksref{iran}
\and Valerio~Bertone\thanksref{pavia}
\and Daniel~Britzger\thanksref{munich}
\and Stefano~Camarda\thanksref{cern}
\and Amanda~Cooper-Sarkar\thanksref{oxford}
\and Achim~Geiser\thanksref{desy}
\and Francesco~Giuli\thanksref{rome}
\and Alexander~Glazov\thanksref{desy}
\and Agnieszka~Luszczak\thanksref{cracow}
\and Ivan~Novikov\thanksref{jinr}
\and Fred~Olness\thanksref{smu}
\and Andrey Sapronov\thanksref{jinr}
\and Oleksandr~Zenaiev\thanksref{hamburg}
}

\institute{Faculty of Physics, Semnan University, 35131-19111 Semnan,
  Iran \label{iran}
  \and Dipartimento di Fisica, Universit\`a di Pavia and INFN, Sezione di Pavia Via Bassi 6, I-27100 Pavia, Italy \label{pavia}
 \and Max-Planck-Institut f\"ur Physik, F\"ohringer Ring 6, D-80805 M\"unchen, Germany \label{munich}
 \and CERN, CH-1211 Geneva 23, Switzerland \label{cern}
  \and Particle Physics, Denys Wilkinson Bdg, Keble Road,
  University of Oxford, OX1 3RH Oxford, UK \label{oxford}
  \and Deutsches Elektronen-Synchrotron (DESY), Notkestrasse 85,
  D-22607 Hamburg, Germany \label{desy}
  \and University of Rome Tor Vergata and INFN, Sezione di Roma 2, Via
  della Ricerca Scientifica 1, 00133 Rome, Italy \label{rome}
  \and T. Kosciuszko Cracow University of Technology, PL-30-084, Cracow, Poland \label{cracow}
  \and Joint Institute for Nuclear Research, Joliot-Curie 6, Dubna, Moscow region, Russia, 141980 \label{jinr}
  \and SMU Physics, Box 0175 Dallas, TX 75275-0175, United States of
  America \label{smu}
  \and Hamburg University, II. Institute for Theoretical Physics, 
  Luruper Chaussee 149, D-22761 Hamburg, Germany \label{hamburg}
}

\date{Received: date / Accepted: date}
\headnote{XXXXXXXXXXXXXXXXXXXXXXXXXX}
\titlerunning{XXXXXXXXXXXXXX}
\authorrunning{Short form of author list} %
\titlerunning{Short form of title} %

\maketitle

\begin{abstract}

  We study charm production in charged-current deep-inelastic
  scattering (DIS) using the \xfitter framework.  Recent results from
  the LHC have focused renewed attention on the determination of the
  strange-quark parton distribution function (PDF), and the DIS charm
  process provides important complementary constraints on this
  quantity.  We examine the current PDF uncertainty and use LHeC
  pseudodata to estimate the potential improvement from this proposed
  facility.  As \xfitter implements both fixed-flavor- and
  variable-flavor-number schemes, we can compare the impact of these
  different theoretical choices; this highlights some interesting
  aspects of multi-scale calculations.
We find that the high-statistics LHeC data covering a
wide kinematic range could substantially reduce the strange PDF uncertainty.
\let\thefootnote\relax\footnotetext{Correspondence: {\tt olness@smu.edu}}
\end{abstract}

\setcounter{tocdepth}{4}
\newpage
\tableofcontents
\newpage
\section{Introduction} \label{sec:intro}

The deep-inelastic-scattering (DIS) experiments traditionally have
provided important tests of perturbative QCD (pQCD) and are essential
to precisely determine the parton distribution functions (PDFs) of the
nucleon.
In addition to the numerous dedicated fixed-target DIS experiments
that have been performed so far, the HERA accelerator used colliding
beams of leptons (electrons and positrons) and protons to investigate
the nucleon structure.
The broad kinematic coverage of the HERA charge-current (CC) and
neutral-current (NC) DIS data in terms of the negative virtuality
$Q^2$ of the exchanged vector boson and the Bjorken variable \xbj is
such that these data have significant impact on the determinations of
the PDFs~\cite{Abdolmaleki:2018jln,Abramowicz:2015mha,Gao:2017yyd,Alekhin:2017kpj,Ball:2017nwa}.

In the Standard Model (SM), the charm quark plays an important role in
the investigation of the nucleon
structure~\cite{Behnke:2015qja,Zenaiev:2016kfl,Abdolmaleki:2017wlg,Abdolmaleki:2019tbb}.
In the NC case, the photon-gluon fusion process for charm production
was calculated at ${\cal O}(\alpha_s^2)$ with the full
heavy-quark mass dependence included in the DIS hard cross
sections~\cite{Laenen:1992zk,Laenen:1992xs}.
The heavy-quark mass effects in the CC process have been calculated to
${\cal O}(\alpha_s)$ in
Refs.~\cite{Gottschalk:1980rv,Gluck:1997sj,Blumlein:2011zu,Buza:1997mg,Blumlein:2014fqa},
and the recent work of Ref.~\cite{Berger:2016inr} provides results up
to ${\cal O}(\alpha_s^2)$. The large-$Q^2$ contributions of heavy flavors to the
$xF_3$ structure function had already been computed in
Ref.~\cite{Behring:2015roa}.
In many of the posited models which extend the SM, the coupling to
``new physics'' is proportional to the particle mass; hence, the heavy
quarks will have an enhanced coupling and provide an optimal testing
ground for these searches.

Heavy quarks also play a critical role in helping us fully
characterize the SM, and the charm quark is especially useful in this
respect as it can provide us direct access to the strange-sea quark
distribution.
The strange sea has been extensively investigated in a number of
experiments including  the associated production of a $W$~boson
with a charm-jet final state, which (at LO)
arises from strange--gluon initial states~\cite{Aaltonen:2007dm,Abazov:2008qz,Abazov:2014fka,Chatrchyan:2013uja,Aad:2014xca,Sirunyan:2018hde,AbdulKhalek:2019mps,Kramer:1995yp}.
Additionally, charm production in neutrino/antineutrino-nucleon DIS
has been studied by a number of experiments including:
CCFR~\cite{Seligman:1997mc},
NuTeV~\cite{Tzanov:2005kr},
CHORUS~\cite{Onengut:2005kv},
CDHSW~\cite{Berge:1989hr}
and
NOMAD~\cite{Samoylov:2013xoa}.
With a sign-selected beam ($\nu/\bar{\nu}$), these experiments can
separately probe the strange $s(x)$ and anti-strange $\bar{s}(x)$
distributions. While the neutrino DIS experiments provide
detailed information on the shape of the strange distribution, the
normalization is a challenge, as that is tied to the beam flux.
Separately, the HERMES collaboration used charged-lepton DIS
production of charged kaons to provide a complementary extraction of
$s(x)+ \bar{s}(x)$ at LO~\cite{Airapetian:2008qf}.
Recently, charm production in CC DIS was measured for the first time in $e^{\pm}p$ collisions by ZEUS~\cite{Abt:2019ngj}.

Additionally, charm production mediated by electroweak gauge boson at
hadron colliders provides important information on the strange- and 
charm-quark distributions, and is complementary to the DIS final-state
charm-quark experiments~\cite{Lai:2007dq}.
The Tevatron measured the charm-quark cross section in association
with a $W$ boson at CDF~\cite{Aaltonen:2007dm,Aaltonen:2012wn,Aaltonen:2015aka} and
D0~\cite{Abazov:2008qz}, but these results were limited by low
statistics.

In lieu of significant experimental constraints, many global QCD
analyses tie the strange distribution to the light-sea quarks via the
relation $s= \bar{s}=r_s \, \bar{d}$.
While in principle $r_s$ depends on both \xbj and $Q^2$, it is often
set to a fixed value~\cite{Kretzer:2003it, Martin:2004ir}.

Using inclusive leptonic decays of $W$ and $Z$ bosons, the ATLAS
experiment
has obtained a value of $r_s=1.19 \pm 0.16$  at
$x= 0.023$ and $Q^2_0 = 1.9$~GeV$^2$~\cite{Aaboud:2016btc}.
Additionally, using the cross section ratio for $W^\pm +c$ final
states they also find a comparably large value for $r_s$~\cite{Aad:2014xca}.
In contrast, CMS results generally prefer lower $r_s$
values~\cite{Chatrchyan:2013uja,Sirunyan:2018hde}.  However, a recent analysis using
both ATLAS and CMS data suggests that the LHC data support
unsuppressed strangeness in the proton. While the result is dominated
by ATLAS, this is not in contradiction with the CMS
data~\cite{Cooper-Sarkar:2018ufj,Aaboud:2016btc,Aad:2014xca,Chatrchyan:2013uja}.

Looking to the future, it is clearly important to reduce the
uncertainty of the strange-quark PDF as we strive to make increasingly
precise tests of the SM and search for what might lie beyond.
The proposed Large Hadron Electron Collider (LHeC) program has the
ability to provide high statistics measurements of electrons on both
protons and nuclei across a broad kinematic range to address many of
these outstanding questions.

In this investigation, we make use of the \xfitter
tools~\cite{Alekhin:2014irh} (version 2.0.0) to study the present constraints on the
strange-quark PDFs, and then use LHeC pseudodata~\cite{AbelleiraFernandez:2012cc} to infer how these
might improve. Furthermore, as \xfitter implements both fixed-flavor-
and variable-flavor-number schemes, we can examine the impact of these
different theoretical choices.

This paper is organized as follows.
In Sect.~\ref{sec:thpred} we outline the theoretical details of the
different heavy-flavor schemes.
In Sect.~\ref{sec:thpred-comparison} we compare the theoretical
predictions of the different schemes across the kinematic range, and
examine the individual partonic contributions.
In Sect.~\ref{sec:PDF} we study the impact of the LHeC pseudodata on
the PDFs using a profiling technique.
In Sect.~\ref{sec:discuss} we provide some discussion and summarize the
results.
Finally, in \ref{sec:appendix} we discuss some of the more subtle
theoretical issues that we encounter at higher orders.

\goodbreak
\section{Theoretical predictions for CC charm production at the LHeC} \label{sec:thpred}

The proposed Large Hadron Electron Collider
(LHeC)~\cite{AbelleiraFernandez:2012cc} would collide a newly built
electron beam with the LHC hadron beam at a center of mass energy of
$\sqrt{s} = \sqrt{4 E_e E_p}$; thus the 7~TeV proton beam on a 60~GeV
electron beam provide $\sqrt{s}\sim 1.3$~TeV.
Compared to HERA, the LHeC extends the covered kinematic range by an
order of magnitude in both \xbj and $Q^2$ with a nominal design
luminosity of $10^{33} cm^{-2} s^{-1}$.

Theoretical predictions are calculated for electroweak charged-current
(CC) charm production in $ep$ collisions at the LHeC at centre-of-mass
energy $\sqrt{s} = 1.3$ TeV, using a variety of heavy-flavor
schemes. The predictions are provided for unpolarized beams in the
kinematic range $100 < Q^2 < 100000$~GeV$^2$, $0.0001 < \xbj < 0.25$.
They are calculated as reduced cross sections at different $Q^2$, \xbj
and inelasticity ($y$) points. 
The covered $y$ range is $0.0024 < y < 0.76$.

Experimentally, however, not charm quarks but charmed hadrons (or
rather their decay products) are registered in the detectors.
Therefore, extrapolation to the inclusive charm-production cross
section has to be carried out in a model-dependent way. Furthermore,
CC production of charm quarks in the final state can happen via both
electroweak and QCD processes.  The former leads to an odd number of
charm quarks in the final state with the $W$ boson having the same
electric charge as the sum of the electric charges of final-state
charm quarks, while the latter creates an even number of charm quarks
with total electric charge equal to zero. If the electric charge of
the tagged charm quark can be accessed experimentally
(\textit{e.g.}~when reconstructing $D$ mesons), the QCD contribution
can be subtracted by taking the difference of the yields in the events
with odd and even numbers of charm quarks, otherwise the QCD
contribution can be estimated only in a model-dependent way.

The CC charm process directly depends on the CKM matrix~\cite{Tanabashi:2018oca}.
Here, the CKM
matrix elements $V_{cd}$ and $V_{cs}$ are particularly relevant
and we use the values $V_{cd} = 0.2252$ and $V_{cs} = 0.9734$.
Three different heavy-flavor schemes are employed, all including a full
treatment of charm-mass effects up to NLO,
\textit{i.e.}~${\cal O}(\alpha_s)$;
in the following we describe them in detail for the  particular
application to CC electron-proton reactions.

\subsection{The heavy-flavor schemes} \label{sec:schemes}

The standard ``A'' variant of the fixed-flavor number scheme (FFNS),
which we identify as {\bf FFNS~A}, uses three light flavors in both
PDFs and $\alpha_s$ evolution for all scales, while heavy flavors
(here, charm) are produced exclusively in the matrix-element part of
the calculation. This scheme has been used for the PDF determinations
and cross section predictions of the ABM(P)
group~\cite{Alekhin:2012ig,Alekhin:2014sya,Alekhin:2017kpj,Alekhin:2018pai},
as well as in the FF3A variant of the HERAPDF
analysis~\cite{Abramowicz:2015mha}, and implemented in \xfitter through
the OPENQCDRAD package~\cite{openqcdrad}.

Next, the ``B'' variant of the FFNS ({\bf \ffnsb}), known as the
``mixed'' or ``hybrid'' scheme~\cite{Behnke:2015qja} is also used. In
this scheme, the number of active flavors is still fixed to three in
the PDFs, relying exclusively on ${\cal O}(\alpha_s)$ fully massive
matrix elements for charm production, while the number of flavors is
allowed to vary in the virtual corrections of the $\alpha_s$
evolution. Corrections to the $\alpha_s$ evolution involving
heavy-flavor loops are thus included and resummed to all orders, while
no resummation is applied to other higher order corrections. This
procedure will catch a fraction of the ``large logs'' which might
spoil the fixed-flavor scheme convergence at very high scales, and is
possible since the masses of the charm and beauty quarks provide
natural cutoffs for infrared and collinear divergences.  This scheme
was used in the HERAPDF FF3B variant~\cite{Abramowicz:2015mha} and in
applications of the HVQDIS program~\cite{,Behnke:2015qja}.  In
general, the transition from the \ffns to the \ffnsb requires a
readjustment of the treatment of matrix elements involving
heavy-flavor loops. In the specific case of CC production, no such
loops occur up to NLO (at NNLO they do), so that the same matrix
elements can be used for both schemes; thus the only difference is in
the $\alpha_s$ evolution.

Finally, for the variable-flavor-number scheme (VFNS) we use the ``B''
variant of the fixed-order-next-to-leading-log scheme ({\bf FONLL-B})~\cite{Forte:2010ta}
which combines the NLO ${\cal O}(\alpha_s)$ massive matrix elements of
the FFNS with the ${\cal O}(\alpha_s)$ massless results of the
zero-mass variable-flavor-number scheme (ZM-VFNS), allowing the number
of active flavors to vary with scale, and all-order next-to-leading
log resummation of (massless) terms beyond NLO.  It thus explicitly
includes charm and beauty both in the PDFs and in the evolution of the
strong coupling constant.  Whenever terms would be double-counted in
the merging of the two schemes, the massless terms are eliminated in
favour of the massive ones. The FONLL scheme is commonly used by the
NNPDF group~\cite{Ball:2017nwa} and implemented in \xfitter through
the APFEL package~\cite{Bertone:2013vaa}.

\goodbreak
\newpage
In summary, the schemes used are:
\begin{itemize}
  \setlength\itemsep{1em}

\item[$\bullet$] {\bf \ffns :} a NLO FFNS with $n_f = 3$ at all
  scales, used with the \abmp~\cite{Alekhin:2018pai} or
  \ffthreea~\cite{Abramowicz:2015mha} NLO PDF sets.

\item[$\bullet$] {\bf \ffnsb :} a NLO FFNS with $n_f = 3$ for the PDFs
  and variable $n_f$ for $\alpha_s$, used with the
  \ffthreeb~\cite{Abramowicz:2015mha} NLO PDF set.

\item[$\bullet$] {\bf FONLL-B :} a VFNS used with the \nnpdf NLO PDF
  set~\cite{Ball:2017nwa}.
\end{itemize}
The PDF sets are available via the \lhapdf interface (version
6.1.5)~\cite{Buckley:2014ana}.
Note that we use the PDFs directly from the LHAPDF tables without any additional fitting,
and we use the default values for $\alpha_s(M_Z)$ and the quark masses $m_{c,b}$.
  We've chosen this collection of PDFs as they are consistently extracted in the appropriate VFNS/FFNS schemes.
  Comparing the PDFs at the inital evolution scale,
  we find a  typical variation of $\lesssim 10\%$ for the quarks, and  a bit larger for the gluon (which enters at NLO).
  These differences will lead to a small shift in the ratio plots,
  but will not affect the general features which are the focus of this paper,
  {\it c.f.}, Fig.~\ref{fig:thpred-q2}.

\subsection{The reduced cross section}\label{sec-redsigma}

The reduced CC charm-production cross sections can be expressed as a
linear combination of structure functions:
\begin{equation}
  \sigma^{\pm}_{\text{charm,CC}} = \frac12\left(Y_{+}F_2^{\pm} \mp
    Y_{-}xF_3^{\pm} - y^2F_L^{\pm}\right)\,,
\end{equation}
with 
\begin{equation}
  Y_{\pm} = 1 \pm (1-y)^2 \,.
\end{equation}
In the quark-parton model,
when we neglect the gluons, the structure
functions become:
\begin{equation}
\begin{split}
    F_2^{+} &= xD + x\overline{U}, \\
    F_2^{-} &= xU + x\overline{D},\\
    F_L &= 0,\\
    xF_3^{+} &= xD - x\overline{U}, \\
    xF_3^{-} &= xU - x\overline{D}.
\end{split}
\end{equation}
The terms $xU$, $xD$, $x\overline{U}$ and $x\overline{D}$ denote the
sum of parton distributions for up-type and down-type quarks and
anti-quarks, respectively.%
\footnote{In these expressions, we neglect the CKM mixing for brevity, but it is fully contained in the calculations.}
The $\pm$ superscript on $\sigma$ and $F$ corresponds to the sign of $W^\pm$. 
Below the $b$-quark mass threshold, these
sums are related to the quark distributions as follows:
\begin{equation}
\begin{split}
 xU &= xu + xc , \\
 x\overline{U} &= x\overline{u} + x\overline{c} , \\
 xD &= xd + xs , \\
 x\overline{D} &= x\overline{d} + x\overline{s}.
\end{split}
\end{equation}
In the FFNS the charm-quark densities are zero. In the phase-space
corners $y \to 0$ and $y \to 1$ and using the same quark-parton model approximation, we have the following asymptotic
relations:
\begin{equation}
\begin{split}
 y \to 0: \quad \sigma^{\pm}_{\text{charm,CC}} &= F_2^{\pm} = xD(x\overline{D}) + x\overline{U}(xU), \\[10pt]
 y \to 1: \quad \sigma^{\pm}_{\text{charm,CC}} &= \frac12(F_2^{\pm} \mp xF_3^{\pm}) = x\overline{U} (xU).
\label{eq:y01}
\end{split}
\end{equation}
Thus the contribution from the strange-quark PDF is suppressed at high $y$.

\subsection{\xfitter implementation}

All calculations are interfaced in \xfitter and available with
$\overline{\mbox{MS}}$ heavy-quark masses. The reference value of the
$\overline{\mbox{MS}}$ charm mass is set to $m_c(m_c) = 1.27$
GeV~\cite{Tanabashi:2018oca}, and $\alpha_s$ is set to the value used
for the corresponding PDF extraction: $\alpha_s(M_Z) = 0.1191$ for
\abmp and $\alpha_s(M_Z) = 0.118$ for \nnpdf.  The renormalization and
factorization scales are chosen to be
$\mu_\mathrm{r}^2 = \mu_\mathrm{f}^2 = Q^2$.

To estimate theoretical scale uncertainties, $\mu_\mathrm{r}$ and
$\mu_\mathrm{f}$ are simultaneously varied up and down by a factor of 
two. In the case of the \fonll calculations, also the independent
$\mu_r$ and $\mu_f$ variations are checked. Furthermore, the PDF
uncertainties are propagated to the calculated theoretical
predictions, while the uncertainties arising from varying the charm
mass $m_c(m_c) = 1.27 \pm 0.03$~GeV by one standard deviation are
smaller than $1\%$ and therefore neglected. In the \fonll scheme, as a
cross check, the calculation was performed with the pole charm mass
$m_c^{\text{pole}} = 1.51$ GeV which is consistent with the conditions
of the \nnpdf extraction~\cite{Ball:2017nwa}. The obtained theoretical
predictions differ from the ones calculated with $m_c(m_c) = 1.27$ GeV
by less than $1\%$. The total theoretical uncertainties are obtained
by adding in quadrature scale and PDF uncertainties.

\newpage
\section{Comparison of theoretical predictions}
\label{sec:thpred-comparison}

We now provide some numerical comparisons of the heavy-flavor schemes
using their separate input conditions and associated PDF sets.
Caution is necessary in these comparisons as the PDF sets are
extracted with different input assumptions, data sets, and tolerance
criteria; this is, in part, why we shall separately display the
$\mu_\mathrm{r}$, $\mu_\mathrm{f}$  and PDF
uncertainties in the following.

\subsection{Comparison of theoretical predictions in the \ffns and \fonll schemes}
\label{sec:compareI}

\begin{figure}
    \centering
    \centering{{\includegraphics[width=0.50\textwidth]{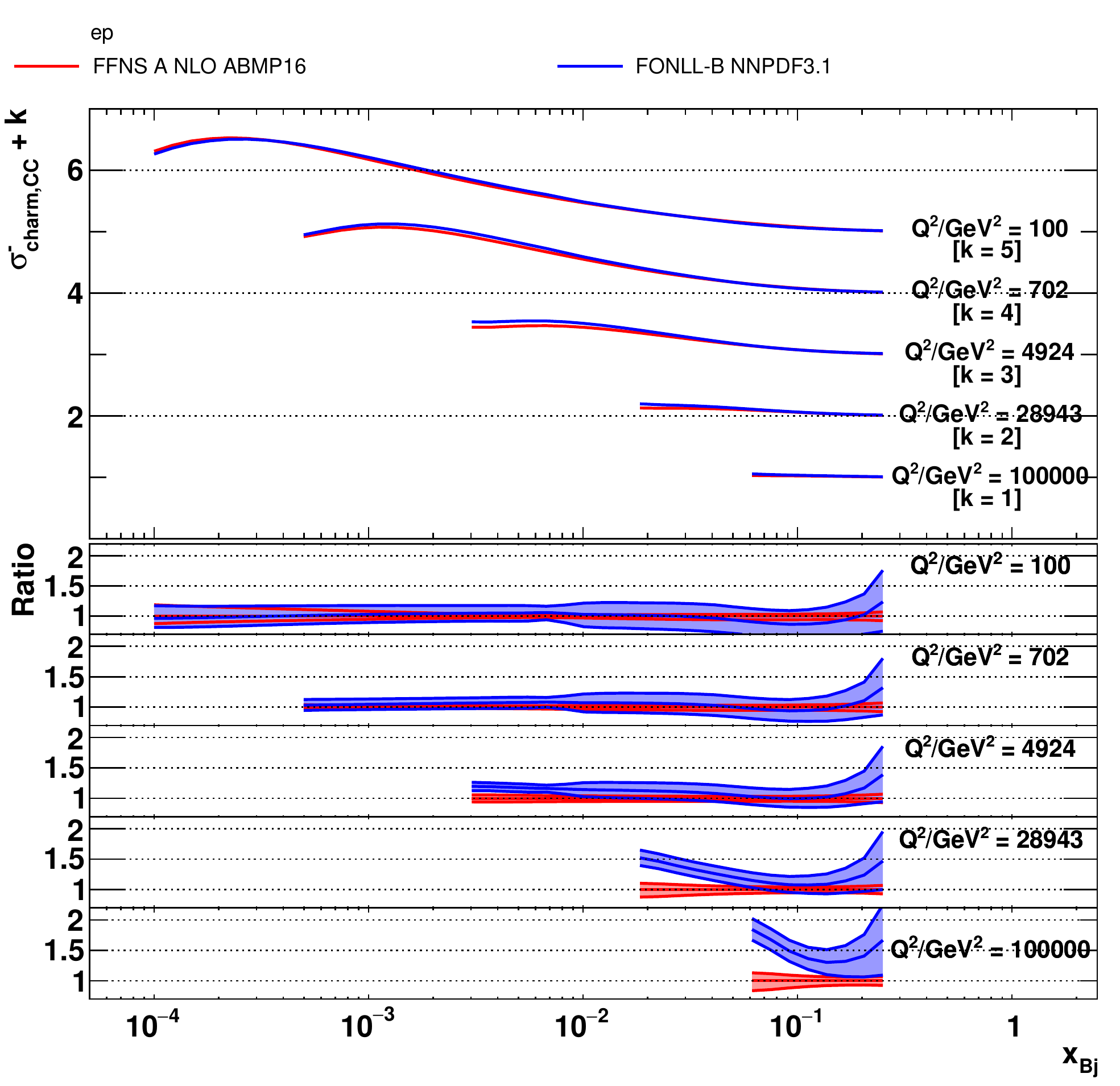}}}
    \caption{The theoretical predictions with their total
      uncertainties for charm CC production at the LHeC as a function
      of \xbj for different values of $Q^2$ calculated in the \ffns
      and \fonll schemes. The bottom panels display the theoretical
      predictions normalized to the nominal values of the \ffns
      predictions.}
    \label{fig:thpred-x}
\end{figure}

\begin{figure}
    \centering
    \centering{{\includegraphics[width=0.50\textwidth]{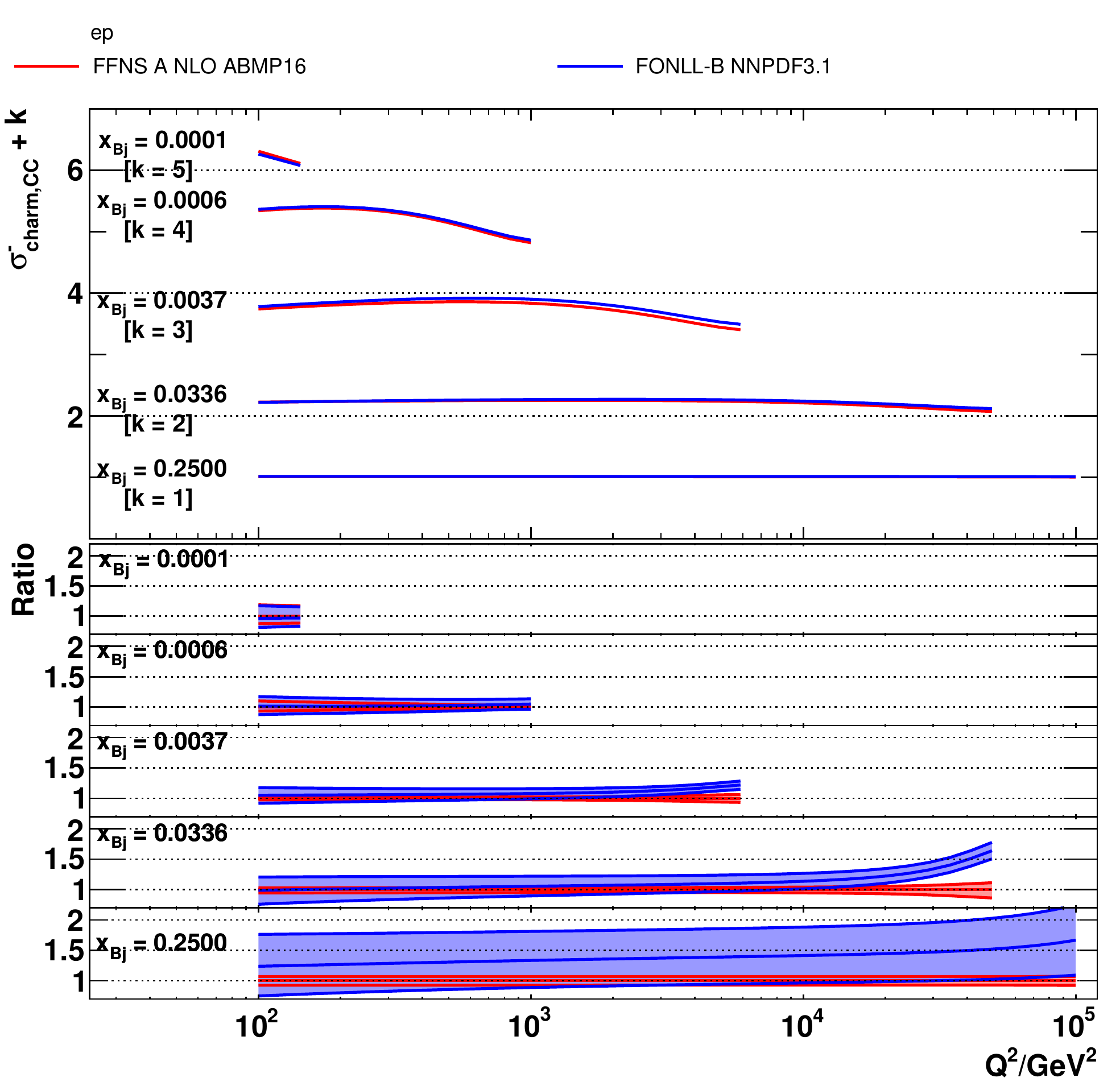}}}
    \caption{The theoretical predictions with their total
      uncertainties for charm CC production at the LHeC as a function
      of $Q^2$ for different values of \xbj calculated in the \ffns
      and \fonll schemes. The bottom panels display the theoretical
      predictions normalized to the nominal values of the \ffns
      predictions.}
    \label{fig:thpred-q2}
\end{figure}

\begin{figure}
    \centering
    \centering{{\includegraphics[width=0.50\textwidth]{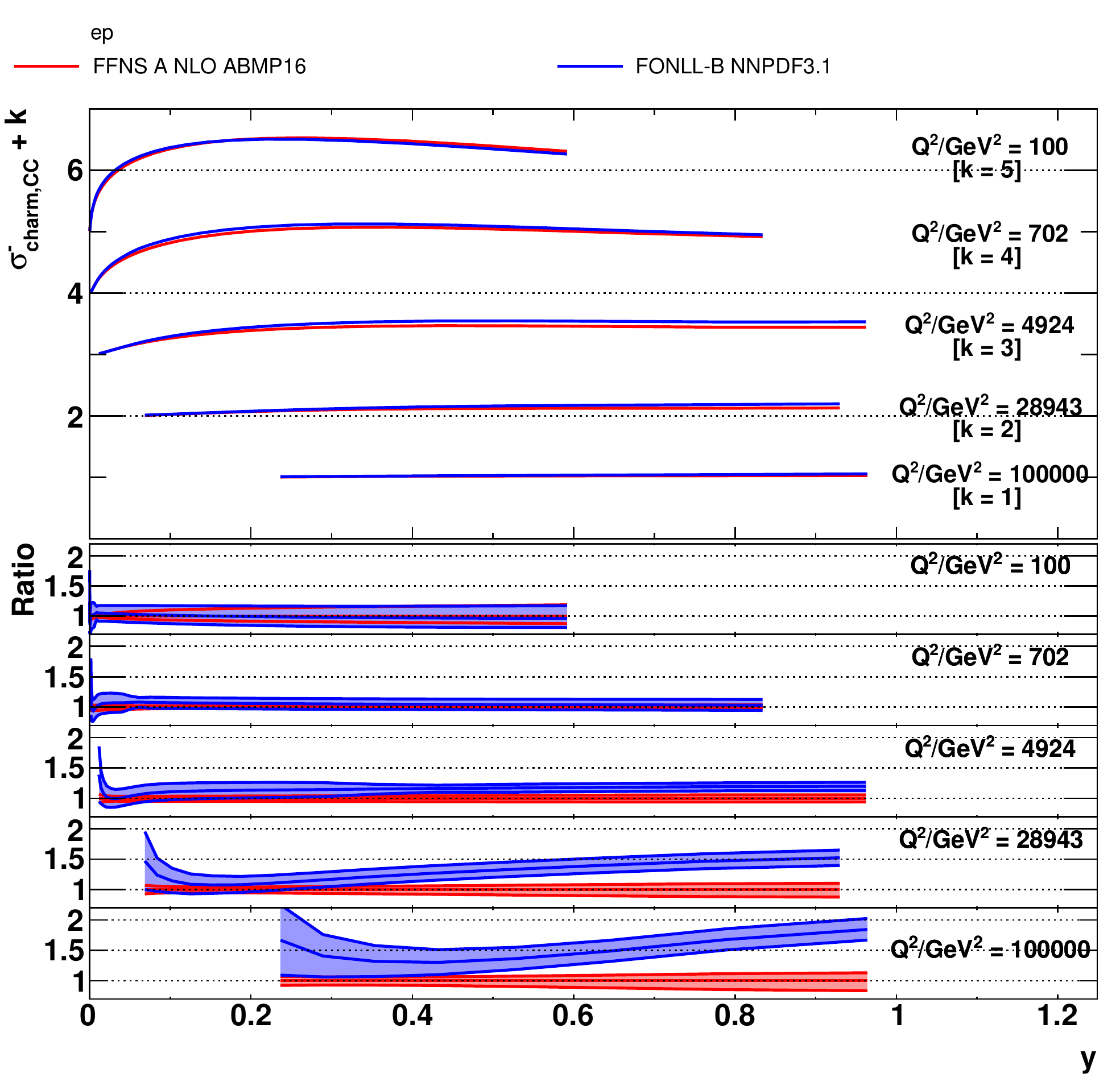}}}
    \caption{The theoretical predictions with their total
      uncertainties for charm CC production at the LHeC as a function
      of $y$ for different values of $Q^2$ calculated in the \ffns and
      \fonll schemes. The bottom panels display the theoretical
      predictions normalized to the nominal values of the \ffns
      predictions.}
    \label{fig:thpred-y}
\end{figure}

Figs.~\ref{fig:thpred-x}, \ref{fig:thpred-q2} and~\ref{fig:thpred-y}
show theoretical predictions for the \ffns and \fonll schemes 
calculated as described in the previous sections 
with their total uncertainties.
The \ffns and \fonll results agree reasonably well within
uncertainties in the bulk of the phase space. However, in phase-space
corners such as $Q^2 \gtrsim 10000$ GeV$^2$ or small $y$
the predictions in the two schemes differ by more than $50\%$, exceeding
the theoretical uncertainties.

\begin{figure*}
    \centering
    \centering{{\includegraphics[width=0.49\textwidth]{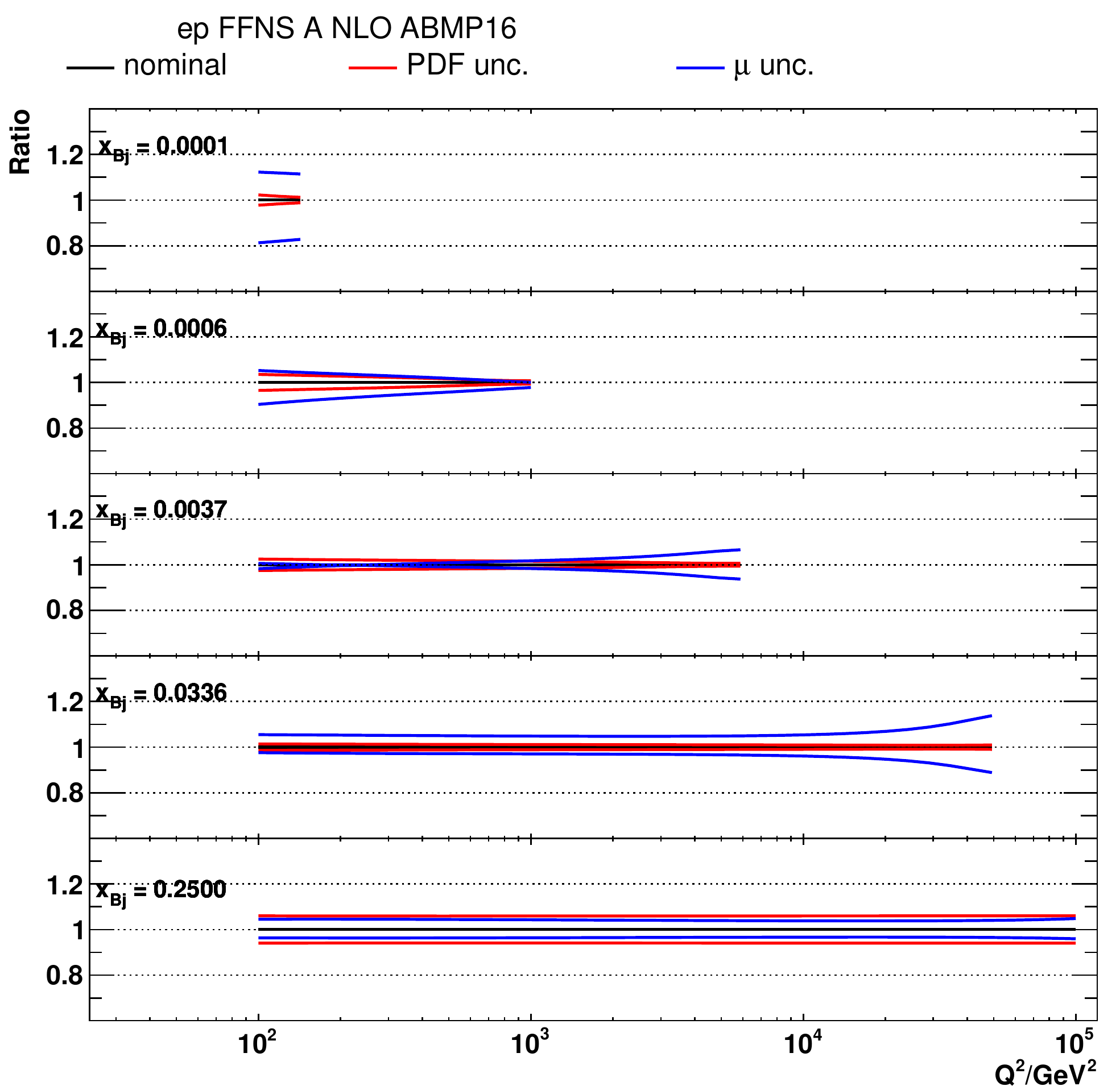}}}
    \centering{{\includegraphics[width=0.49\textwidth]{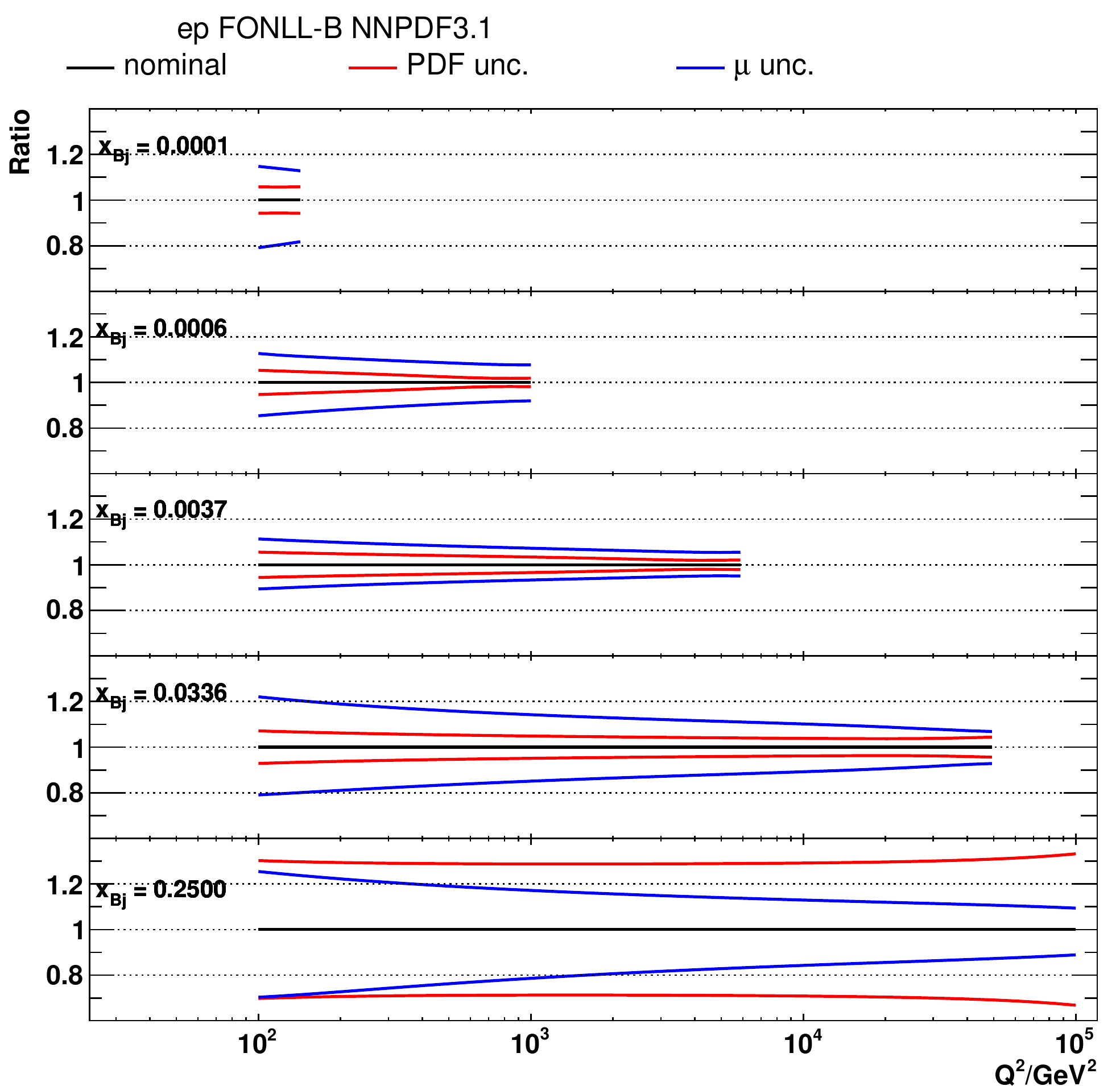}}}
    \caption{Relative theoretical uncertainties of charm CC
      predictions for the LHeC as a function of $Q^2$ for different
      values of \xbj calculated in the \ffns and \fonll schemes. The
      PDF and scale uncertainties are shown separately.}
    \label{fig:thpred-q2-unc}
\end{figure*}

To examine these differences further, in Fig.~\ref{fig:thpred-q2-unc}
we separately compute PDF and scale uncertainties (setting
$\mu_\mathrm{r}=\mu_\mathrm{f}=\mu$) of the charm CC cross section as
a function of $Q^2$ for different values of \xbj calculated in the
\ffns and \fonll scheme.

Comparing the two schemes, the larger variation of the \fonll scheme
reflects the larger PDF uncertainty of the underlying PDF sets used:
ABMP16 for \ffns and NNPDF3.1 for \fonll.
This difference is most evident in Fig.~\ref{fig:thpred-q2-unc} which
specifically separates out the PDF uncertainty, and reflects the
independent inputs and assumptions used in the different PDF
extractions.

\begin{figure*}
    \centering
    \centering{{\includegraphics[width=0.49\textwidth]{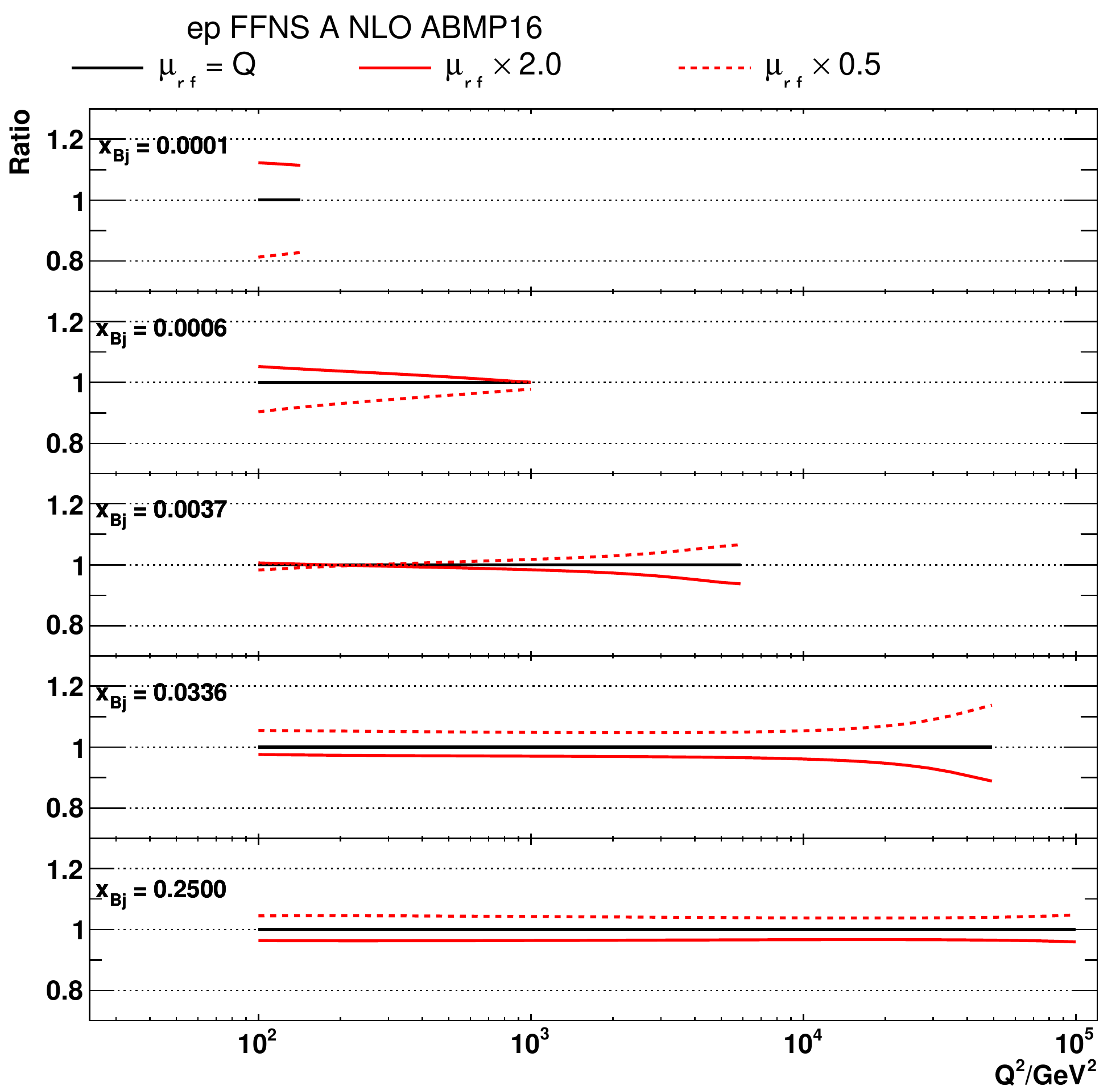}}}
    \centering{{\includegraphics[width=0.49\textwidth]{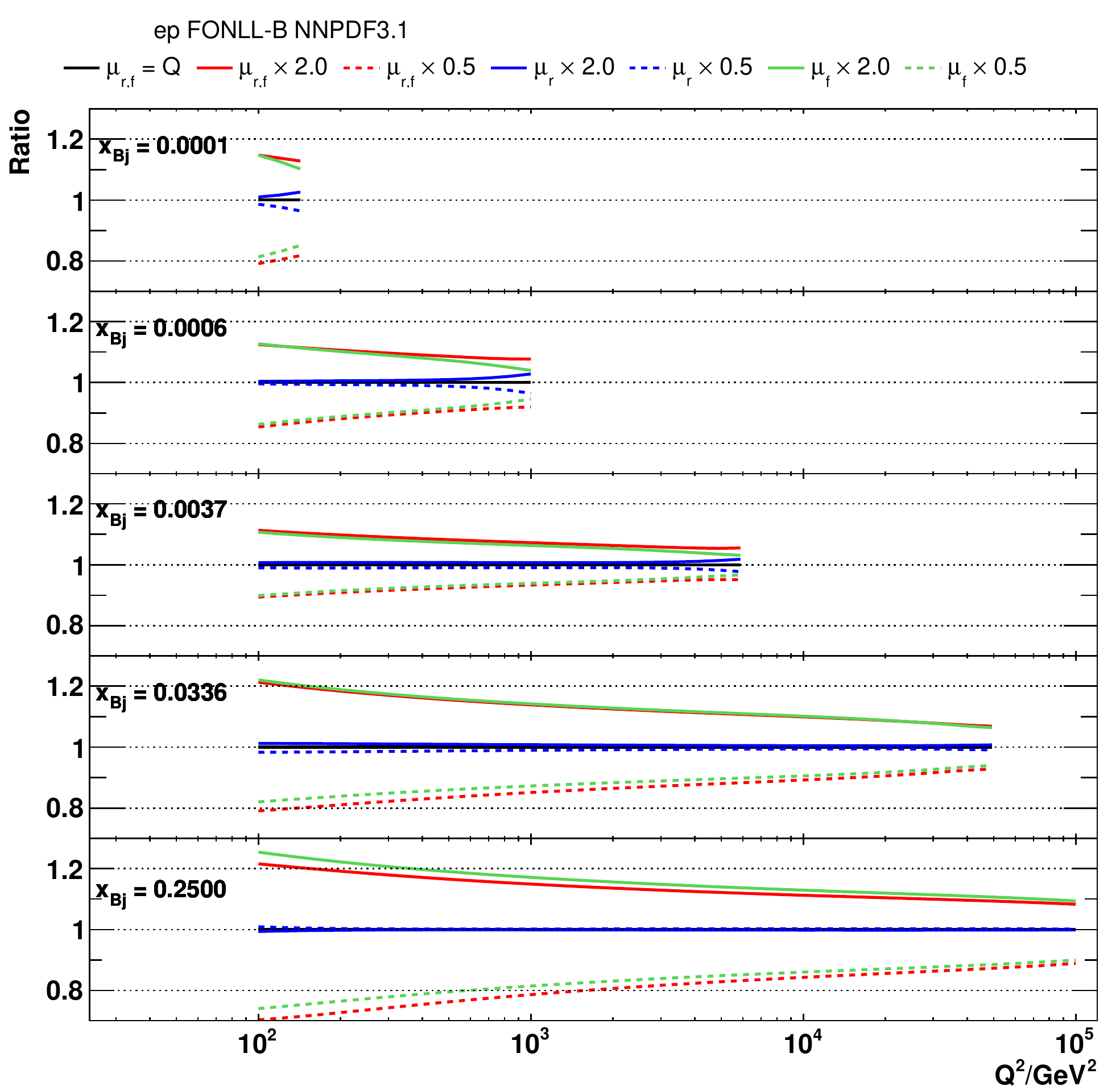}}}
    \caption{The impact of separate scale variations on charm CC
      predictions for the LHeC as a function of $Q^2$ for different
      values of \xbj calculated in the \ffns and \fonll schemes.}
    \label{fig:thpred-q2-varmu}
\end{figure*}

Examining the results of Fig.~\ref{fig:thpred-q2-unc}, we also observe
some other interesting features.
For both of the calculations, the PDF uncertainties are relatively stable
across the  $Q^2$ range for fixed \xbj, but tend to increase at larger \xbj values.
As is well known, in pQCD calculations the effect of scale variations
is indicative of the convergence of the series.
We observe that the scale uncertainties for the \fonll scheme
uniformly decrease with increasing $Q^2$.
For the \ffns scheme, the scale uncertainties decrease for small \xbj
values but increase with $Q^2$ at intermediate values of \xbj.
Additional details are shown in Fig.~\ref{fig:thpred-q2-varmu} where
we separately vary $\mu_\mathrm{r}$ and $\mu_\mathrm{f}$ for the
\fonll scheme. Here we note that the uncertainty associated to
$\mu_\mathrm{r}$ is very small and the total scale uncertainty is
dominated by the variations of $\mu_\mathrm{f}$ which is tied to the
PDFs, $f_i(x,\mu_\mathrm{f})$.
For the \ffns in \xfitter, it is not possible to separately vary
$\mu_\mathrm{r}$ and $\mu_\mathrm{f}$ in the current implementation,
so the separate uncertainties can only be inferred by comparison to
the \fonll case.

\begin{figure*}
  \centering
  \centering{{\includegraphics[width=0.49\textwidth]{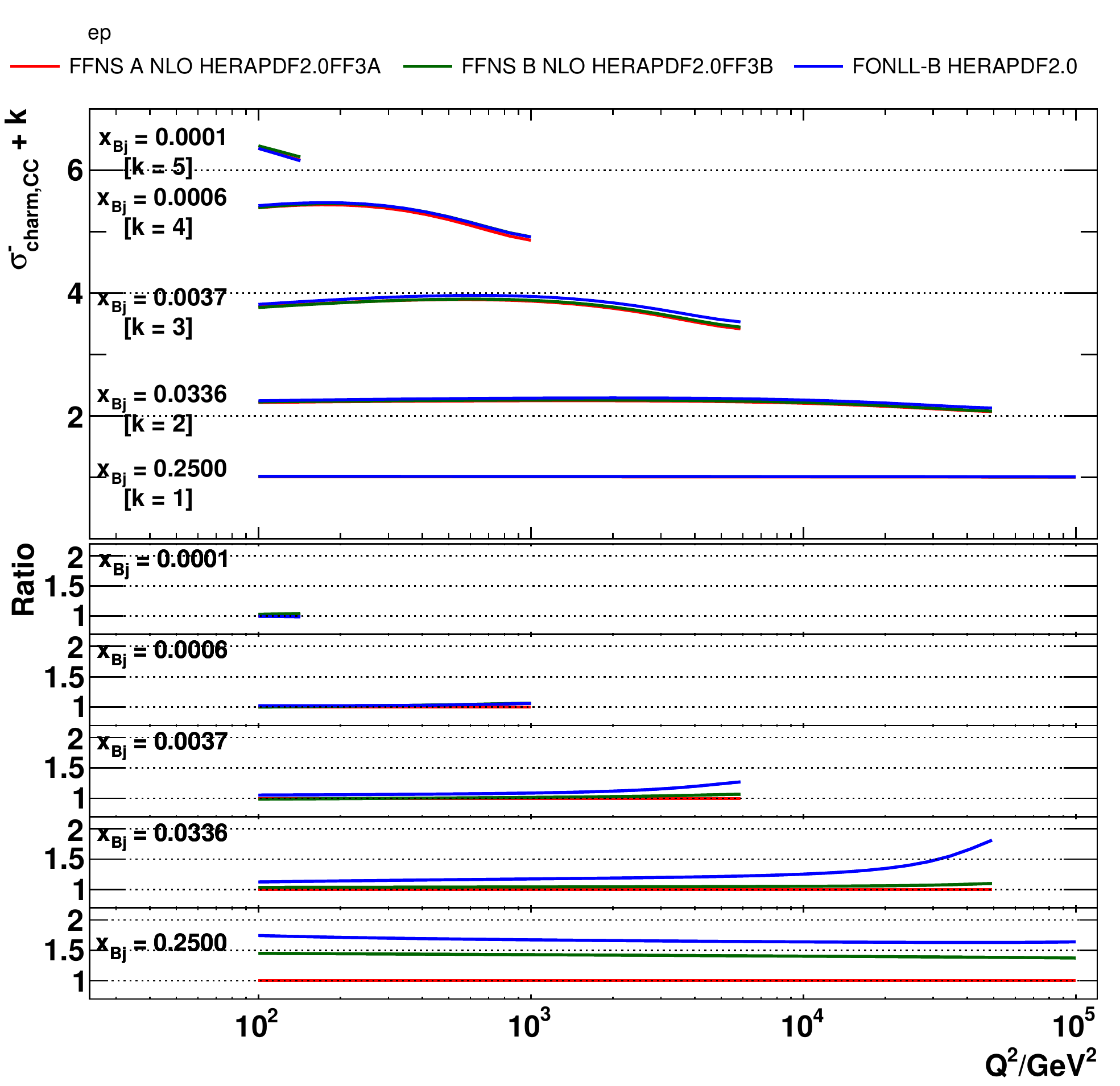}}}
  \centering{{\includegraphics[width=0.49\textwidth]{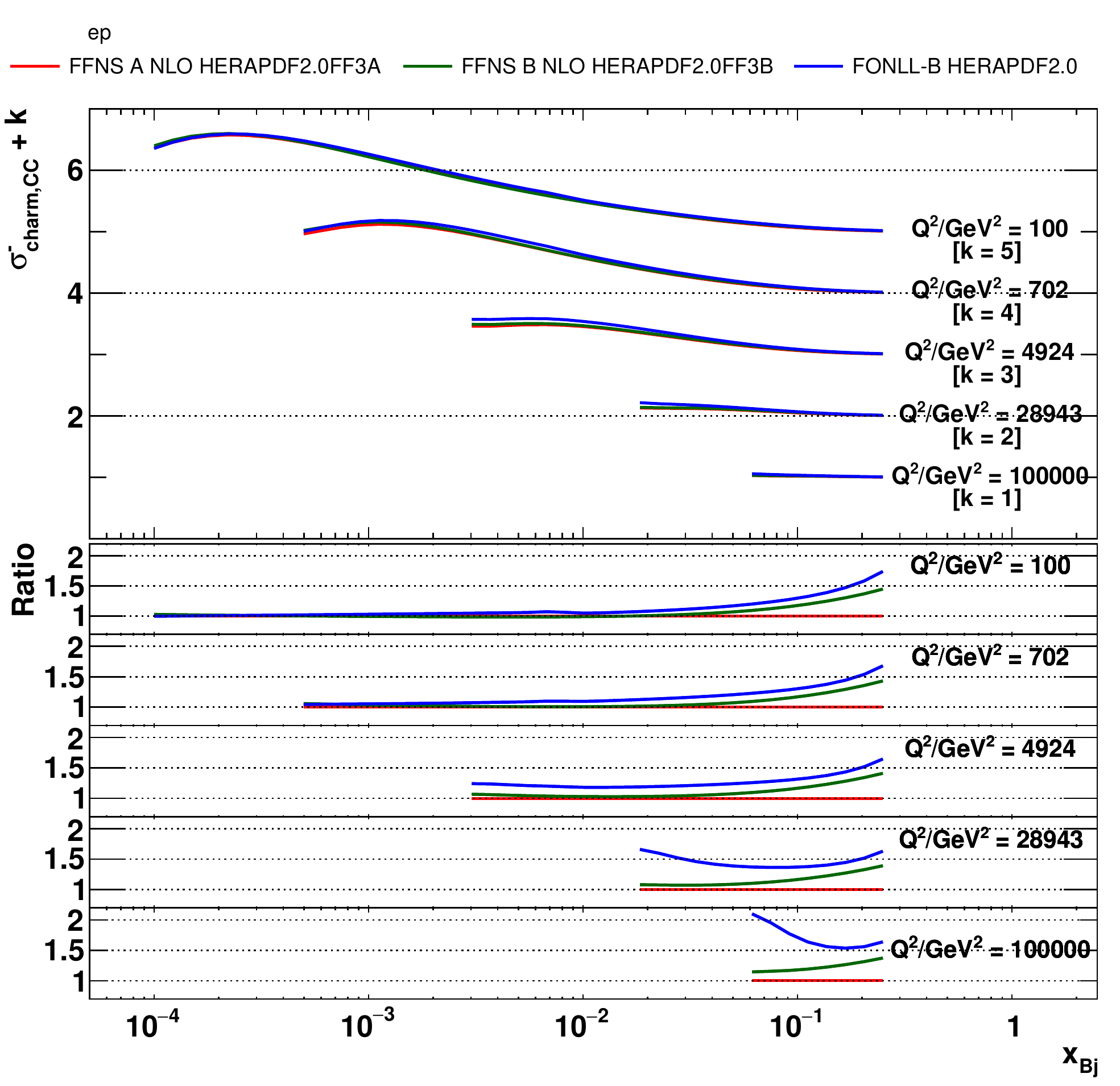}}}
  \centering{{\includegraphics[width=0.49\textwidth]{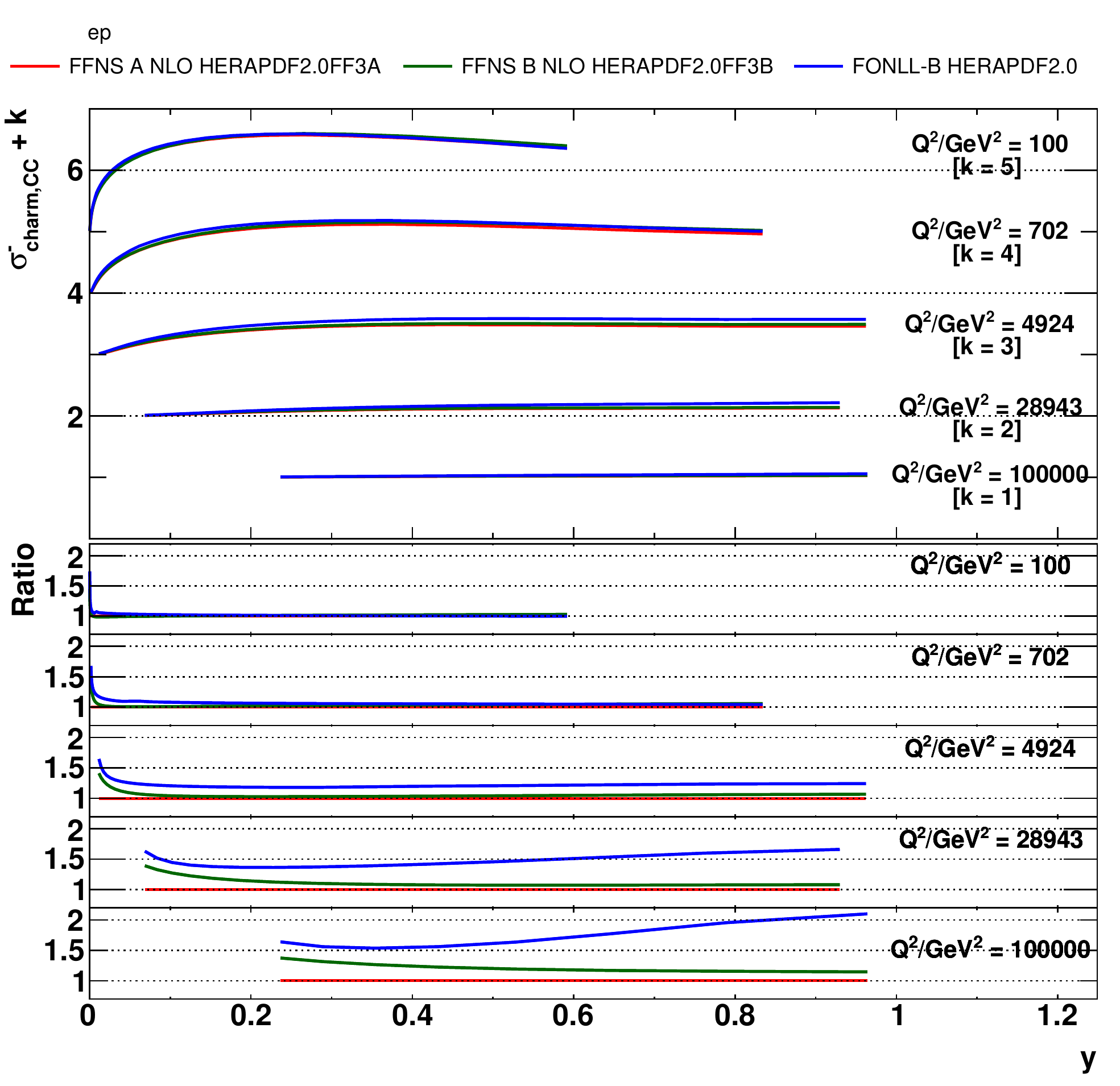}}}
  \caption{The theoretical predictions for CC charm production at the
    LHeC as a function of $Q^2$ ($\xbj$, $y$) for different values of
    $\xbj$ ($\xbj$, $Q^2$) obtained using the HERAPDF2.0 PDF sets in
    the \ffns, \ffnsb and \fonll schemes. The bottom panels display the
    theoretical predictions normalized to the nominal values of the
    \ffns predictions.}
  \label{fig:thpred-ff3abfonll}
\end{figure*}

\subsection{Additional comparisons}
\label{sec:compareII}

To further explore whether the differences between the two sets of
theoretical predictions are due to the different treatment of heavy
quarks or to the different PDF sets, theoretical calculations in \ffns
and \fonll are repeated with the HERAPDF2.0 PDF sets extracted from
the HERA DIS data~\cite{Abramowicz:2015mha}.
Predictions in the \ffnsb scheme are also produced using the \ffthreeb
PDF set and the \ffnsb matrix elements, which are equivalent to the
\ffns matrix elements at NLO for CC charm production. The results are
displayed in Fig.~\ref{fig:thpred-ff3abfonll}. The differences between
\ffns and \fonll are similar to those displayed in
Figs.~\ref{fig:thpred-x}-\ref{fig:thpred-y} and demonstrate that these
differences arise from the different treatment of the heavy quarks in
the two schemes. The \ffnsb predictions lie between the \ffns and
\fonll predictions, indicating that a large part of the difference
is due to the different treatment of heavy quarks in the running of
$\alpha_s$ at high \xbj or low $y$.

\begin{figure}
  \centering
  \centering{{\includegraphics[width=0.50\textwidth]{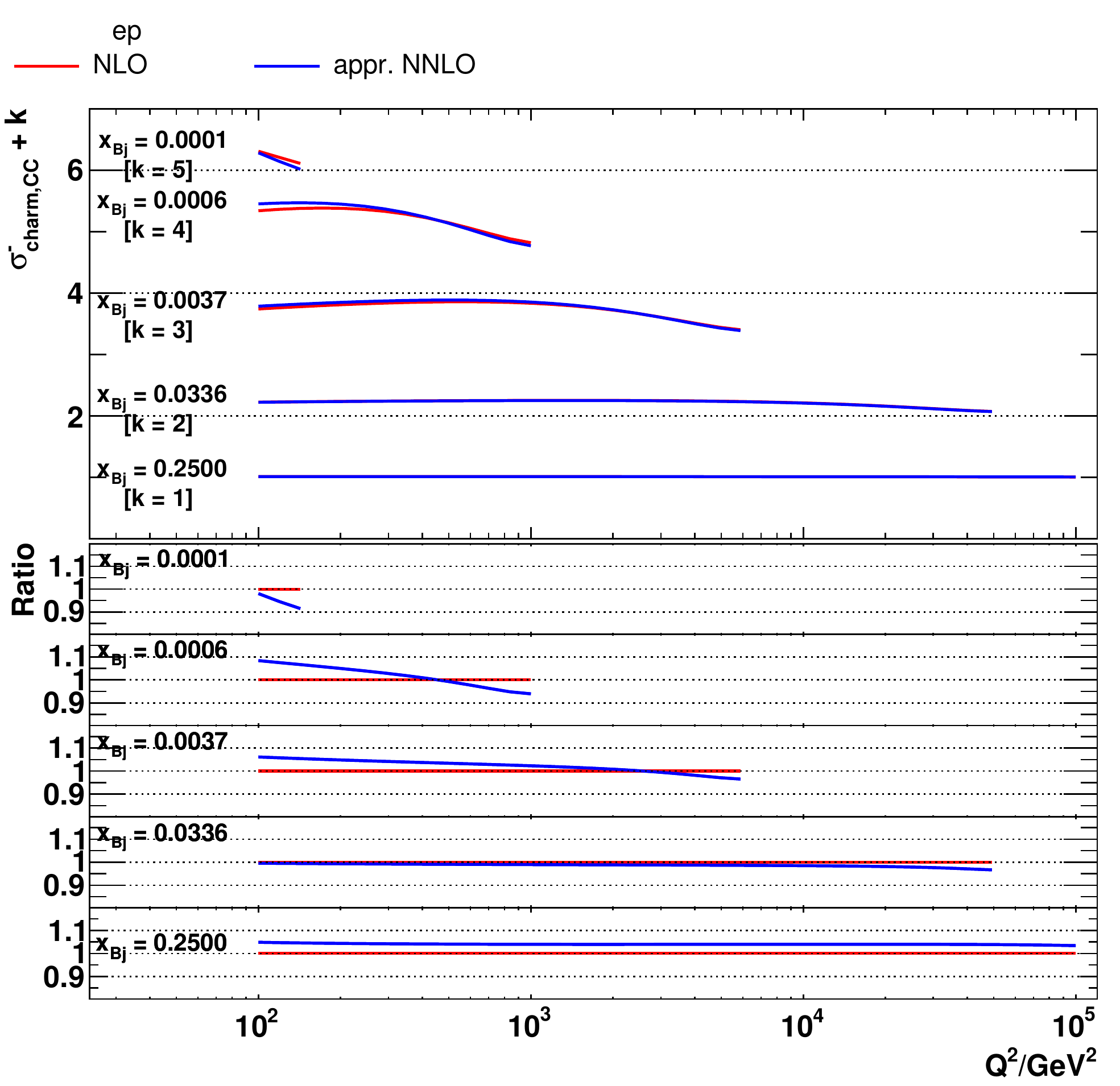}}}
  \caption{The theoretical predictions
    for CC charm production at the LHeC as a function of $Q^2$ for
    different values of \xbj calculated in the \ffns scheme at NLO and
    approximate NNLO. The bottom panels display the theoretical
    predictions normalized to the nominal values of the \ffns NLO
    predictions.}
  \label{fig:thpred-q2-nnlo}
\end{figure}

Furthermore, to investigate the impact of the NNLO corrections
available at $Q \gg m_c$ for the FFNS calculation, approximate NNLO
predictions are obtained using the \abmp NNLO PDF
set~\cite{Alekhin:2017kpj}. The results for the cross section as a
function of $Q^2$ for different values of \xbj are shown in
Fig.~\ref{fig:thpred-q2-nnlo}, where they are compared to the NLO
\ffns predictions from Fig.~\ref{fig:thpred-q2}.
The approximate NNLO corrections
do not exceed $\sim 10\%$ and thus cannot account for the differences
between the \ffns and \fonll theoretical predictions.
Similar results
are observed for the cross sections as functions of other kinematic
variables.

To better understand the differences between the FFNS and VFNS
calculations, Fig.~\ref{fig:thpred-ff3abfonll} is particularly
instructive.
We see that at low $Q^2$ the FFNS (\ffns and \ffnsb) and VFNS
(\fonll) results agree within uncertainties (as demonstrated in 
Fig.~\ref{fig:thpred-q2}). When the scale $\mu$ is
below the charm-threshold scale $\mu_c$ (typically taken to be equal
to $m_c(m_c)$) the charm PDFs vanish and the FFNS and VFNS reduce to
the same result.\footnote{Note that while the charm-threshold scale
  $\mu_c$ is commonly set to the charm quark mass $m_c(m_c)$, the
  choice of $\mu_c$ is arbitrary and amounts to a renormalization
  scheme choice~\cite{Bertone:2017ehk}.}
For increasing scales, the VFNS resums the $\alpha_s\ln(\mu^2/\mu_c^2)$
contributions via the DGLAP evolution equations and the FFNS and VFNS
will slowly diverge logarithmically. This behavior is observed in
\hbox{Fig.~\ref{fig:thpred-ff3abfonll}} and is consistent with the
characteristics demonstrated in Ref.~\cite{Kusina:2013slm}.

More precisely, Ref.~\cite{Kusina:2013slm} used a matched set of
$n_f=3$ and $n_f=5$ PDFs to study the impact of the scheme choice at
large scales. They found that the resummed contributions in the VFNS
yielded a larger cross section than the FFNS (the specific magnitude
was $x$-dependent), and that for $Q^2$ scales more than a few times
the quark mass, the differences due to scheme choice exceeded the
differences due to (estimated) higher-order
contributions.
Thus, we have identified the source of the scheme differences at large $Q^2$.

The  source of the scheme differences at large $\xbj$ is a bit more subtle. 
The VFNS includes a resummation of higher-order logarithms of the form $\alpha_s \ln(\mu^2/\mu_c^2 )$.
In Fig.~\ref{fig:acot} of the Appendix we display the separate contributions of the VFNS for a choice of
$\{\xbj,Q^2\}$; the difference  between the LO and SUB curves is indicative of the
additional contribution of the resummed logarithms.
This contribution depends on the particular $\xbj$ value as well as $Q^2$
({\it c.f.,} Fig.~11 of Ref.~\cite{Kusina:2013slm}).
Thus, it is a balance between the LO and SUB contributions which determines the
difference between the VFNS and FFNS; hence,  the behavior is not just a simple factor.
In 
Figure~\ref{fig:thpred-ff3abfonll}-b)
we observe that for $Q^2$ values not too large ($Q^2 \lesssim 4924~{\rm GeV}^2$),
the VFNS is above the \ffns\ result for $\xbj \gtrsim 0.1$.
Interestingly, we see the \ffnsb scheme is quite similar to the VFNS in this kinematic region. 
For larger scales the patterns are not so clear, as the large $Q^2$ effects discussed previously
now complicate the situation. 
%
%

%
\subsection{Contributions from different partonic subprocesses}
\label{sec:thpred-partonic}

The fundamental difference between the FFNS and the VFNS is the
treatment of the heavy partons, the charm in particular. In the FFNS
the charm is not included in the PDFs as an active parton, so charm
quarks only arise from gluon splitting, $g\to c \bar{c}$. In contrast,
the VFNS does include the charm as an active partonic flavor, and thus
allows for charm-initiated subprocesses.
To better appreciate these differences, we will study the individual
partonic contributions to the cross section as functions of the
kinematic variables \xbj, $Q^2$, and $y$.

\begin{figure*}
  \centering
  \centering{{\includegraphics[width=0.49\textwidth]{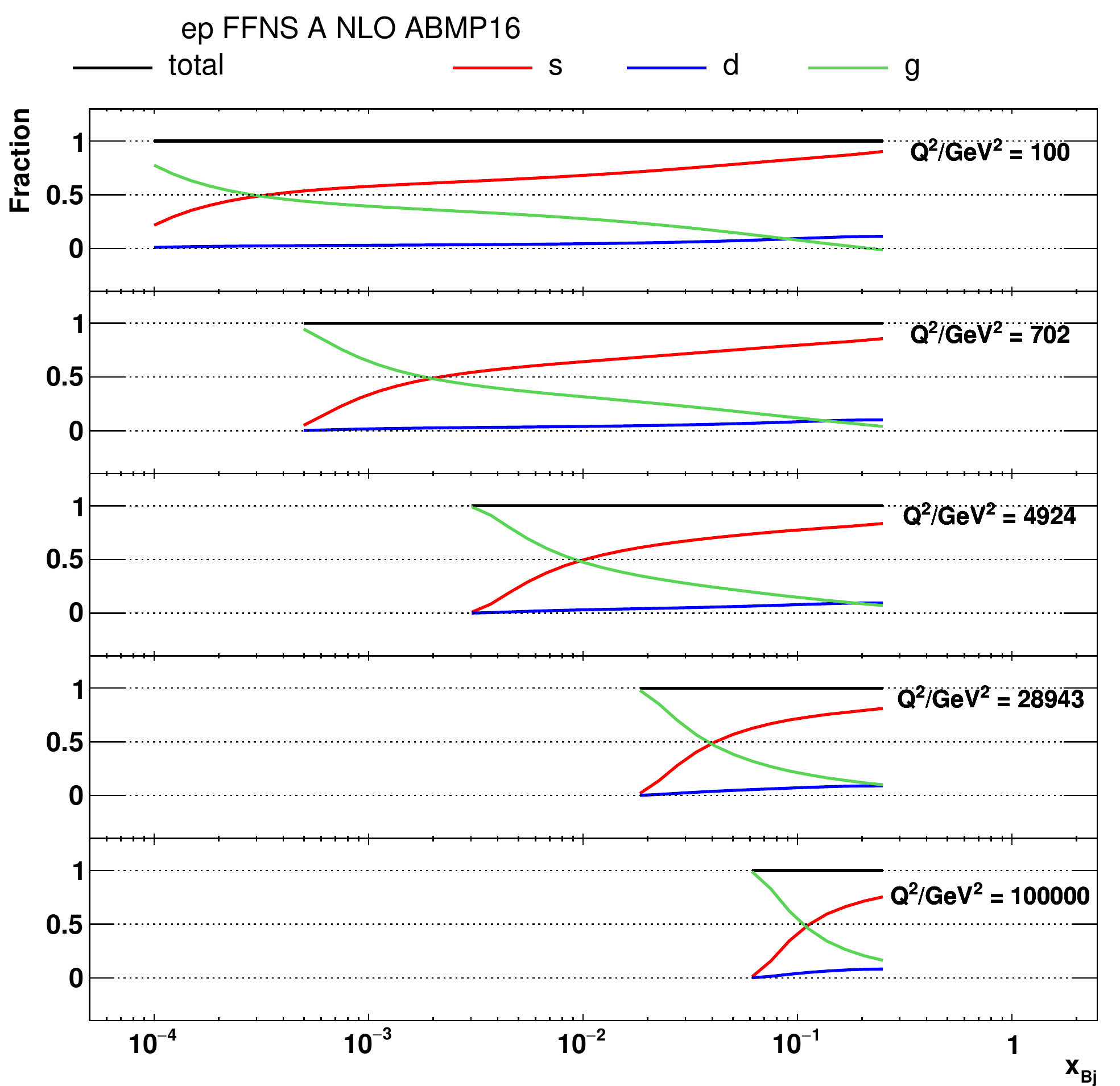}}}
  \centering{{\includegraphics[width=0.49\textwidth]{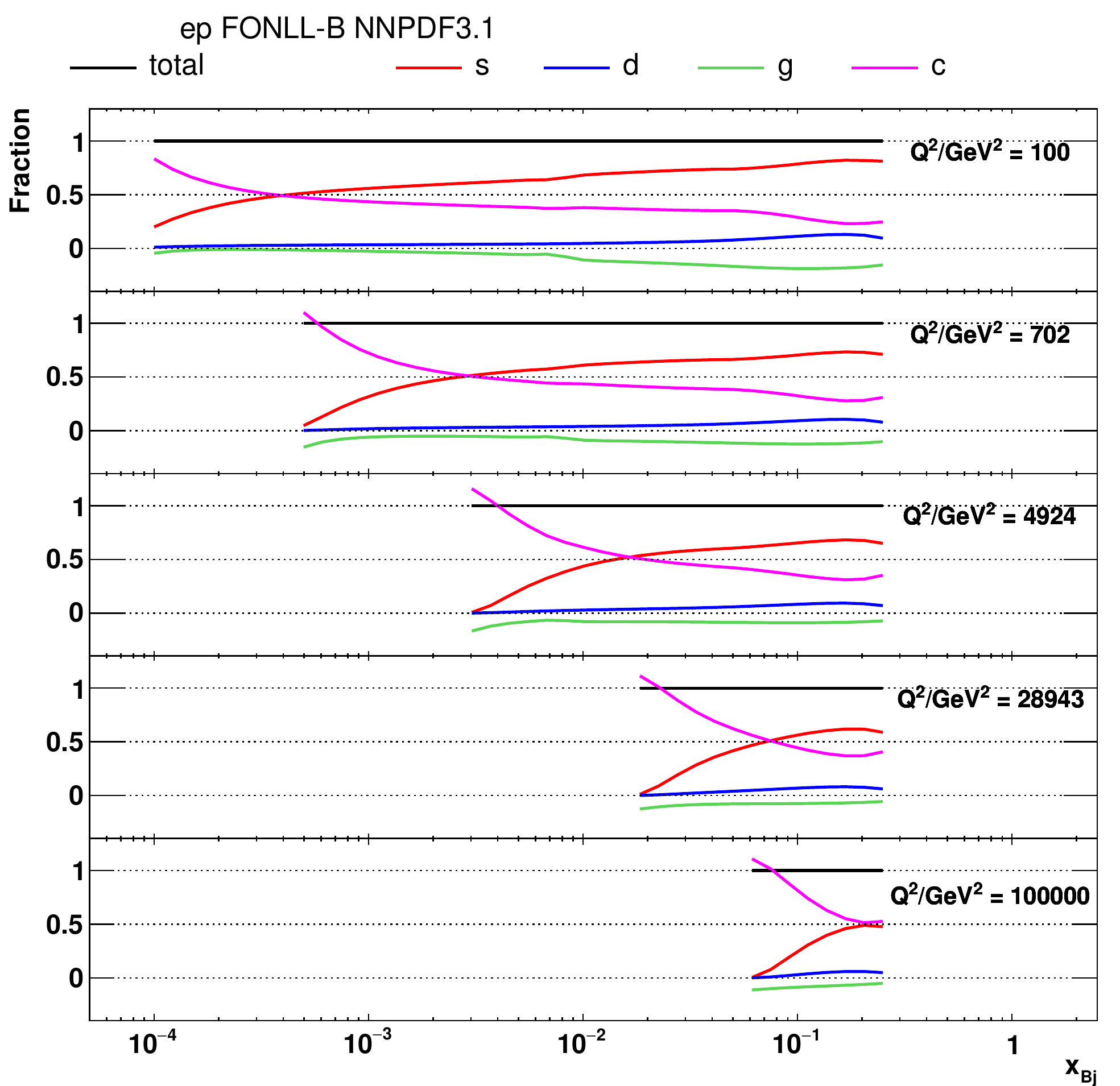}}}
  \caption{The partonic subprocesses for charm CC production cross
    sections in the \ffns (left) and \fonll (right) schemes as a
    function of \xbj for different values of $Q^2$.}
  \label{fig:partonic-x}
\end{figure*}

\begin{figure*}
  \centering
  \centering{{\includegraphics[width=0.49\textwidth]{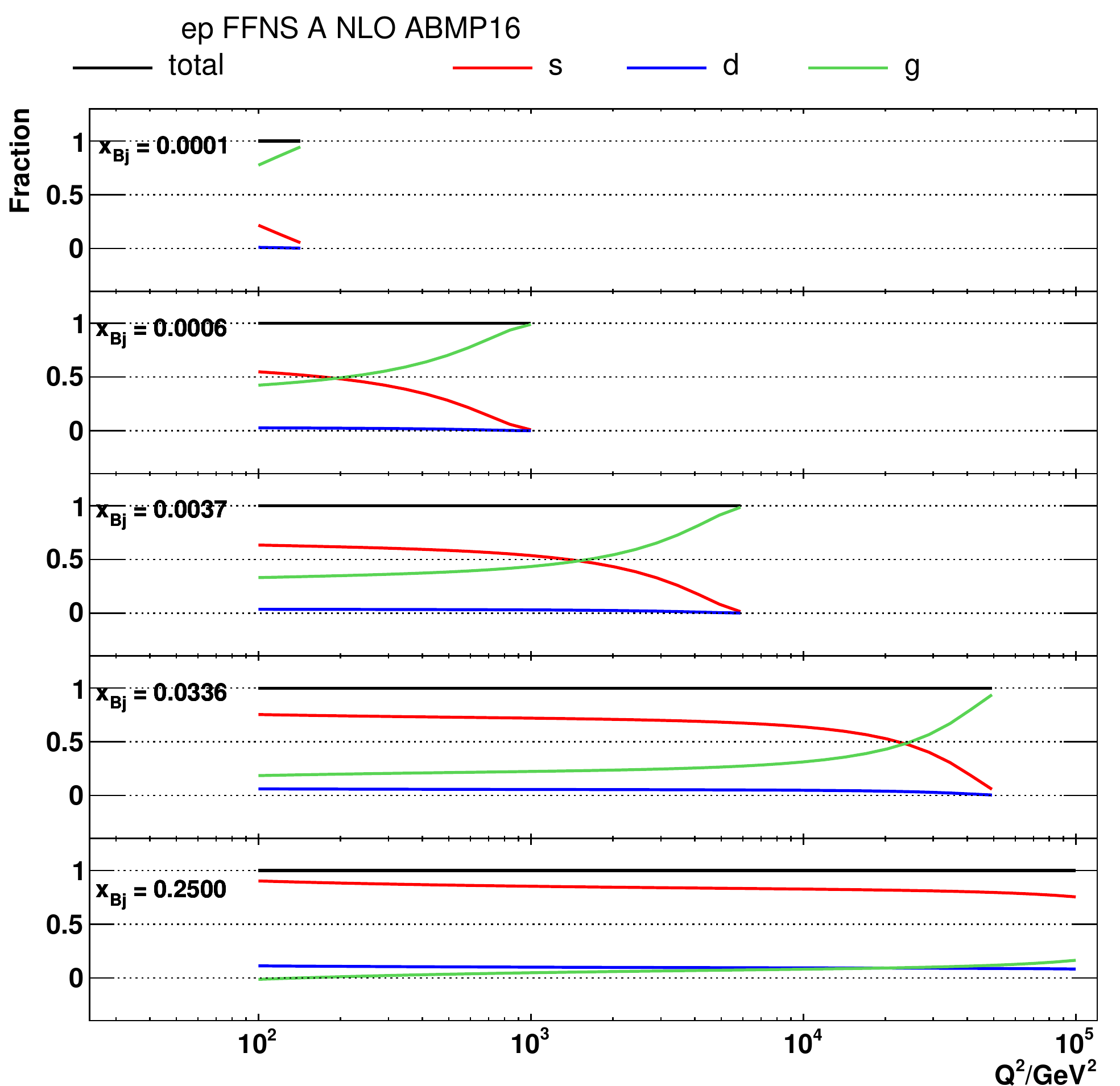}}}
  \centering{{\includegraphics[width=0.49\textwidth]{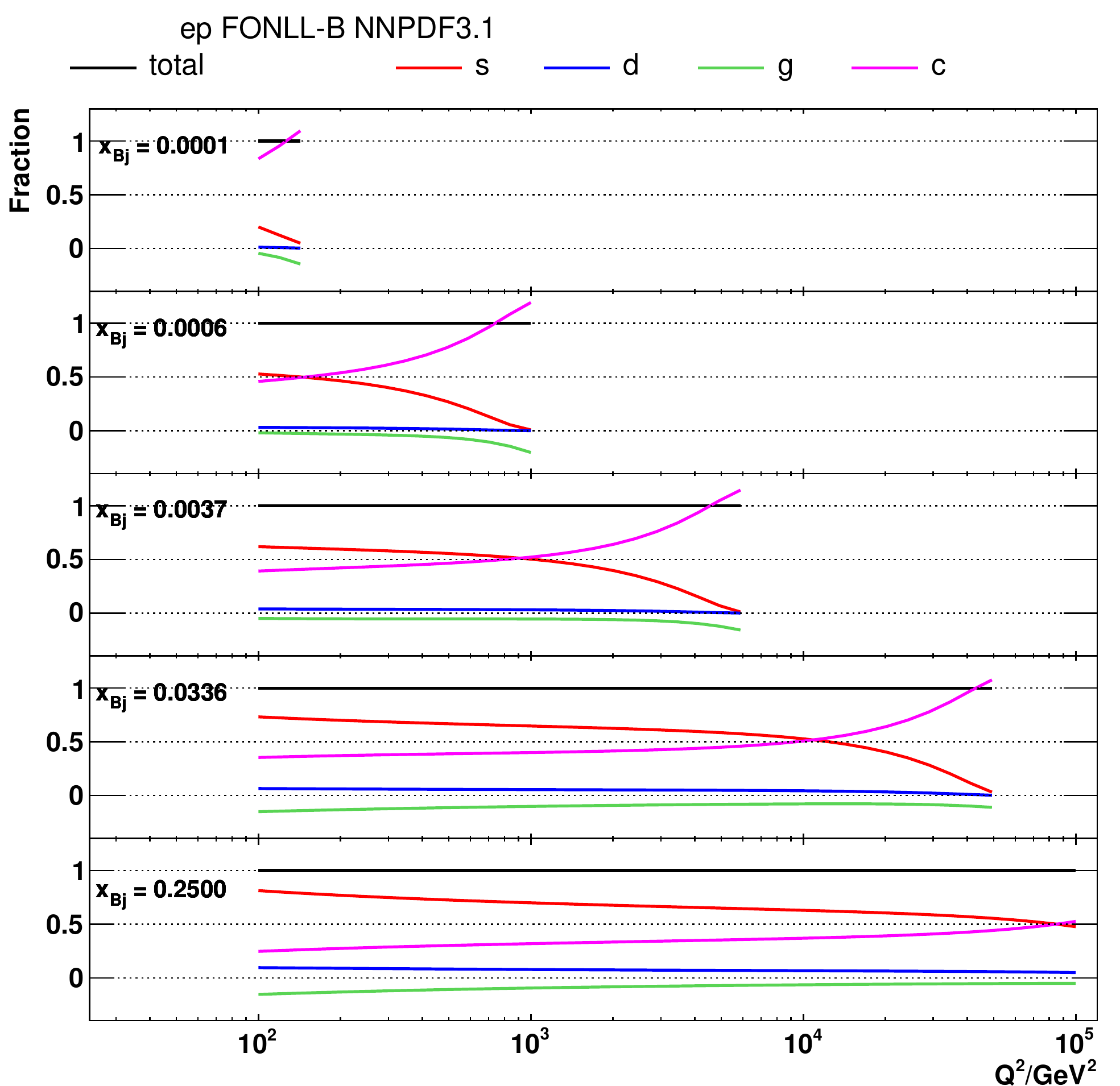}}}
  \caption{The partonic subprocesses for charm CC production cross
    sections in the \ffns (left) and \fonll (right) schemes as a
    function of $Q^2$ for different values of \xbj.}
  \label{fig:partonic-q2}
\end{figure*}

\begin{figure*}
  \centering
  \centering{{\includegraphics[width=0.49\textwidth]{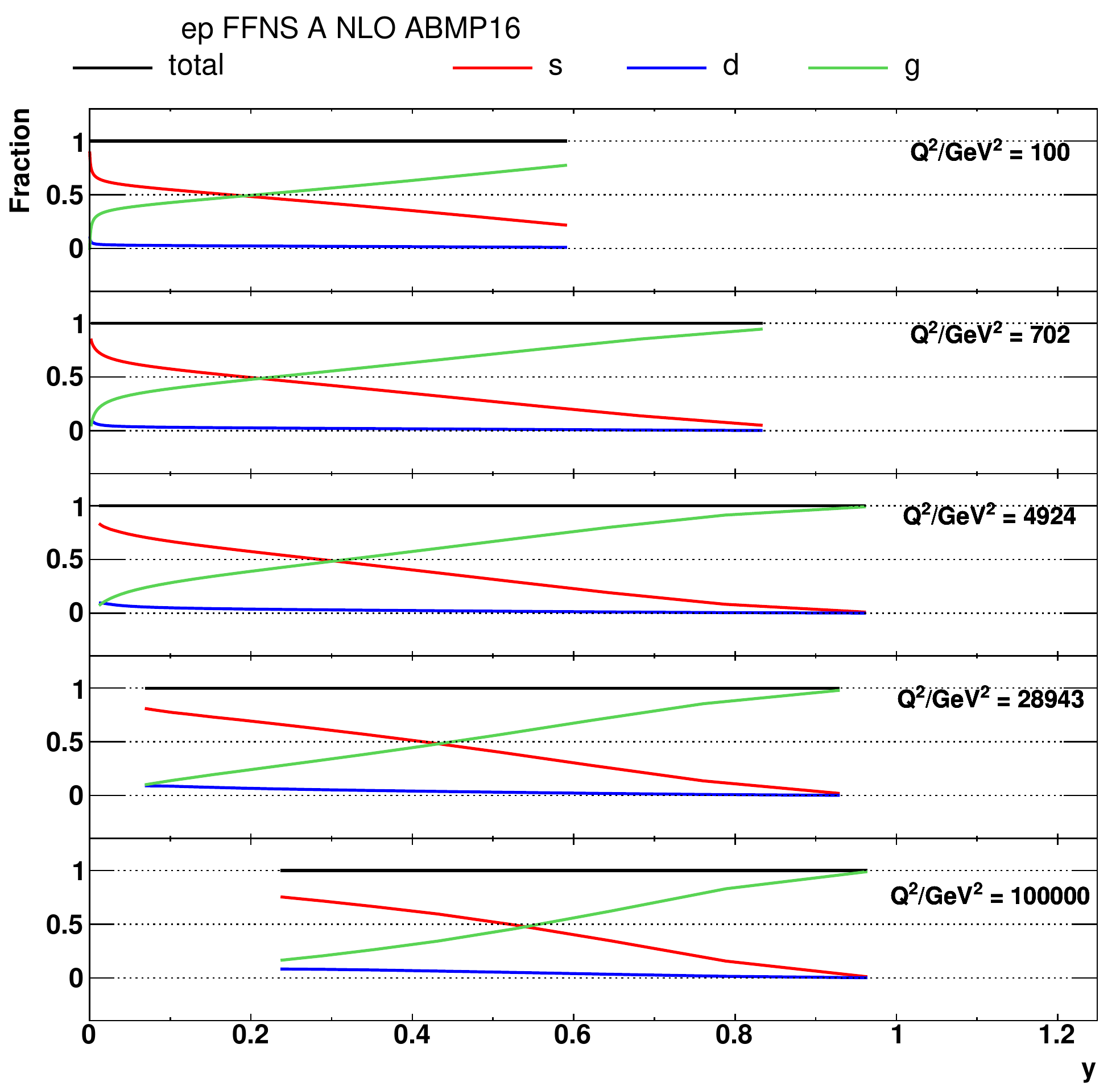}}}
  \centering{{\includegraphics[width=0.49\textwidth]{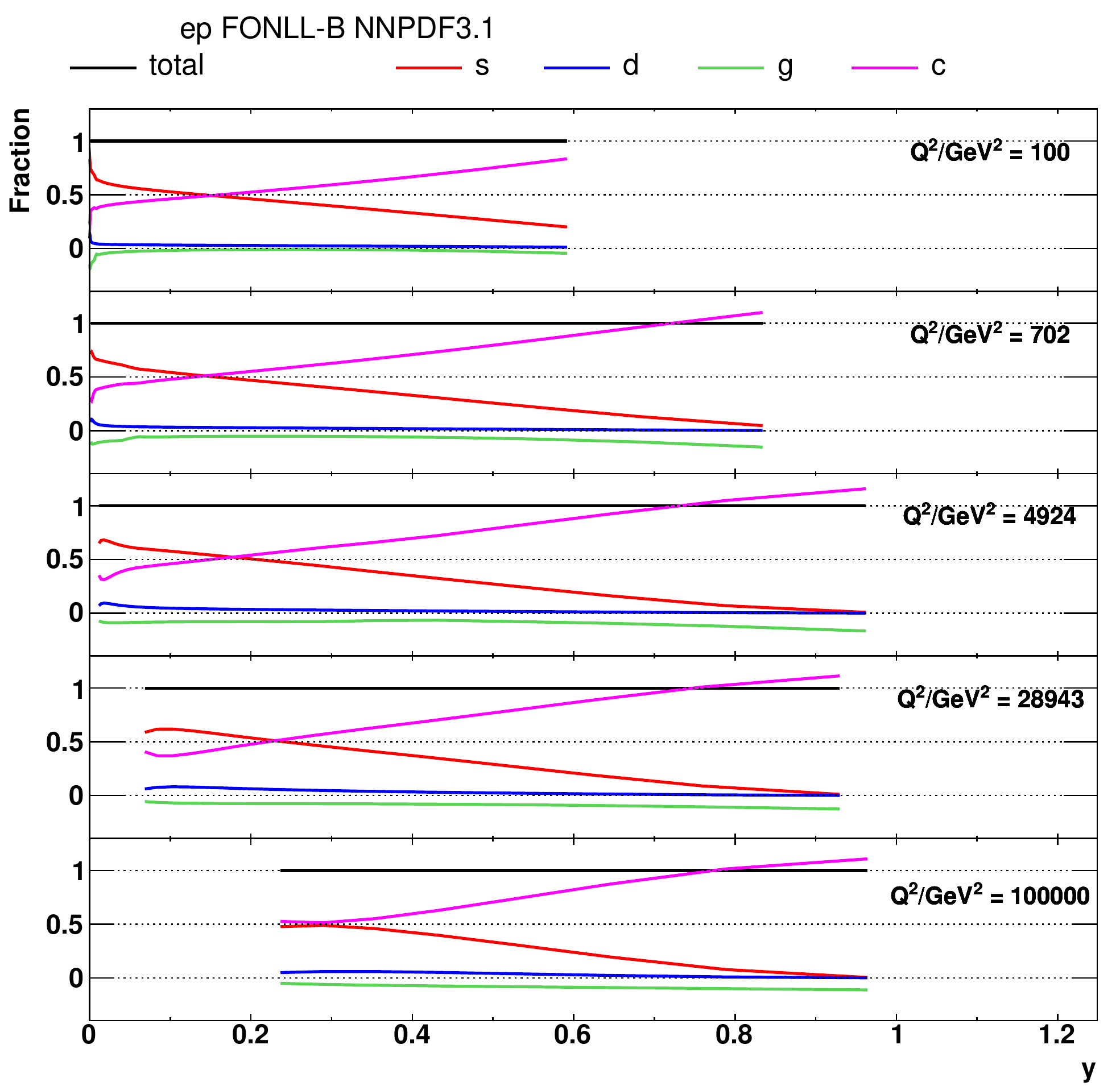}}}
  \caption{The partonic subprocesses for charm CC production cross
    sections in the \ffns (left) and \fonll (right) schemes as a
    function of $y$ for different values of $Q^2$.}
  \label{fig:partonic-y}
\end{figure*}

Figs.~\ref{fig:partonic-x}, \ref{fig:partonic-q2}
and~\ref{fig:partonic-y} show the contributions from separate partonic
subprocesses to the CC charm production cross section in the \ffns and
\fonll schemes as a function of: \xbj for different values of $Q^2$,
$Q^2$ for different values of \xbj, and $y$ for different values of
$Q^2$, respectively.

In these figures we observe that the gluon contribution to the FFNS is
strikingly similar to the charm contribution to the VFNS.
This is explained by the fact that in the FFNS the charm is present
only in the final state and produced predominantly in the hard process
$W^+ g \to c\bar{s}$.
In contrast, in the VFNS the charm is present also
in the initial state and mainly produced by $g\to c\bar{c}$ collinear
splitting through DGLAP evolution.
The fundamental underlying process is (and has to be) the same in both
the FFNS and VFNS, but the factorization boundary between PDFs and
hard scattering cross section, $\hat{\sigma}\otimes f$, (determined by
the scale $\mu$ and the scheme choice) is different.\footnote{%
  Note
  there is a ``subtraction'' term which closely matches the LO
  process, but this ${\cal O}(\alpha_s)$ process is contained in the
  NLO gluon-initiated contribution.  For details,
  see~\ref{sec:appendix} 
}

These figures highlight another interesting feature of the QCD theory;
we observe that for the VFNS the gluon contribution (green curves) can
become negative in particular kinematic regions.\footnote{%
  Note, the FFNS can also have  negative contributions at higher orders due
  to a similar ``subtration'' term for the strange PDF~\cite{Gao:2017kkx}.
  }
This is because in the VFNS we combine the gluon-boson fusion process
(the NLO terms of Figs.~\ref{fig:tchannel} and~\ref{fig:uchannel})
with the counter-term  (the SUB terms),
and this combination can be negative.
This behavior underscores the fact that the renormalization scale $\mu$
is simply ``shuffling'' contributions among the separate sub-pieces,
but the total physical cross section remains positive and stable,
{\it cf.}, Fig.~\ref{fig:acot} and Ref.~\cite{Aivazis:1993pi}.
This is a triumph of the QCD theory.

Next, turning our attention to the strange PDF contribution, it is
notable that the FFNS and VFNS behave qualitatively very similar as
functions of $Q^2$, \xbj, and $y$. In particular, we observe that the
strange fraction increases for \xbj and decreases for $Q^2$ and $y$.
In particular, at high $y$ the strange PDF contribution drops to zero
in favor of the gluon or charm quark PDFs (see
Fig.~\ref{fig:partonic-y} and Eq.~(\ref{eq:y01})). Similar phenomena
(although less pronounced) are observed at low \xbj and/or high
$Q^2$. In these phase-space regions, the dominant contributions to the
cross section are proportional to the gluon PDF in the FFNS or to the
charm-quark PDFs in the VFNS.

  Finally, we note that in Fig.~\ref{fig:partonic-q2} for the  \ffns,
  the gluon contribution at high \xbj ($\xbj\sim 0.25$) is minimal throughout
  the $Q^2$ range. To ensure this is not an artifact of either the \ffns or the \abmp set,
  we regenerated these curves (not shown) in the \ffnsb with the \ffthreeb set and found the same behaviour;
  hence, this feature is truly a characteristic of the FFNS in the hi-\xbj region.

\section{PDF constraints from charm CC pseudodata}
\label{sec:PDF}

Now we turn to examine how the LHeC can reduce the PDF uncertainties
and thus improve our predictive power.

The impact of charm CC cross section measurements at the LHeC on the
PDFs is quantitatively estimated using the profiling
technique~\cite{Paukkunen:2014zia}. This technique is based on
minimizing the \chisq between data and theoretical predictions taking
into account both experimental and theoretical uncertainties arising
from PDF variations.
Two NLO PDF sets were chosen for this study:
\abmp~\cite{Alekhin:2018pai} and \nnpdf~\cite{Ball:2017nwa}. All PDF sets are provided with
uncertainties in the format of eigenvectors.
In the presence of strong constraints (the LHeC data is very precise),
it is preferable to use the eigenvector representation as only a few MC
replicas would survive the Bayesian reweighting.

\subsection{The CC  charm pseudodata}
\label{sec:pseudodata}

For this study, pseudodata for charm CC production cross section
differential in $Q^2$ and \xbj and corresponding to an integrated
luminosity of $100{\rm~fb}^{-1}$~\cite{AbelleiraFernandez:2012cc,Blumlein:1992we} and
polarization $P=-0.8$ are used.
Theoretical predictions are calculated at NLO
in pQCD both in the \ffns with number of active flavors $n_f = 3$ and
in the \fonll scheme. The charm-mass reference value in the
$\overline{\mbox{MS}}$ scheme is set to $m_c(m_c) = 1.27$ GeV and
$\alpha_s$ is set to the value used for the corresponding PDF
extraction. The renormalization and factorization scales are chosen to
be $\mu_\mathrm{r}^2 = \mu_\mathrm{f}^2 = Q^2$.

The \chisq value is calculated as follows:
\begin{equation}
\begin{split}
  \chisq = \mathbf{R}^{T}_{} \mathbf{Cov}^{-1}_{} \mathbf{R}_{} + \sum_{\beta} b_{\beta,\rm th}^2 \ , \\[5pt]
  \mathbf{R} = \mathbf{D} - \mathbf{T} - \sum_{\beta} \Gamma^{}_{\beta,\rm th} b_{\beta,\rm th} \ ,
\label{eq:chisq}
\end{split}
\end{equation}
where $\mathbf{D}$ and $\mathbf{T}$ are the column vectors of the
measured (data) and predicted (theory) values, respectively.  The
correlated theoretical PDF uncertainties are included using the
nuisance parameters $b_{\beta, \rm th}$ with their influence on the
theory predictions described by $\Gamma_{\beta,\rm th}$, where the
index $\beta$ runs over all PDF eigenvectors. For each nuisance
parameter a penalty term is added to the \chisq, representing the
prior knowledge of the parameter.
No theoretical uncertainties except
the PDF uncertainties are considered; the PDF Hessian uncertainties are treated symmetrically.
The full covariance matrix ${\bf Cov }$
representing the statistical and systematic uncertainties of the data
is used in the fit. The statistical and systematic uncertainties are
treated as additive, \textit{i.e.} they do not change in the fit.
The systematic uncertainties are assumed uncorrelated between bins.

The values of the nuisance parameters at the minimum,
$b^{\rm min}_{\beta,\rm th}$, are interpreted as optimized, or
profiled, PDFs, while  uncertainties of $b^{\rm min}_{\beta,\rm th}$   determined using the
tolerance criterion of $\Delta\chi^2 = 1$ correspond to the new PDF
uncertainties. The profiling approach assumes that the new data are
compatible with the theoretical predictions using the existing PDFs,
such that no modification of the PDF fitting procedure is
needed. Under this assumption, the central values of the measured
cross sections are set to the central values of the theoretical
predictions.

\subsection{The profiled PDFs}
\label{sec:profile}

\begin{figure}
  \centering
  {{\includegraphics[width=0.5\textwidth]{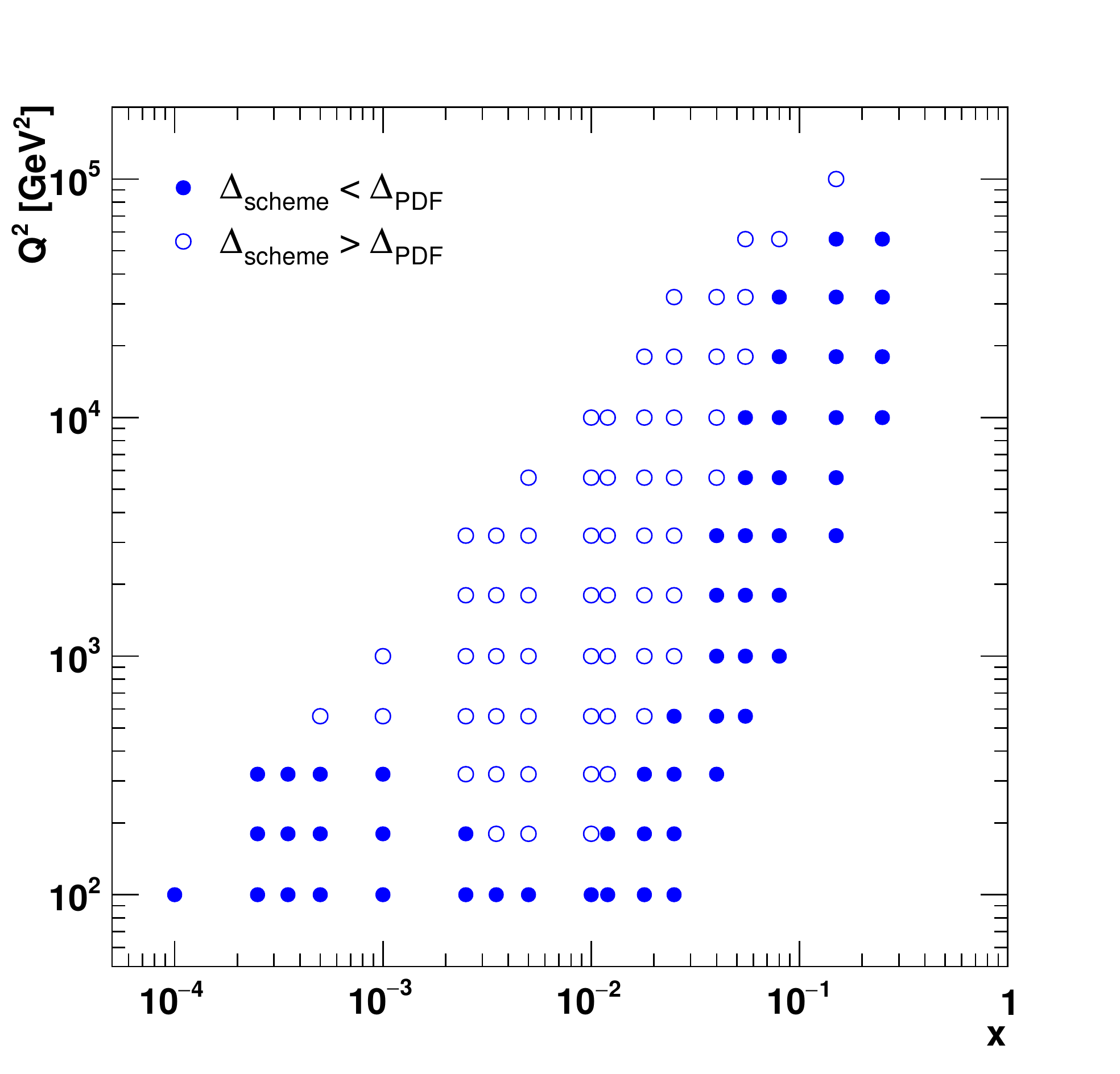}}}
  \caption{The full ($\Delta_{\rm scheme} < \Delta_{\rm PDF}$, $\Delta_{\rm scheme} > \Delta_{\rm PDF}$) and restricted ($\Delta_{\rm scheme} < \Delta_{\rm PDF}$) sets of data points which are used for PDF profiling.}
  \label{fig:data}
\end{figure}

\new{
The profiling study is performed using two sets of LHeC charm CC pseudodata:
\begin{itemize}
 \item the full set,
 \item a restricted set with data points for which the difference between the \ffns and \fonll are smaller than the present PDF uncertainties. 
 The latter is taken for simplicity as the sum of the \abmp and \nnpdf uncertainties, but for the most data points it is dominated by the \nnpdf uncertainties (see Fig.~\ref{fig:thpred-q2-unc}).
\end{itemize}
Given the sizable differences observed between the \ffns and \fonll predictions, the study with the restricted data set (also referred to as `with cuts') aims to check whether or not model independent constraints on the strange PDF can be extracted using the charm CC reaction at LHeC. The two sets of data points are shown in Fig.~\ref{fig:data} as functions of $Q^2$ and \xbj.
}

%
The comparison between  \abmp  and \nnpdf is insightful as \abmp represents a more restricted parametrization.
For the HERAPDF2.0 set, the strange PDF was not fit directly, but computed via the relation
$f_s = \bar{s}/(\bar{s}+\bar{d})=0.4\pm 0.1$ and the uncertainty was approximated using the variation on $f_s$;
hence, these uncertainties are not the same as the Hessian diagonalized eigenvectors, so we will not
profile the HERAPDF2.0 PDF set.

The original and profiled \abmp and \nnpdf PDF uncertainties are shown
in Figs.~\ref{fig:pdf-abmp}--\ref{fig:pdf-nnpdf-100000}.  The
uncertainties of the PDFs are presented at the scales
$\mu_\mathrm{f}^2=100$ GeV$^2$ and $\mu_\mathrm{f}^2=100000$ GeV$^2$.
A strong impact of the charm CC pseudodata on the PDFs is observed for
both PDF sets.  In particular, the uncertainties of the strange PDF
are strongly reduced once the pseudodata are included in the fit.
Also the gluon PDF uncertainties are decreased. Furthermore, in the
case of the NNPDF3.1 set, the charm PDF uncertainties are reduced
significantly.
\new{For all PDF sets, only small differences can be noticed 
between the PDF constraints obtained using the full or restricted set 
because the whole \xbj range is covered in both cases (see Fig.~\ref{fig:data}) 
despite the fact that the number of data points in the restricted set
is roughly half of the  total number of data points.}

\new{
Additionally, in the case of the NNPDF3.1 set, it is possible to check 
the constraints on the strange quark and anti-quark distributions 
separately, because no assumption $s=\bar{s}$ is used in NNPDF3.1. 
The LHeC $e^{-}p$ pseudodata provide direct constraints only on $\bar{s}$. 
Nevertheless due to the apparently strong correlation between $s$ and 
$\bar{s}$ in the NNPDF3.1 fit, quite strong constraints are present on both 
the $s$ and $\bar{s}$ distributions once the direct constraints on $\bar{s}$ 
are provided by the LHeC pseudodata. However, only mild constraints 
are put on the ratio $s/\overline{s}$. This indicates that for precise 
determination of $s/\overline{s}$ both $e^{-}p$ and $e^{+}p$ data will be needed.

\begin{figure}
  \centering
  {{\includegraphics[width=0.235\textwidth]{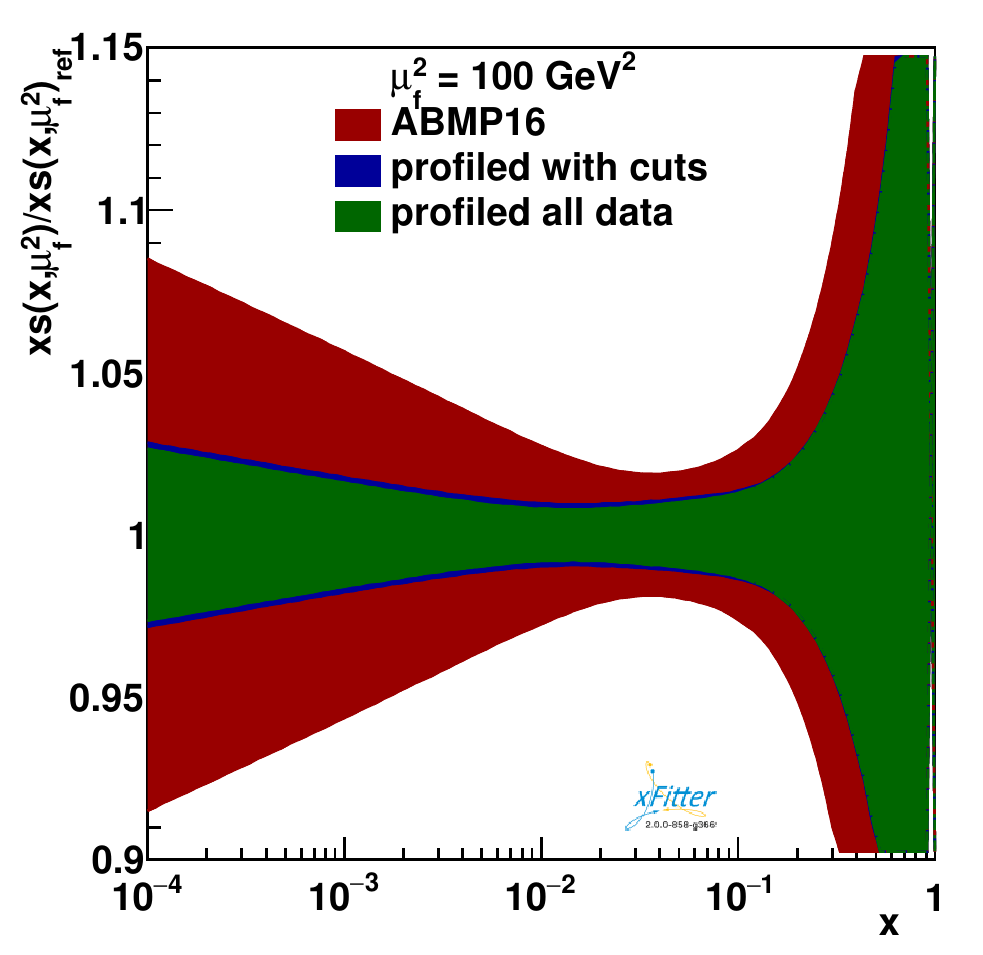}}}
  {{\includegraphics[width=0.235\textwidth]{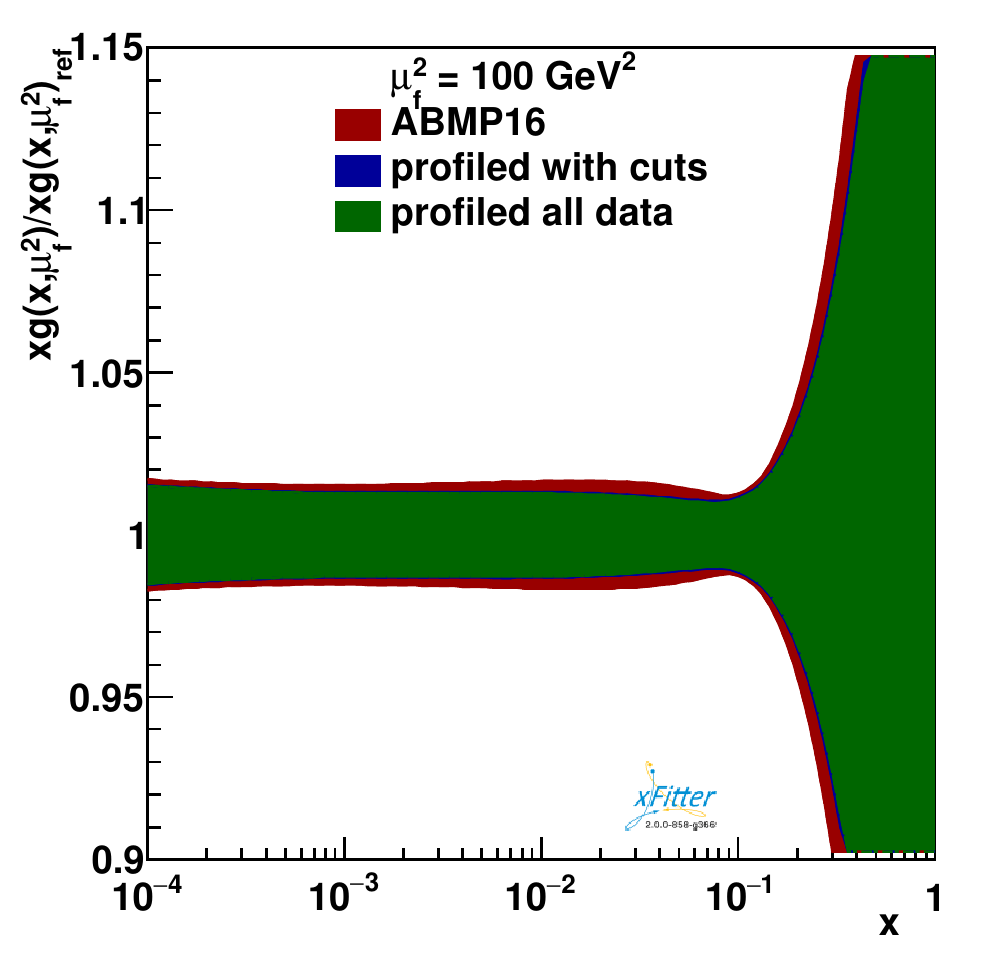}}}\\
  {{\includegraphics[width=0.235\textwidth]{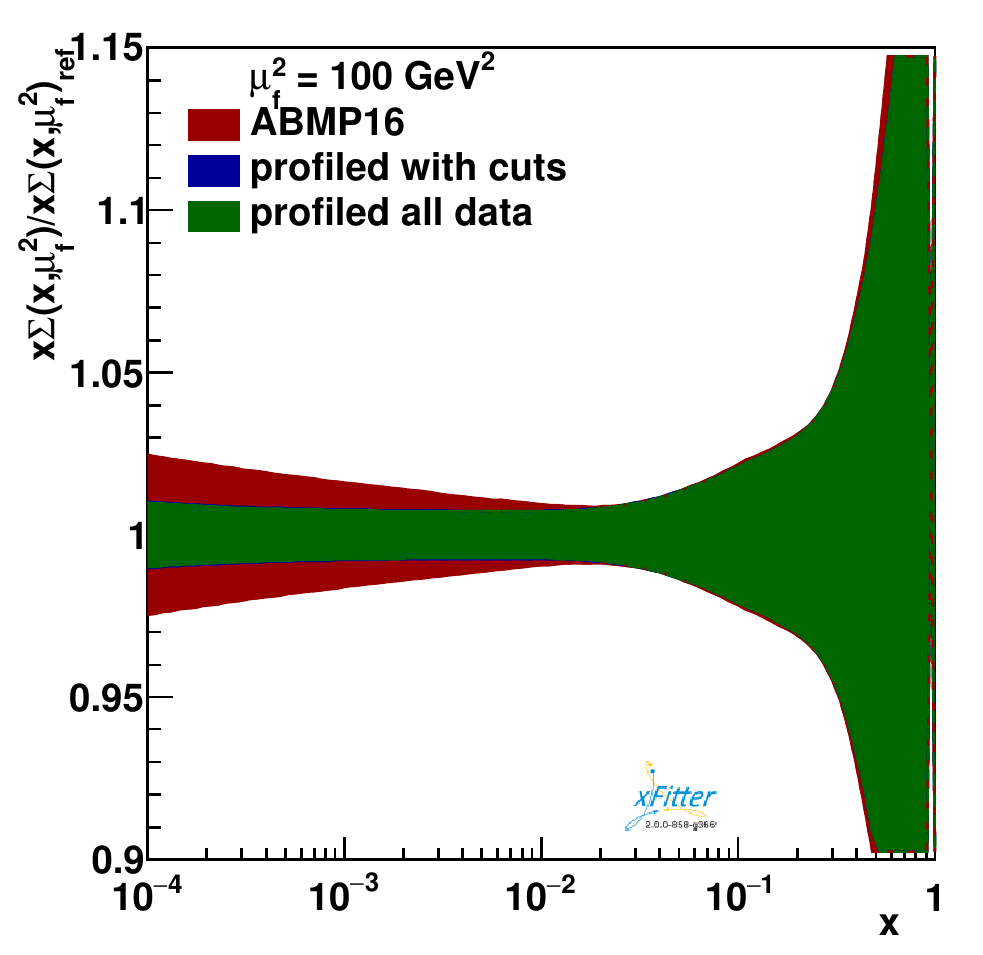}}}
  {{\includegraphics[width=0.235\textwidth]{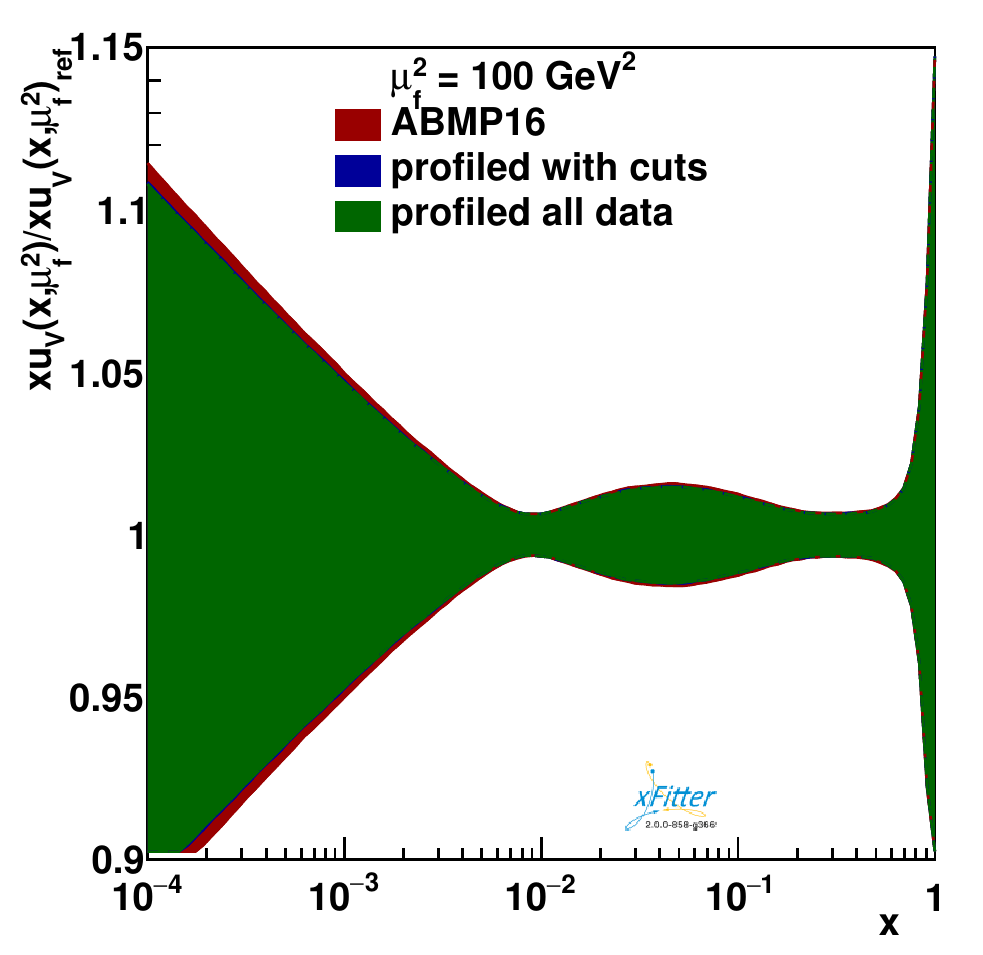}}}
  {{\includegraphics[width=0.235\textwidth]{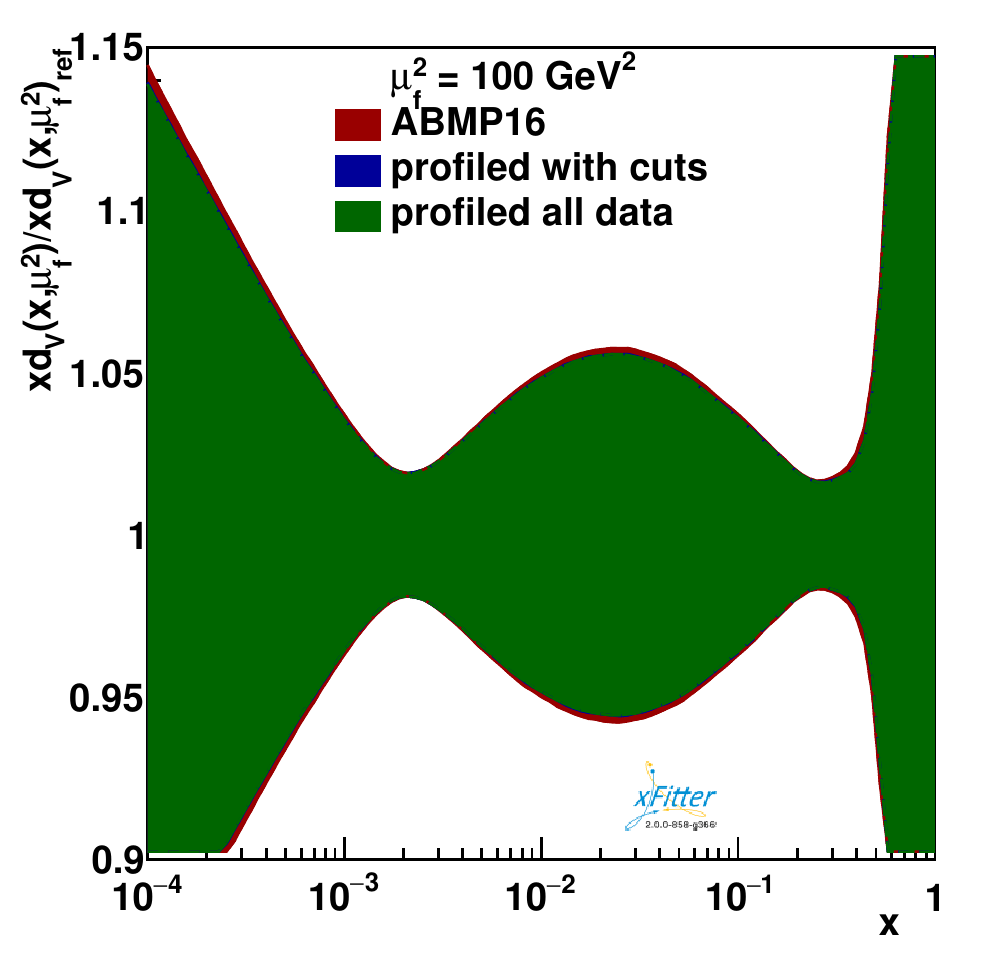}}}
  \caption{The relative strange (top left), gluon (top right), sea
    quark (middle left), u valence quark (middle right) and d valence
    quark (bottom) PDF uncertainties at $\mu_\mathrm{f}^2=100$ GeV$^2$
    of the original and profiled \abmp PDF set.}
  \label{fig:pdf-abmp}
\end{figure}

\begin{figure}
  \centering
  {{\includegraphics[width=0.235\textwidth]{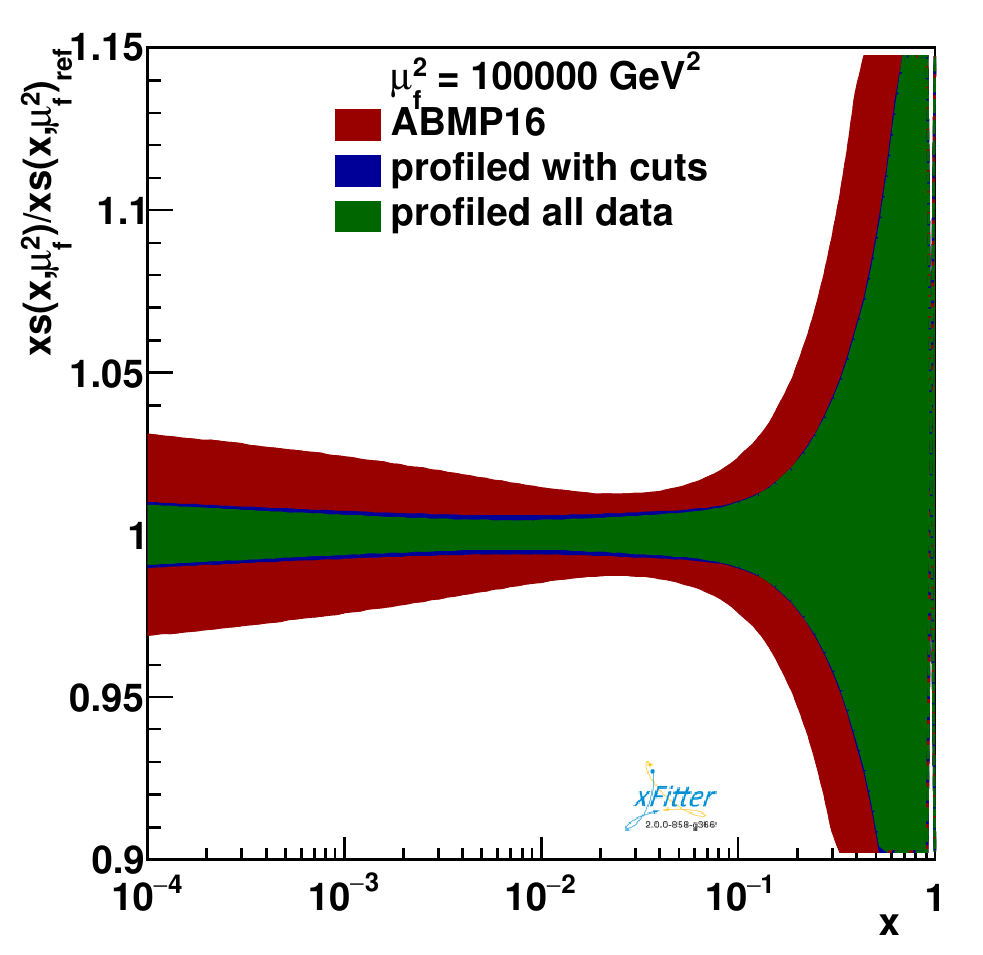}}}
  {{\includegraphics[width=0.235\textwidth]{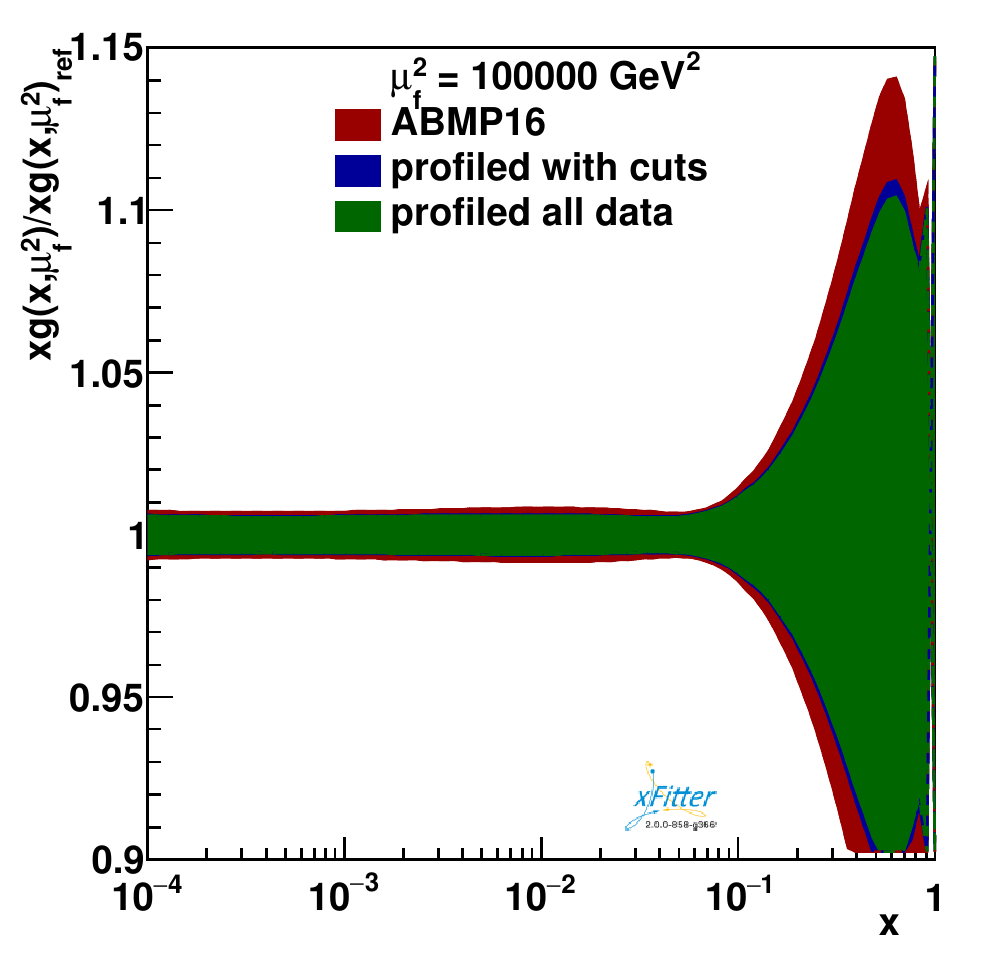}}}\\
  {{\includegraphics[width=0.235\textwidth]{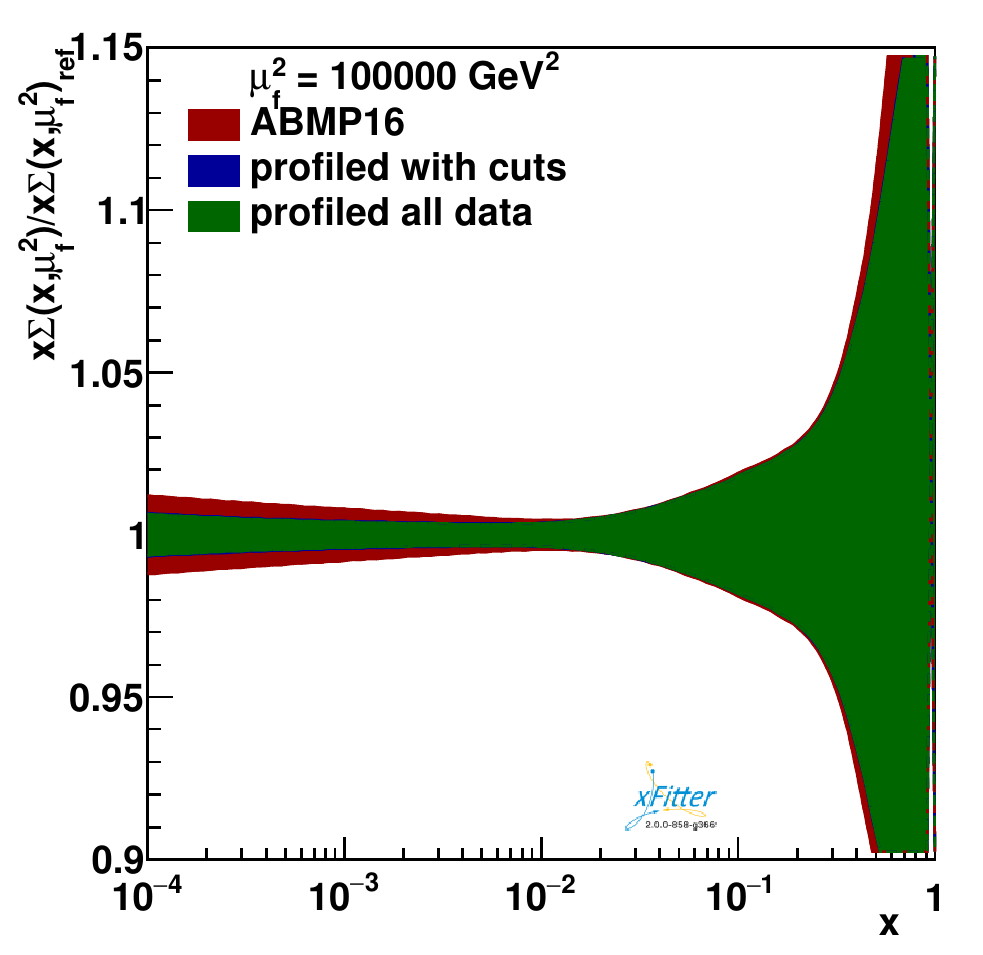}}}
  {{\includegraphics[width=0.235\textwidth]{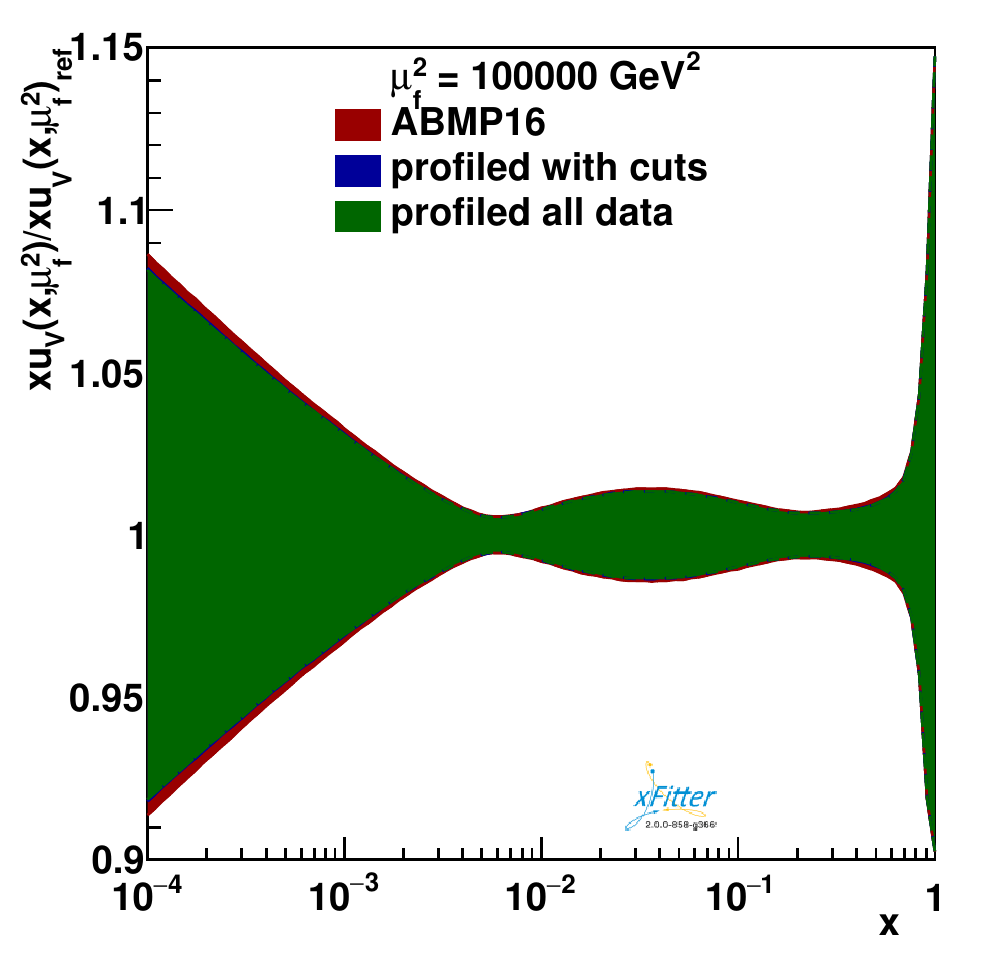}}}
  {{\includegraphics[width=0.235\textwidth]{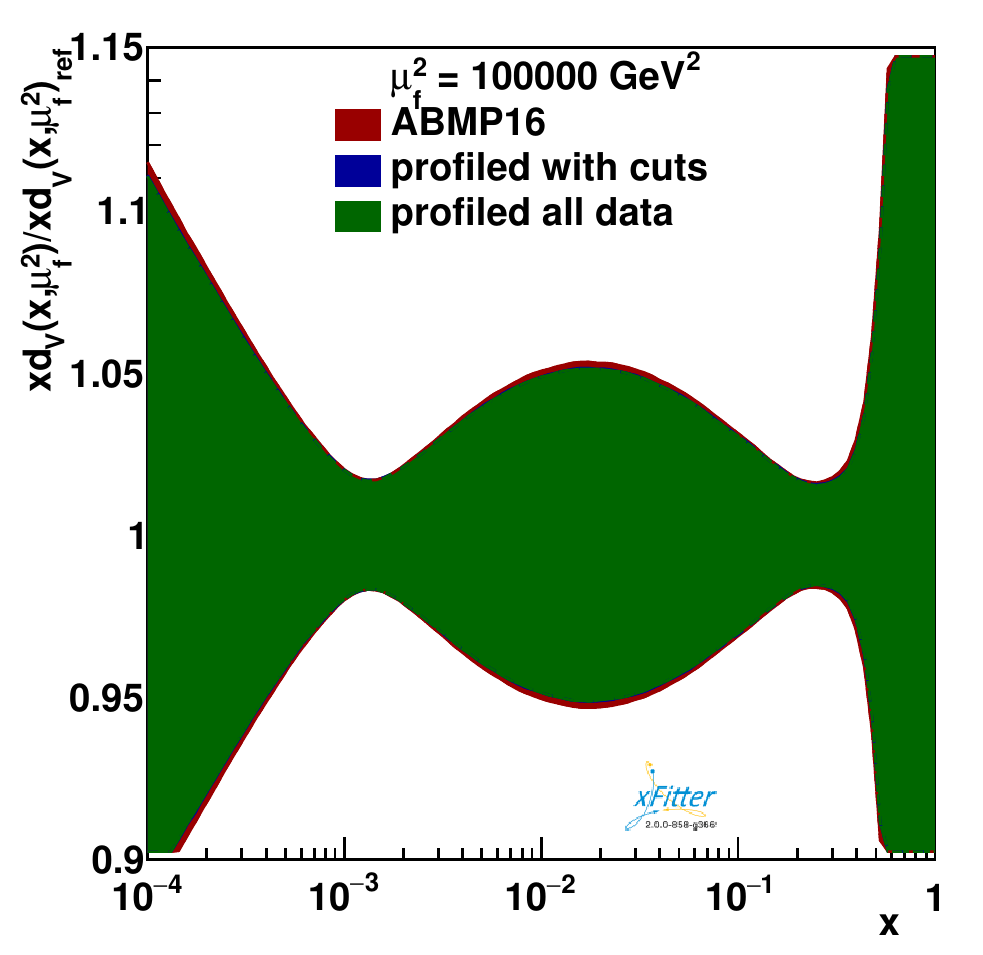}}}
  \caption{The relative strange (top left), gluon (top right), sea
    quark (middle left), u valence quark (middle right) and d valence
    quark (bottom) PDF uncertainties at $\mu_\mathrm{f}^2=100000$
    GeV$^2$ of the original and profiled \abmp PDF set.}
  \label{fig:pdf-abmp-100000}
\end{figure}

\begin{figure}
  \centering
  {{\includegraphics[width=0.235\textwidth]{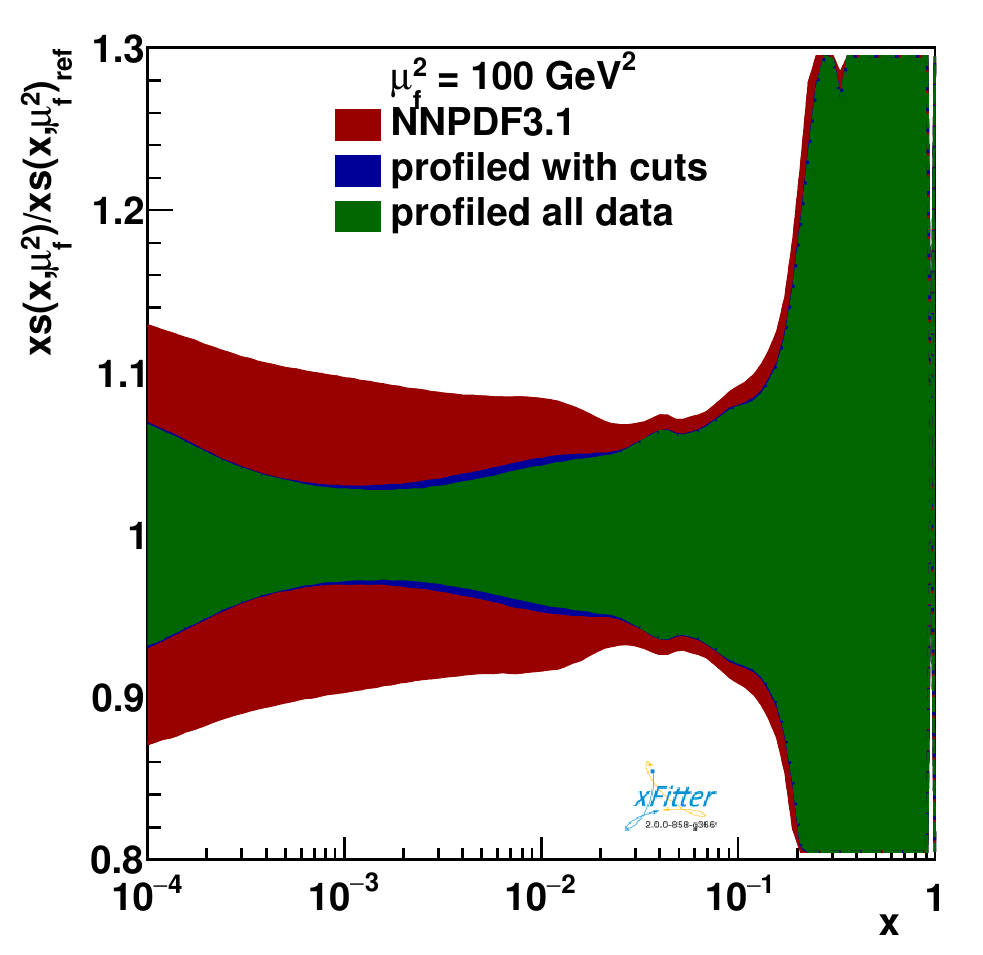}}}
  \put(-95,95){(a)}
  {{\includegraphics[width=0.235\textwidth]{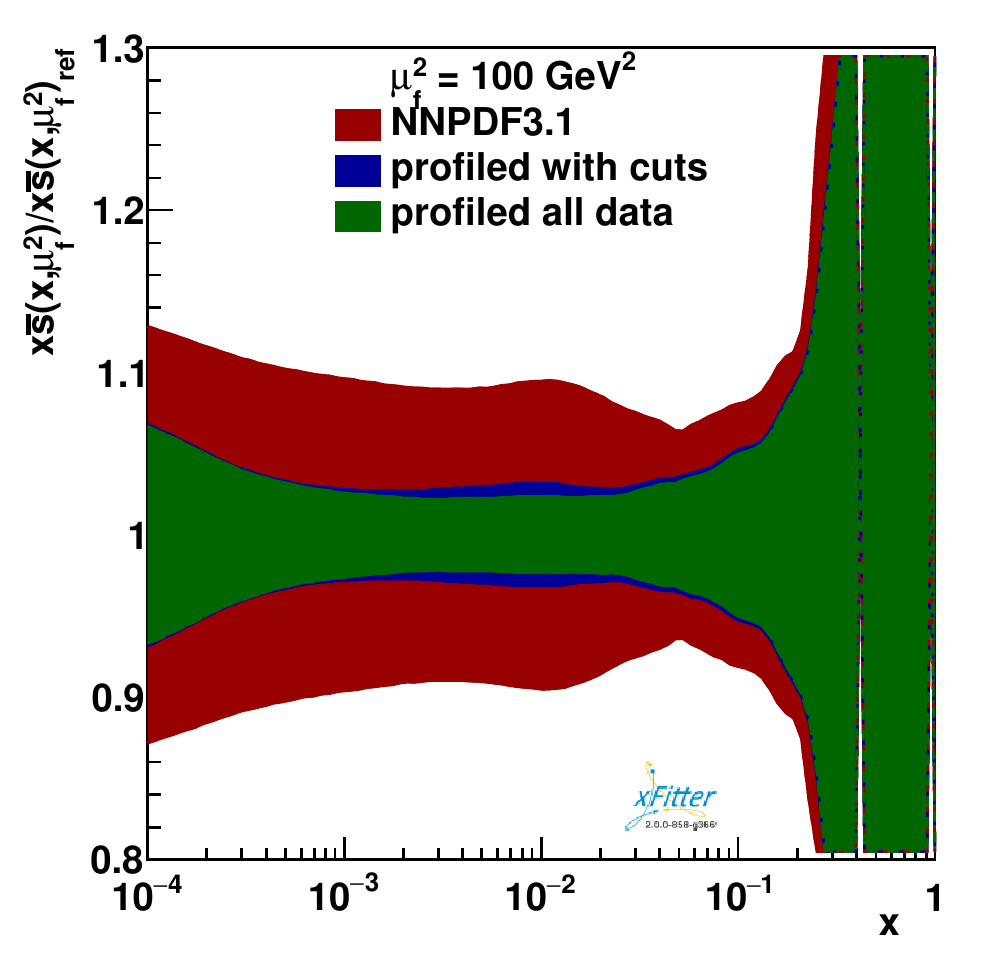}}}
  \put(-95,95){(b)}\\
  {{\includegraphics[width=0.235\textwidth]{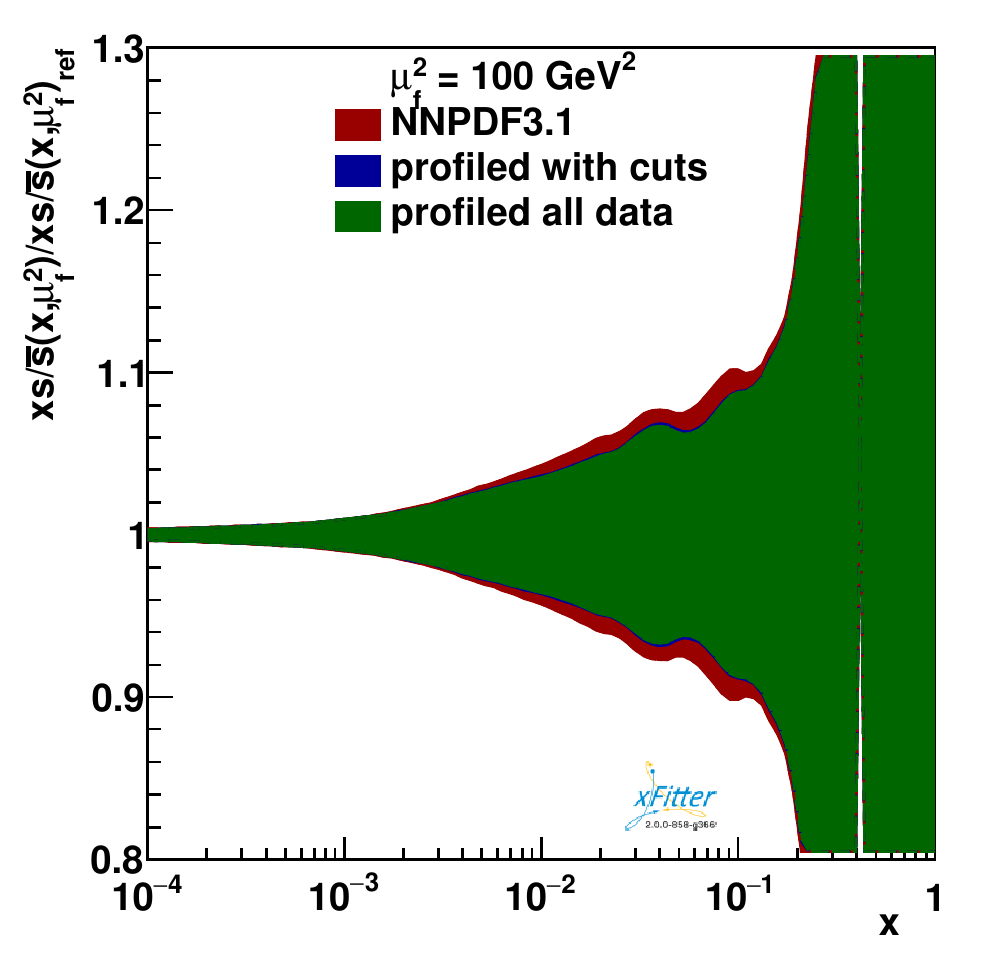}}}
  \put(-95,95){(c)}
  {{\includegraphics[width=0.235\textwidth]{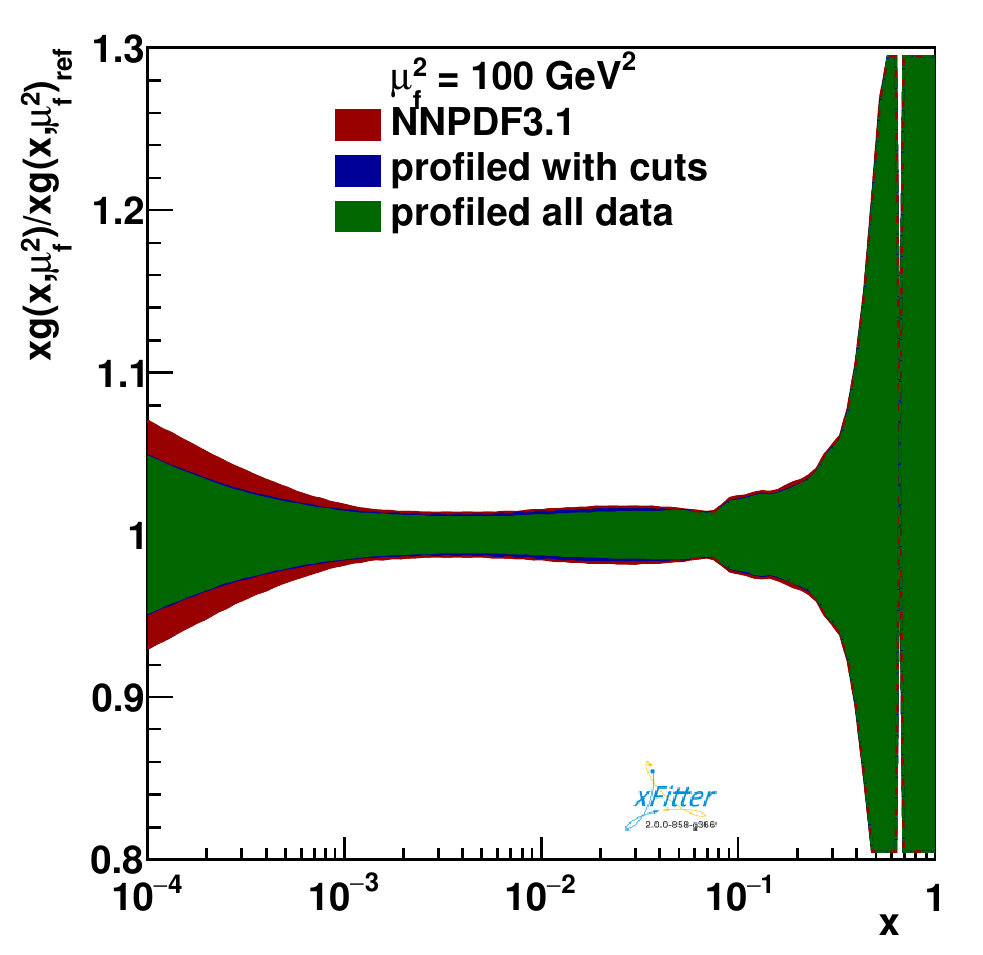}}}
  \put(-95,95){(d)}\\
  {{\includegraphics[width=0.235\textwidth]{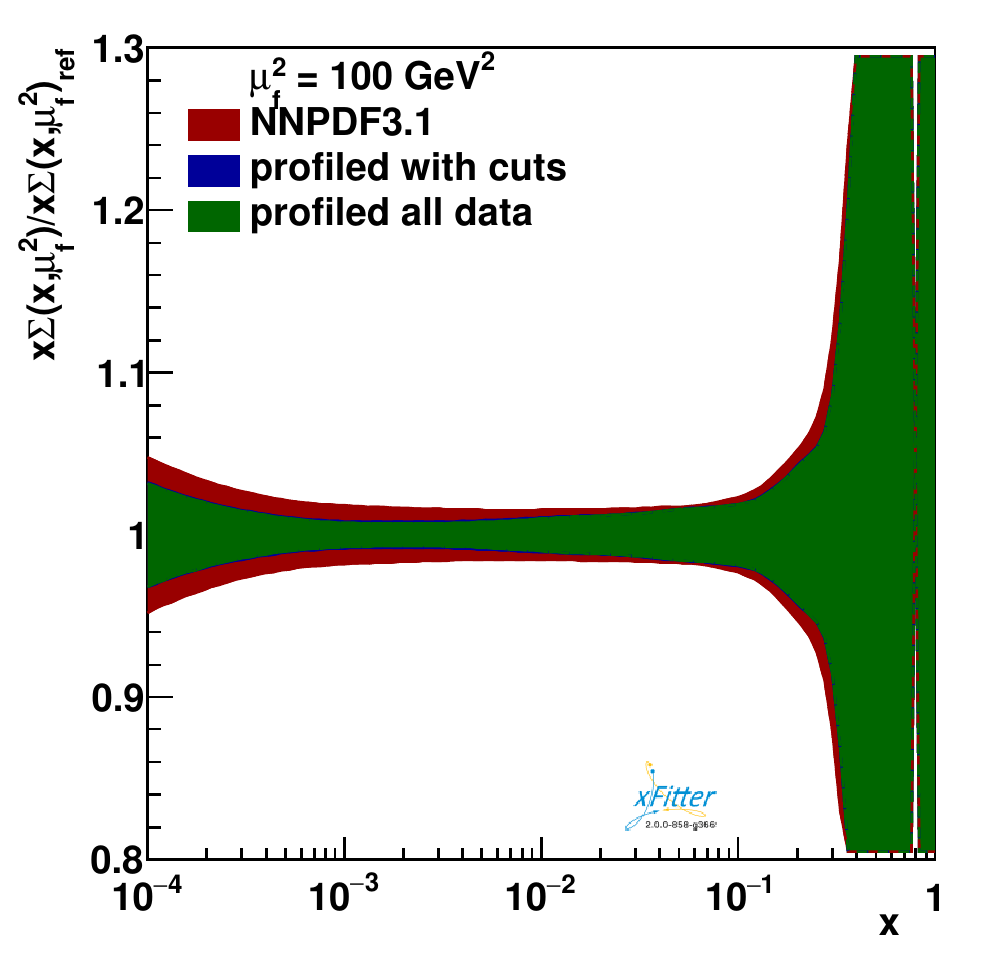}}}
  \put(-30,95){(e)}
  {{\includegraphics[width=0.235\textwidth]{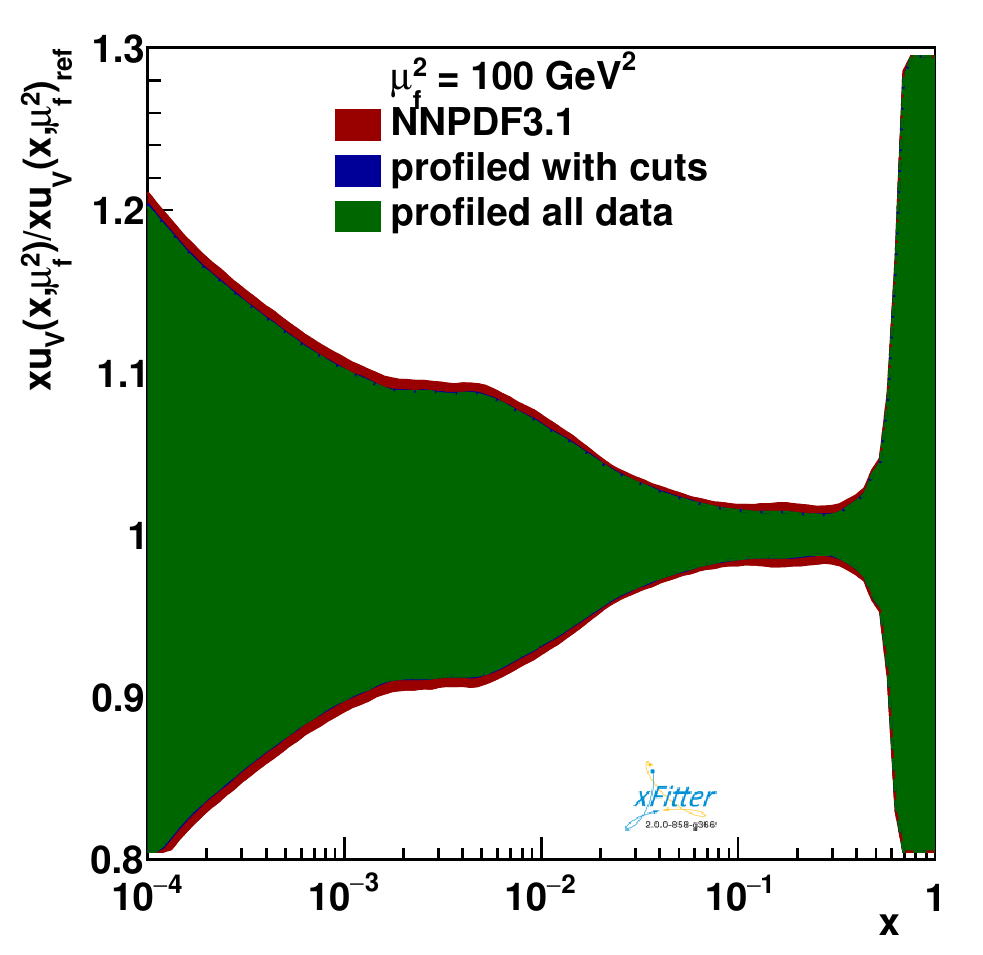}}}
  \put(-95,95){(f)}\\
  {{\includegraphics[width=0.235\textwidth]{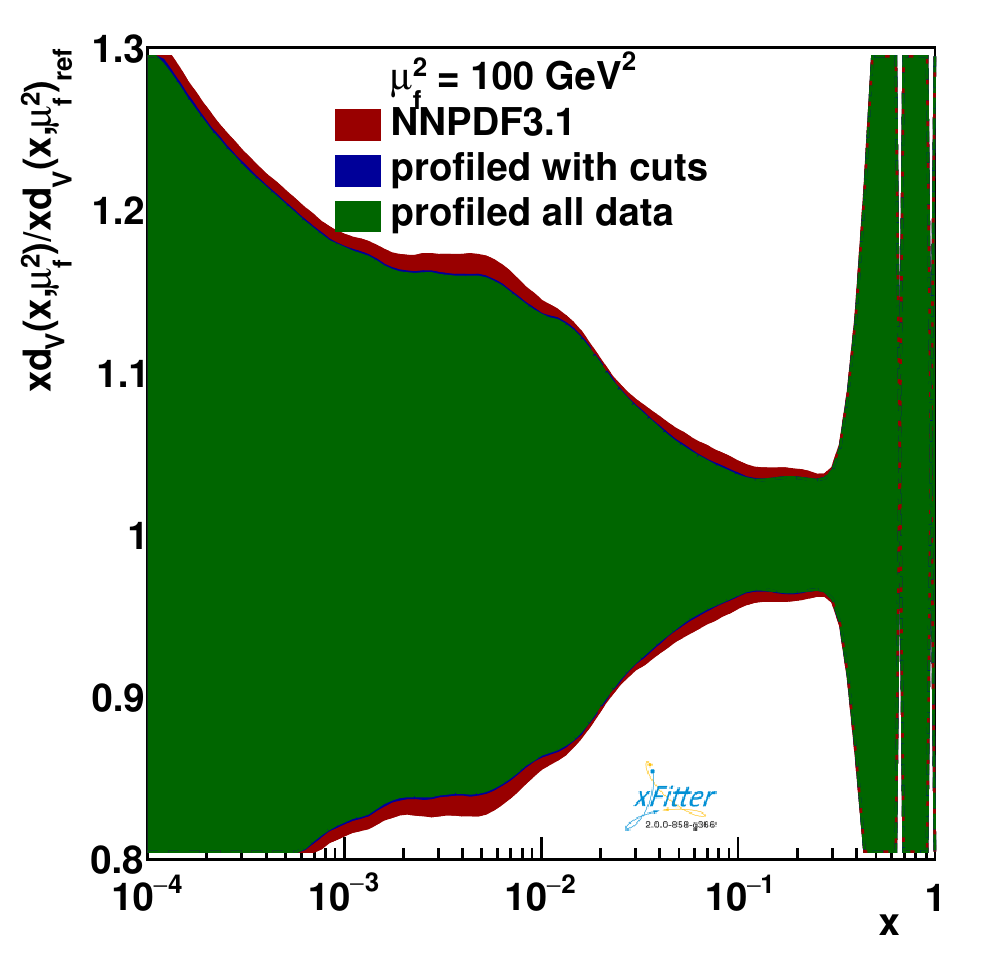}}}
  \put(-30,95){(g)}
  {{\includegraphics[width=0.235\textwidth]{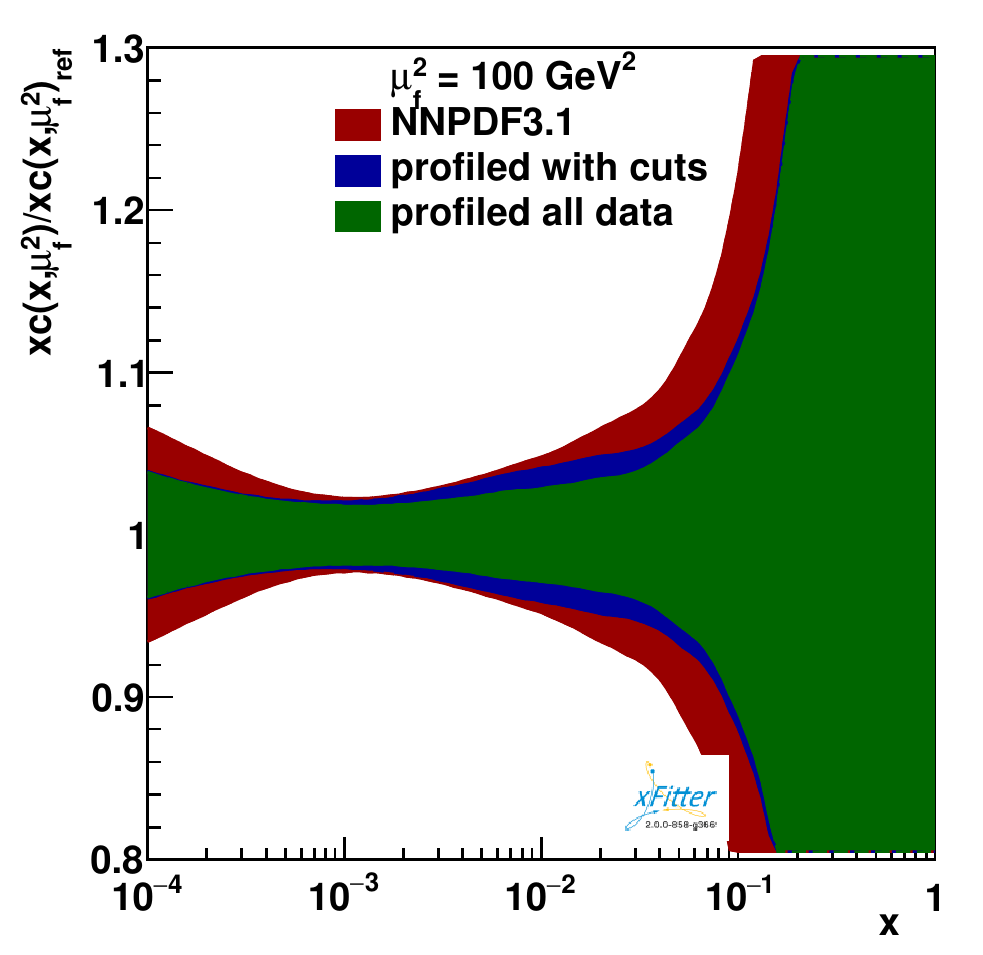}}}
  \put(-95,95){(h)}
  \caption{The relative strange quark (a), strange anti-quark (b), and ratio $s/\overline{s}$ (c), gluon (d), sea
    quark (e), u valence quark (f), d valence quark (g) and charm quark (h) PDF
    uncertainties at $\mu_\mathrm{f}^2=100$ GeV$^2$ of the original
    and profiled \nnpdf PDF set.}
  \label{fig:pdf-nnpdf}
\end{figure}

\begin{figure}
  \centering
  {{\includegraphics[width=0.235\textwidth]{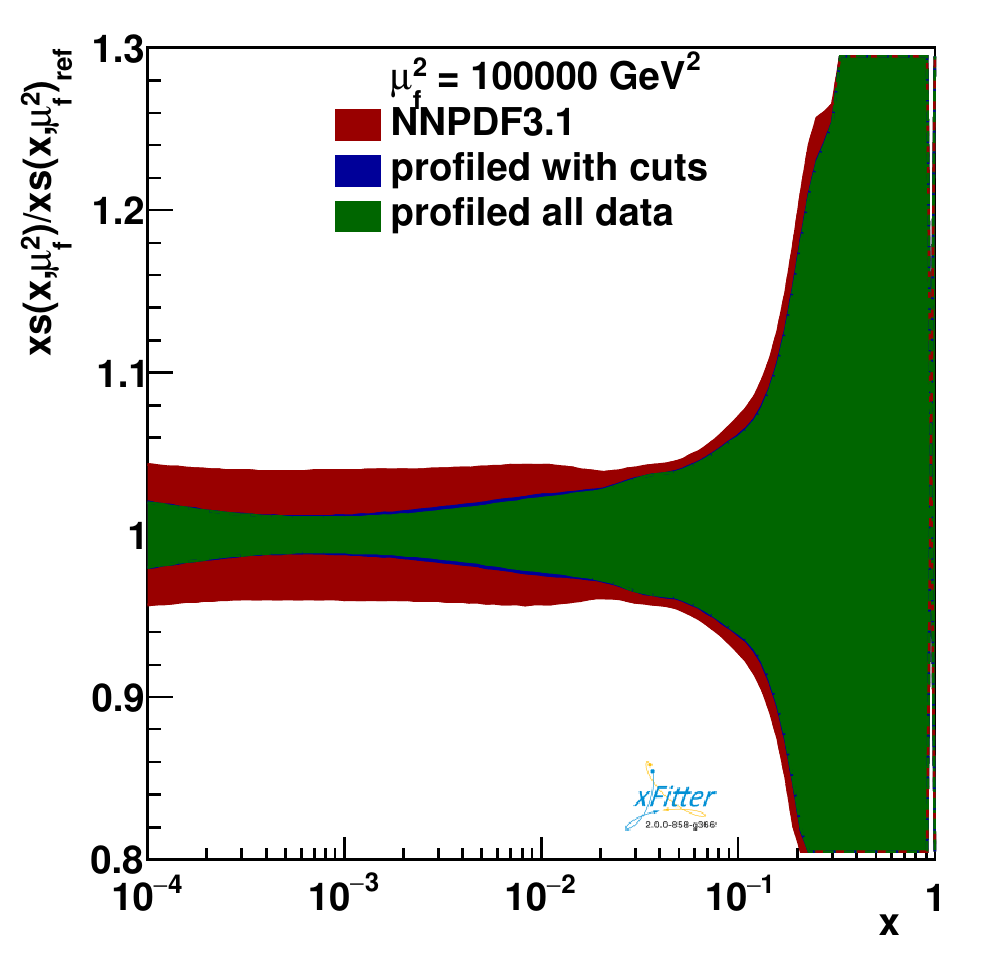}}}
  \put(-95,95){(a)}
  {{\includegraphics[width=0.235\textwidth]{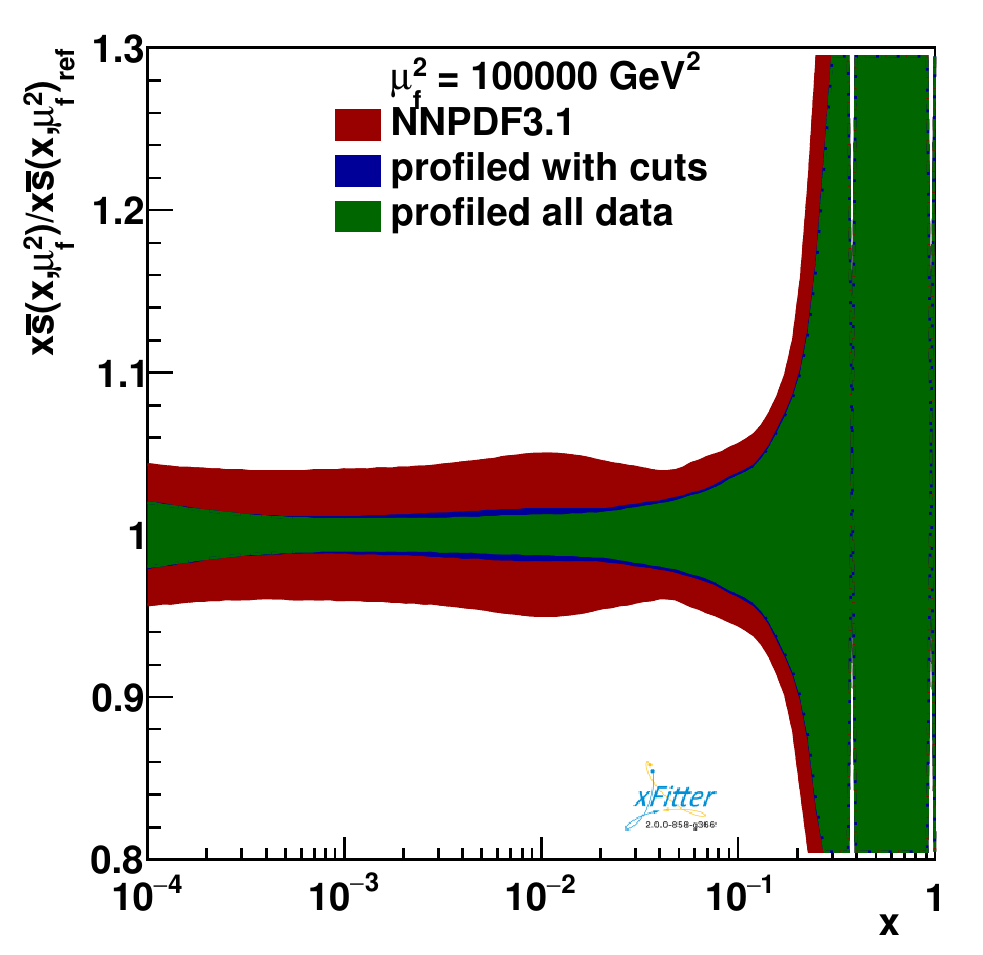}}}
  \put(-95,95){(b)}\\
  {{\includegraphics[width=0.235\textwidth]{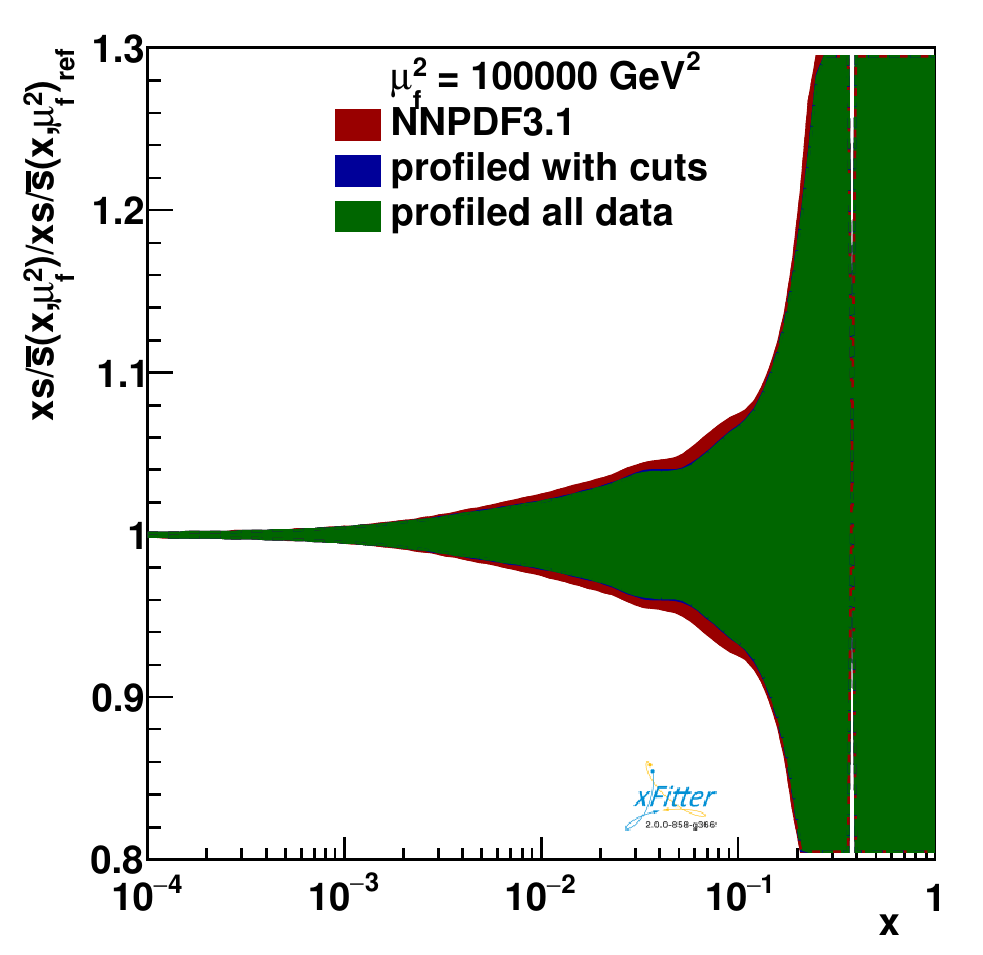}}}
  \put(-95,95){(c)}
  {{\includegraphics[width=0.235\textwidth]{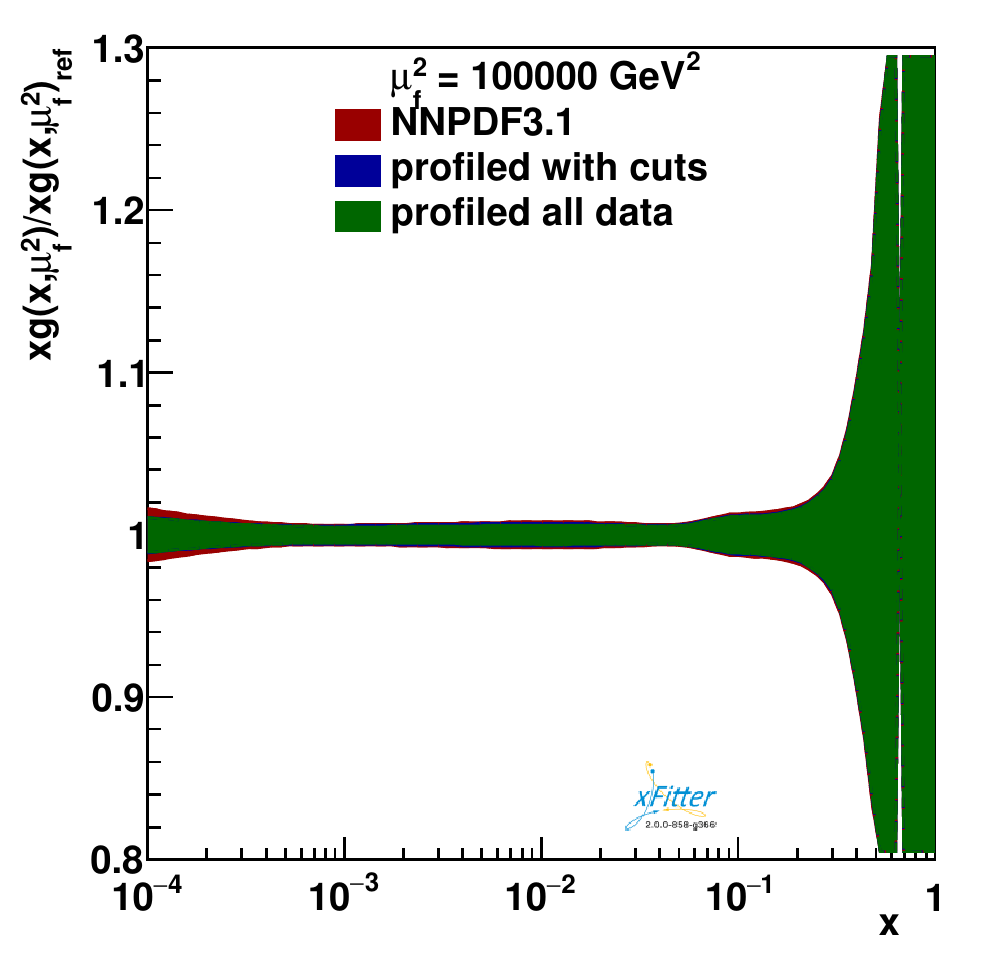}}}
  \put(-95,95){(d)}\\
  {{\includegraphics[width=0.235\textwidth]{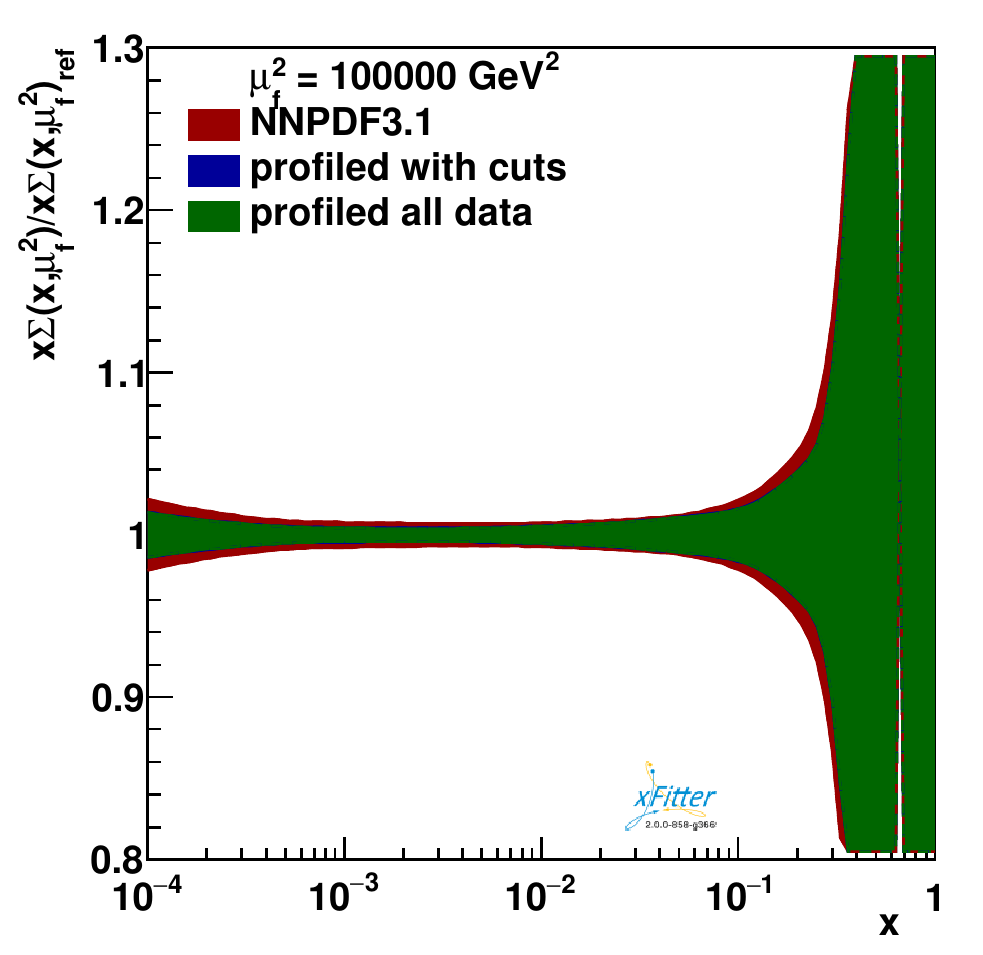}}}
  \put(-30,95){(e)}
  {{\includegraphics[width=0.235\textwidth]{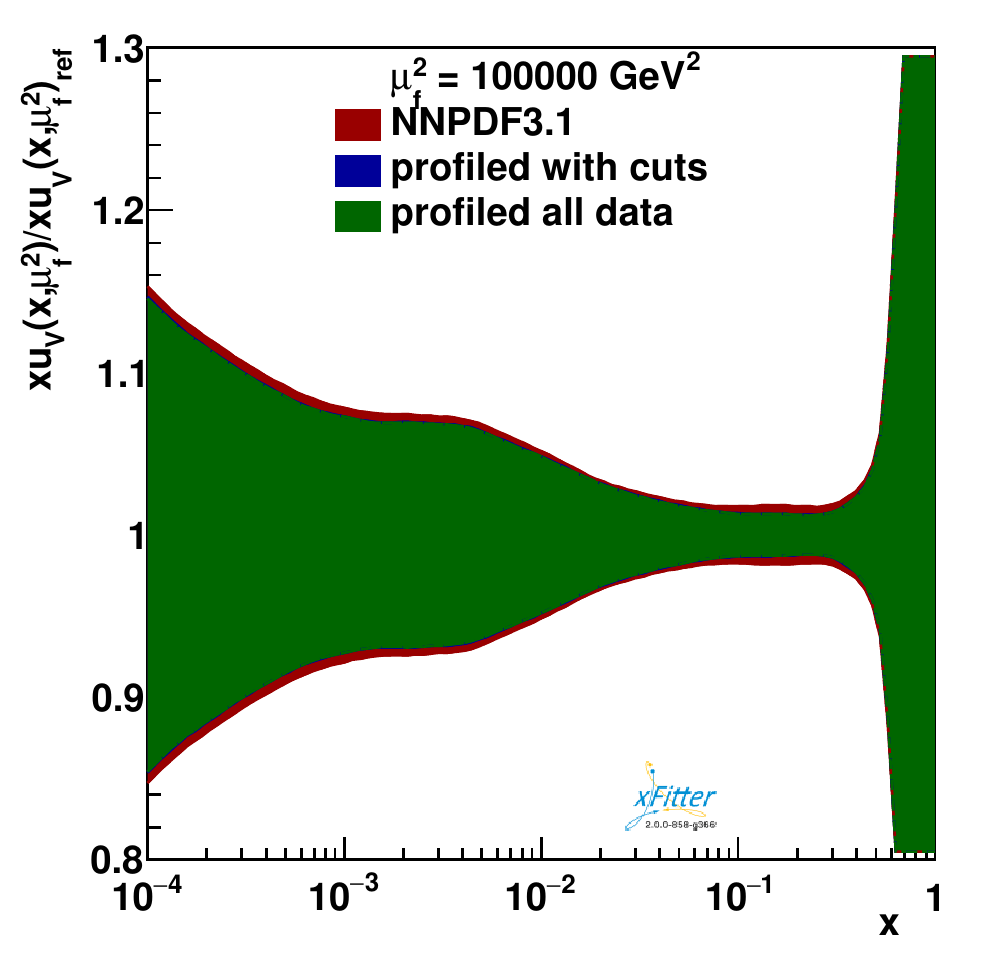}}}
  \put(-95,95){(f)}\\
  {{\includegraphics[width=0.235\textwidth]{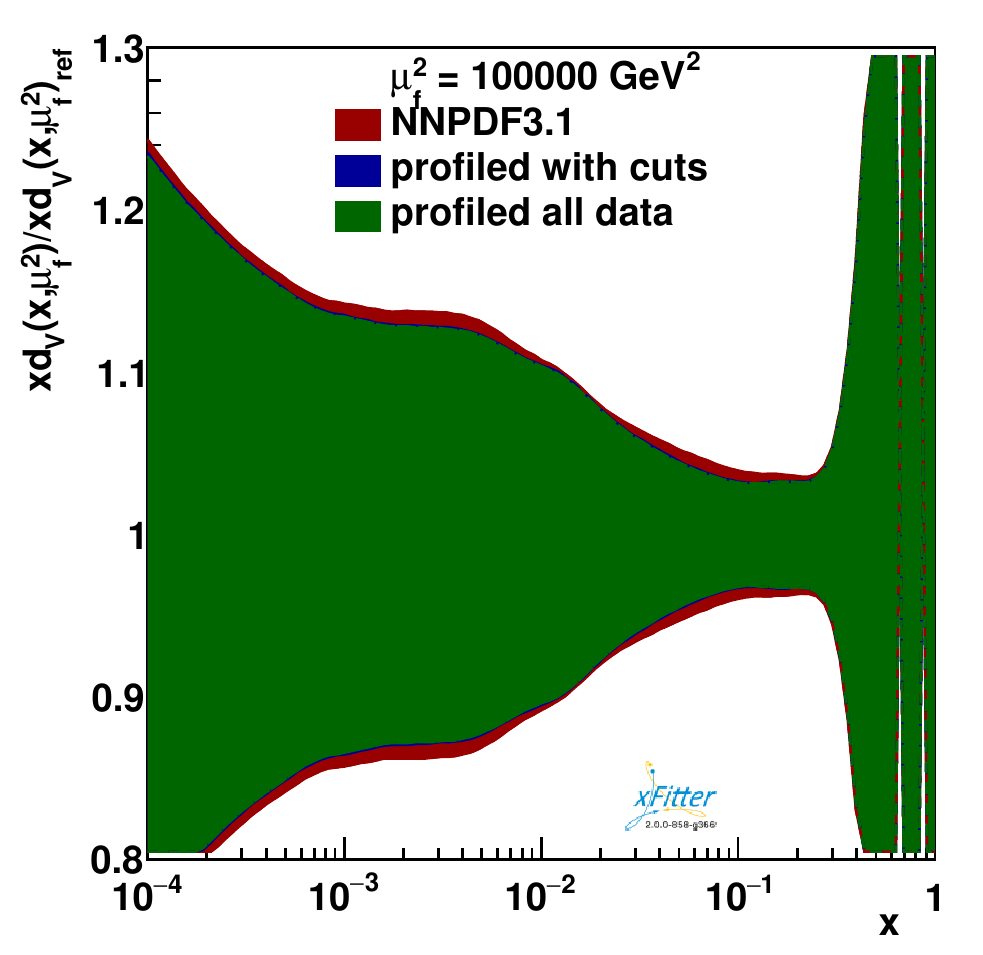}}}
  \put(-30,95){(g)}
  {{\includegraphics[width=0.235\textwidth]{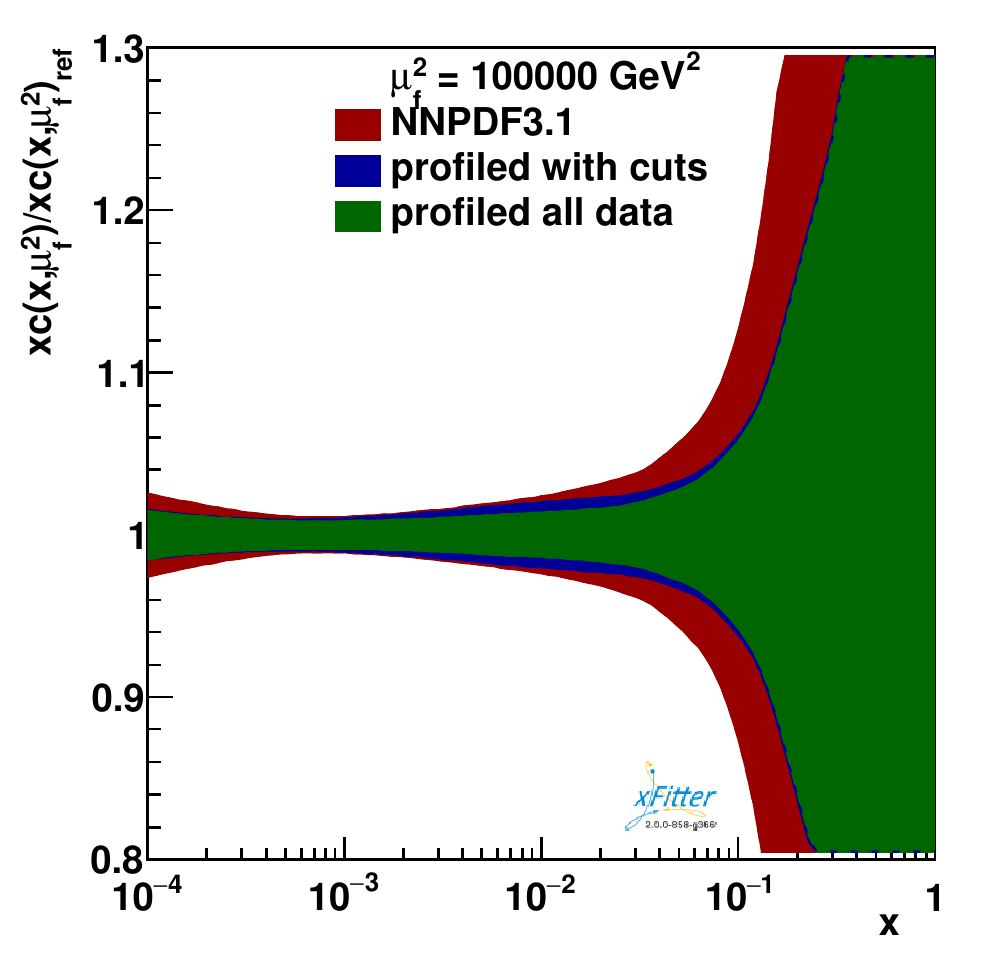}}}
  \put(-95,95){(h)}
  \caption{The relative strange quark (a), strange anti-quark (b), and ratio $s/\overline{s}$ (c), gluon (d), sea
    quark (e), u valence quark (f), d valence quark (g) and charm quark (h) PDF
    uncertainties at $\mu_\mathrm{f}^2=100000$ GeV$^2$ of the original
    and profiled \nnpdf PDF set.}
  \label{fig:pdf-nnpdf-100000}
\end{figure}
}

A recent study~\cite{AbdulKhalek:2019mps}
has examined potential improvments in the PDFs
for both the LHeC and HL-LHC facilities with both CC and NC.
Our improvement in the strange PDF is comparable,
while  additional channels of Ref.~\cite{AbdulKhalek:2019mps}
yields improved constraints on the gluon.

  Comparing the results of profiled PDFs in the FFNS and the VFNS, we
find both analyses are able to significantly improve the constraints
on the strange quark PDF.  This result gives us confidence that the
general features we observe here are independent of the details of the
heavy flavor scheme.

\goodbreak
\newpage
\section{Discussion and summary}
\label{sec:discuss}

The recent performance of the LHC has exceeded expectations and
produced an unprecedented number of precision measurements to be
analyzed; thus, it is essential to improve the theoretical
calculations to match.
The uncertainty for many of these precision measurements stems
primarily from the PDFs.
Hence, our ability to measure fundamental parameters of the Standard Model (SM),
such as the $W$ boson mass and $\sin^2\theta_W$, ultimately comes down to how
accurately we determine the underlying PDFs~\cite{Atlas:2019qfx}. 
Additionally, our ability to characterize and constrain
SM processes can indirectly impact beyond-standard-model (BSM) signatures.

We have focused on the  strange-quark distribution which,
at the LHC, can have a significant impact on the $W/Z$ cross section:
one of the ``standard candle'' measurements.
If we can reduce the uncertainty for these predictions, we can set
stringent limits on any admixture of physics at higher scales.
Unfortunately, at present the strange PDF has a comparably large uncertainty
because measurements from the LHC and HERA, as well as older
fixed-target experiments, do not seem to provide a definitive result
for this flavor component.

This situation  has  prompted us to examine the CC DIS charm production at
the LHeC to determine the impact of this data set on the PDF
uncertainty.
We considered the LHeC  as this  high-energy $ep/A$
facility  could potentially run in parallel with the LHC and
provide insights into these issues at low $x$ and high $Q^2$ in advance of a FCC program.

This case study of the CC DIS charm production at the LHeC provides a
practical illustration of the many features of \xfitter.
As the \xfitter framework is designed to be a versatile open-source
software framework for the determination of PDFs and the analysis of
QCD physics, we can readily adapt this tool to address the impact and
influence of new data sets.
Furthermore, as both FFNS and VFNS calculations are implemented, we
can use \xfitter as a theoretical ``laboratory'' to study the
resummation of large logarithms and multi-scale issues.
We have outlined some of these issues in the Appendix.
In particular, the CC DIS charm production involves a flavor-changing
$W^\pm$ boson, multiple quark masses enter the calculation, and this
introduces some subtle theoretical issues to properly address the
disparate mass and energy scales.

Using the \xfitter framework, 
we find that the LHeC can provide strong constraints on the
strange-quark PDF, especially in the previously unexplored small-\xbj
region.\footnote{In this study we have focused exclusively on the LHeC
  result for CC DIS charm production; however, the LHeC has a broad
  multi-faceted program which is described in
  Refs.~\cite{AbelleiraFernandez:2012ty,Klein:2018rhq}.}
\new{A large reduction of uncertainties is observed also when restricting 
the input data to the kinematic range where the differences between the \ffns and 
\fonll schemes are not larger than the present PDF uncertainties, indicating 
that the obtained  PDF constraints are stable and 
independent of the particular heavy-flavor scheme.}
As noted above, a reduction of the strange-PDF uncertainties
influences the $W/Z$ production, and thus the Higgs production; hence,
the LHeC CC DIS charm production data represent a valuable addition
for the future global PDF fits.

\new{However, since charm CC production in $e^{-}p$ 
collisions mostly probe $\bar{s}$, only mild constraints are put on
the ratio $s/\overline{s}$ 
using the NNPDF3.1 PDF set as reference; therefore for a precise 
determination of this ratio, both $e^{-}p$ and $e^{+}p$ data will be needed.}

In conclusion, we find that CC DIS charm production at the LHeC can
provide strong constraints on the strange PDF which are complementary
to the current data sets.
As the PDF uncertainty is the dominant factor for many precision
analyses, a reduction of these uncertainties will allow for more
accurate predictions which can be used to constrain both SM and BSM
physics processes.

\begin{acknowledgements}
\begin{hyphenrules}{nohyphenation}
    
We would like to thank
Max Klein for providing the pseudodata, and 
John~C.~Collins,
Aleksander~Kusina,
Pavel~Nadolsky,
Ted~C.~Rogers,
Ingo~Schienbein,
George~Sterman,
for useful discussion.
The work of O.\,Z. has been supported by Bundesministerium f\"ur Bildung und Forschung (contract 05H18GUCC1).
This work F.\,O. was supported by the U.S. Department of Energy under Grant No. DE-SC0010129.

\end{hyphenrules}
\end{acknowledgements}

%
\null\vfill\eject  
\newcommand{\xsec}[1]{\vskip 6pt \noindent {\bf #1} \quad }
\appendix
\addtocontents{toc}{\setlength{\cftsecnumwidth}{14ex}}

\section{$F_{2}^{c}$ Beyond leading-order}\label{sec:appendix}

\begin{figure*}
  \centering \includegraphics[clip,width=0.8\textwidth]{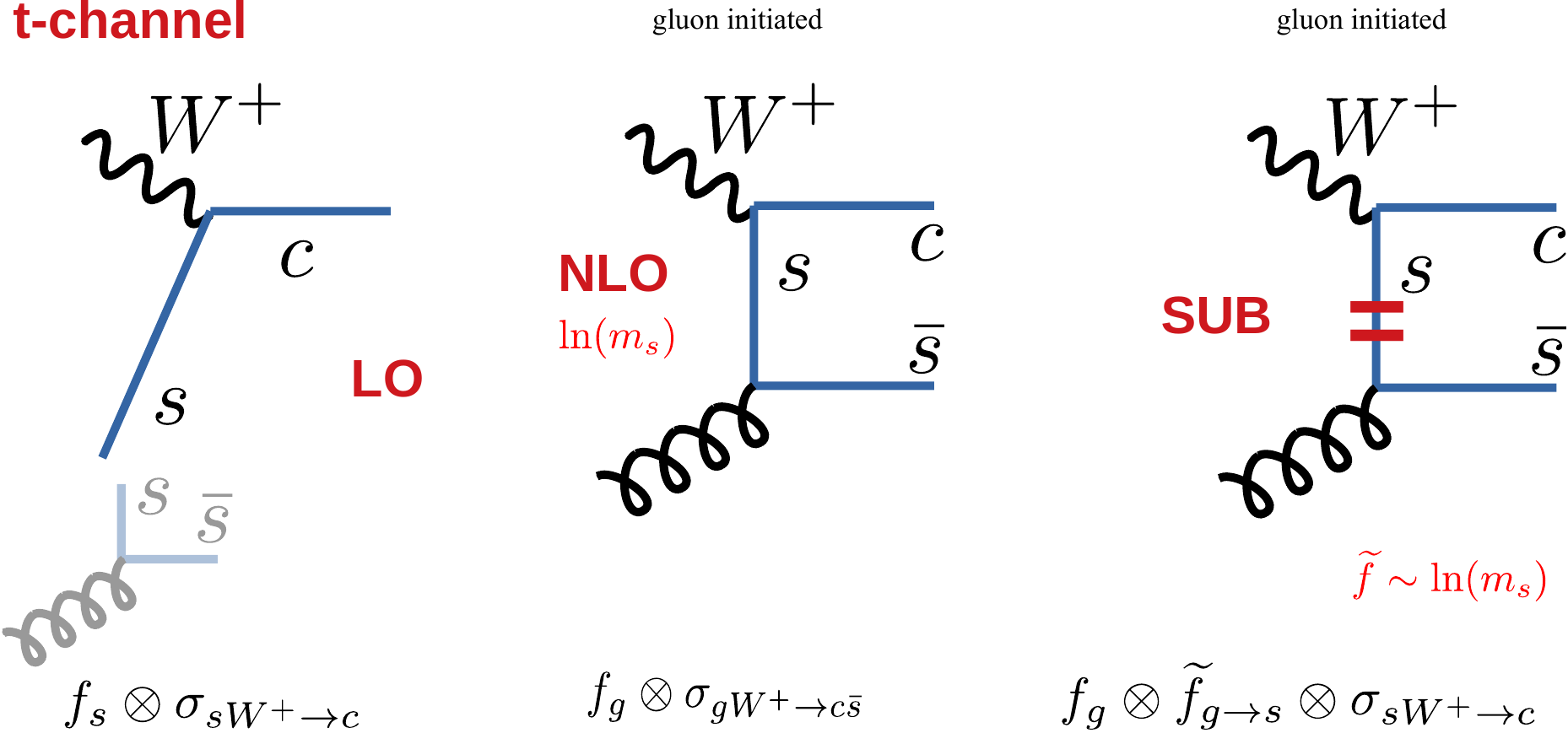}
  \caption{The $t$-channel processes up to ${\cal  O}(\alpha_S^1)$.
Note we sum the combination (NLO$-$SUB) to obtain the complete ${\cal  O}(\alpha_S^1)$ correction;
we find it useful to study these terms separately.
The higher-order quark-initiated contributions are not show, but are included in the calculation. 
\label{fig:tchannel}}
\end{figure*}

\begin{figure*}
  \centering \includegraphics[clip,width=0.8\textwidth]{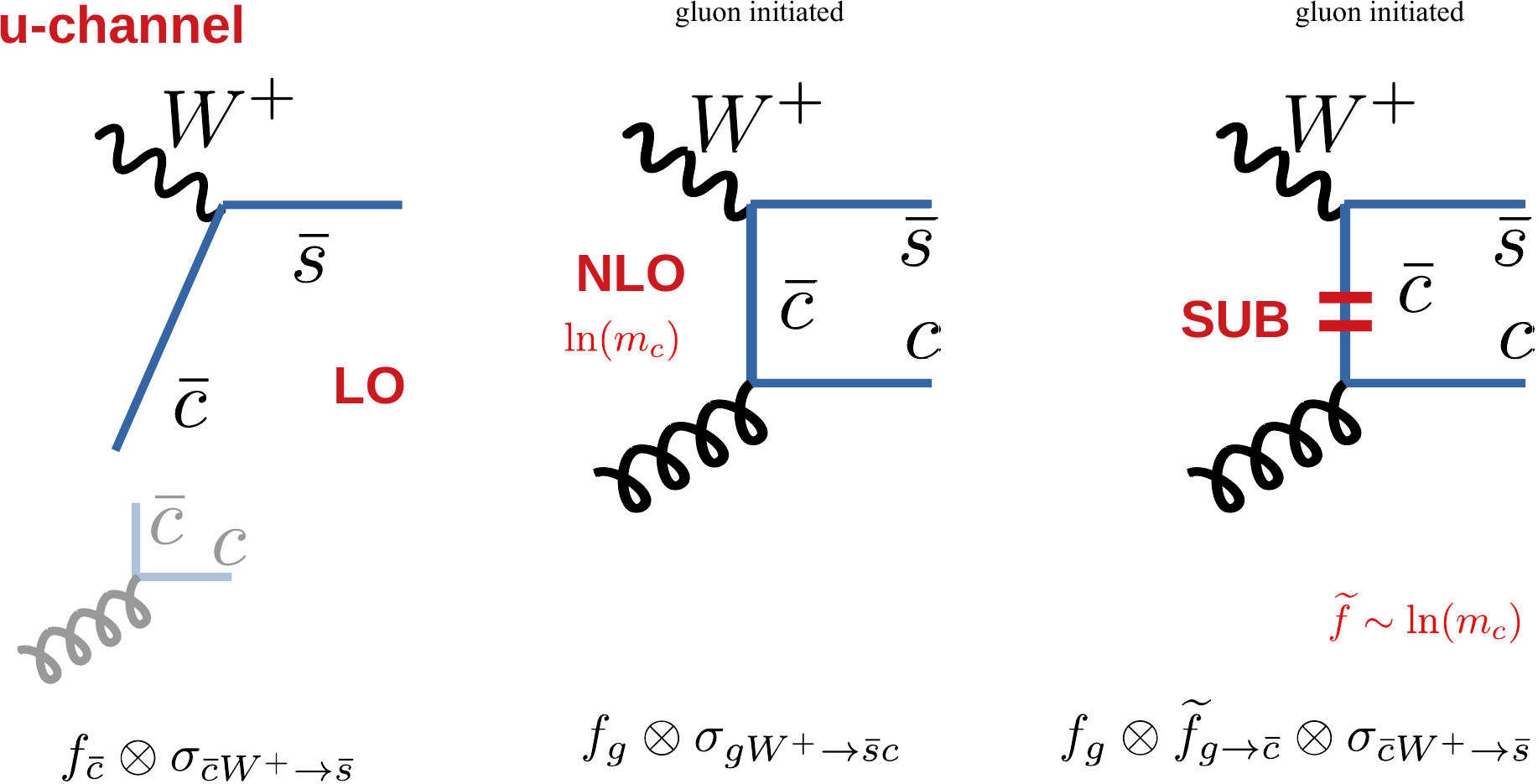}
\caption{The $u$-channel processes  up to ${\cal  O}(\alpha_S^1)$.
Note the NLO  $t$-channel and $u$-channel terms are combined coherently at the amplitude  level. 
The higher-order quark-initiated contributions are not show, but are included in the calculation. 
\label{fig:uchannel}}
\end{figure*}

\xsec{The multi-scale problem:}
The CC DIS charm production process involves some interesting issues
that we will explore here in detail. In particular, there are multiple
mass and energy scales which span a wide kinematic range, and it
becomes an intricate puzzle to treat them all properly.

For this current illustration, we will focus on the contribution to
the DIS $F_{2}^{c}$ structure function from the process involving the
strange and charm quark; other quark combinations can be addressed in
a similar manner.
The fully inclusive $F_2$ can be studied using the energy and angle of
the outgoing lepton; in contrast, $F_{2}^{c}$ also requires
information about the final hadronic state, and this introduces some
subtleties.
In particular, we will show that as we go to higher orders the
$F_{2}^{c}$ structure function must be defined carefully so that:
i)~theoretically it is free of divergences and independent of the
renormalization scales when calculated to all orders, 
and ii)~experimentally it matches what is
measured by the detector.

\xsec{The mass scales:}
What makes this process complex is that we encounter a number of
different mass scales. Furthermore, there is no fixed hierarchy for
the mass scales, and we will need to compute both in the low-$Q$
region, where $Q\lesssim m_{c}$, as well as in the high-$Q$ region,
$Q\gg m_{c}$.

The $Q$ scale is related to the invariant mass of the virtual-boson probe
($W^{+}$ in this case), and can be expressed in terms of the energy and angle of
the lepton; this is a physically measurable kinematic variable.

In contrast, the scale $\mu$ is an unphysical scale which implements
the separation between the PDF and the hard-scattering cross section, 
and the scale at which $\alpha_{s}$ is evaluated;
thus, the physics should be insensitive to a variation of $\mu$. As
our calculations typically involve the dimensionless combination
$\ln(\mu/Q)$, we generally choose $\mu\sim Q$ to avoid large
logarithms.

The strange quark is a ``light'' active parton with an associated PDF
$s(x)$ and mass $m_{s} < \Lambda_{\rm QCD}$. The strange-quark mass is
comparable to or less than other hadronic scales which are neglected;
as such, it serves only as a regulator and plays no physical
role. Effectively, we can take $m_{s}\to 0$ if we choose.
We treat the up and down quarks masses $m_{u,d}$ in a similar manner.

The charm quark is a ``heavy'' object; its associated mass
$m_{c}> \Lambda_{\rm QCD}$ does play a physical role and cannot
generally be neglected. There may or may not be a PDF associated with
the charm. In a $n_f=3$ FFNS scheme, we will assume the charm PDF to
be zero.\footnote{It is possible to extend this to incorporate an
  intrinsic-charm PDF.} In a VFNS there is a charm PDF only when the
$\mu$ scale is above the scale where the charm PDF is activated; we
call this the matching scale, $\mu_{c}$.  It is common\footnote{The
  choice of matching scale $\mu_{c}=m_{c}$ is common because at NLO
  the $\overline{\mbox{MS}}$ matching conditions on the PDFs are
  proportional to the DGLAP kernel times $\ln(\mu/m_{c})$. As an
  explicit calculation shows, the constant term vanishes. Therefore,
  by choosing $\mu_{c}=m_{c}$ we have the simple boundary condition
  $f_{c}(x,\mu=m_{c})=0$. At NNLO, the constant term is non-zero and
  this yields $f_{c}(x,\mu=m_{c})\not=0$. See
  Ref.~\cite{Stavreva:2012bs} and references therein.}
to set $\mu_{c}=m_{c}$, but this is not required.\footnote{By
  displacing the matching scale to larger values $\mu_{c} > m_{c}$,
  one can have the advantage of avoiding delicate cancellations in the
  region $\mu\sim m_{c}$; this flexibility was explored in
  Refs.~\cite{Bertone:2017ehk,Bertone:2018ids}.}
In this study, however, we will adopt this common choice.

Because there are two different quark masses involved ($m_{s}$ and
$m_{c}$) in the CC DIS process, we can examine the mass singularities
of the $t$-channel and $u$-channel separately.
This separation is particularly useful to understand how the
individual mass singularities are addressed, and how the FFNS and the
VFNS organize the contributions to the total structure function.

\xsec{The $n_f=3$ FFNS:}
To be specific, we will consider CC DIS production of a charm
quark. We first compute this in the $n_f=3$ FFNS where $\{u,d,s\}$ are
light ``active'' partons in the proton, and the charm $c$ is
considered an external ``heavy'' particle. This can be implemented in
the ACOT scheme~\cite{Aivazis:1993pi} for example by using a CWZ
renormalization~\cite{Collins:1978wz} where the light ``active''
partons are renormalized with normal $\overline{\mbox{MS}}$, and the
``heavy'' quarks use a zero-momentum subtraction. In this scheme, the
\textbf{leading-order} (LO) process is \mbox{$sW^{+}\to c$} as
illustrated in Fig.~\ref{fig:tchannel}. At \textbf{next-to-leading-order}
(NLO), we then include \mbox{$gW^{+}\to c\bar{s}$} which has both
$t$-channel (Fig.~\ref{fig:tchannel}) and $u$-channel  (Fig.~\ref{fig:uchannel}) contributions.\footnote{Note, there are
  also corresponding quark-initiated processes; we will focus on the
  gluon-initiated processes as this is sufficient to illustrate our
  points. Both the gluon- and quark-initiated contributions are
  included in our calculations.}

\xsec{$t$-channel:}
The $t$-channel process has an intermediate $s$-quark exchanged, and
if we use the strange quark mass $m_{s}$ to regulate the
singularities, this will yield a contribution proportional to
$\ln(Q/m_{s})$. This mass singularity arises from the region of phase
space where the exchanged $s$-quark becomes collinear and close to the
mass shell; that is, when the phase space of the
\mbox{$gW^{+}\to c\bar{s}$} process begins to overlap with that of the
\mbox{$sW^{+}\to c$} process. This ``double counting'' is resolved by
a \textbf{subtraction} (SUB) counter-term
given by:
\[
(SUB)\sim f_{g}\otimes\widetilde{f}_{g\to s}\otimes\sigma_{sW^{+}\to
  c}\,.
\]
Here, $\widetilde{f}_{g\to s}$ is the perturbative splitting of the
gluon into an $s\bar{s}$ pair; the leading term is proportional
to:\footnote{The scale of the SUB term is $\mu$ as the relevant scale
  here is the renormalization scale of the PDF:
  $f(x,\mu)\otimes\hat{\sigma}(x,Q,\mu)$.}
\[
\widetilde{f}_{g\to s}(x,\mu)\sim\frac{\alpha_{S}(\mu)}{2\pi}P_{g\to
  s}^{(1)}(x)\:\ln\left(\frac{\mu^{2}}{m_{s}^{2}}\right)+
  {\cal   O}(\alpha_{s}^{2})
\]
where $P_{g\to s}^{(1)}(x)$ is the $\mathcal{O}(\alpha_{s})$ DGLAP
splitting kernel for \hbox{$g\to s$}.

The complete contribution to the structure function is given by:
\[
F_{2}^{c}\sim TOT=LO+(NLO-SUB)
\]
The complete ${\cal O}(\alpha_{s})$ contribution is the combination
$(NLO-SUB)$; our separation into $NLO$ and $SUB$ is simply to
illustrate the interplay of these components. Both the NLO and SUB
terms have $\ln(m_{s})$ divergences, but these precisely cancel and
yield a well-defined result even if we take the $m_{s}\to 0$
limit.\footnote{In fact, we could have taken $m_{s}=0$ initially and
  used dimensional regularization to compute the contributions.}

\xsec{$u$-channel:}
We next examine the $u$-channel NLO contribution to the
\mbox{$gW^{+}\to c\bar{s}$} process. This has an intermediate
$c$-quark exchanged and is proportional to $\ln(Q/m_{c})$. In the FFNS
where the charm is a ``heavy'' non-parton, there is no counter-term
for this graph, and the resulting observables will retain the
$\ln(Q/m_{c})$ dependence. In principle, this means that when we go to
large $Q$ scales, these terms will begin to degrade the convergence of
the perturbative series. In practice, while this degradation only
grows logarithmically, at large scales (such as at the LHC energies)
we do find it convenient to treat the charm on an equal footing as the
${u,d,s}$ partons.

\xsec{The VFNS:}
We now turn to the VFNS scheme where we include the charm quark as an
``active'' parton and compute its associated PDF.

In this case, there is a $u$-channel counter-term (SUB) given by
$f_{g}\otimes\widetilde{f}_{g\to\bar{c}}\otimes\sigma_{\bar{c}W^{+}\to\bar{s}}$
which is proportional to $\ln(\mu/m_{c})$. The NLO $u$-channel
contribution will have a $\ln(Q/m_{c})$ factor, so the combination
$(NLO-SUB)$ is also free of mass singularities.\footnote{Specifically,
  the combination $(NLO-SUB)$ is free of mass singularities and finite
  in the limit $m_{c}\to0$. Note that the VFNS fully retains the charm
  quark mass $m_{c}$ and (in contrast to some claims in the
  literature) the factorization holds up to
  ${\cal O}(\Lambda^{2}/Q^{2})$ corrections; all terms of order
  $(m_{c}^{2}/Q^{2})$ are fully included~\cite{Collins:1998rz}.}

What is less obvious is that we must also include the LO process
\mbox{$\bar{c}W^{+}\to\bar{s}$}. There are two ways we can understand
why this is necessary.

\begin{figure}
\centering \includegraphics[clip,width=0.45\textwidth]{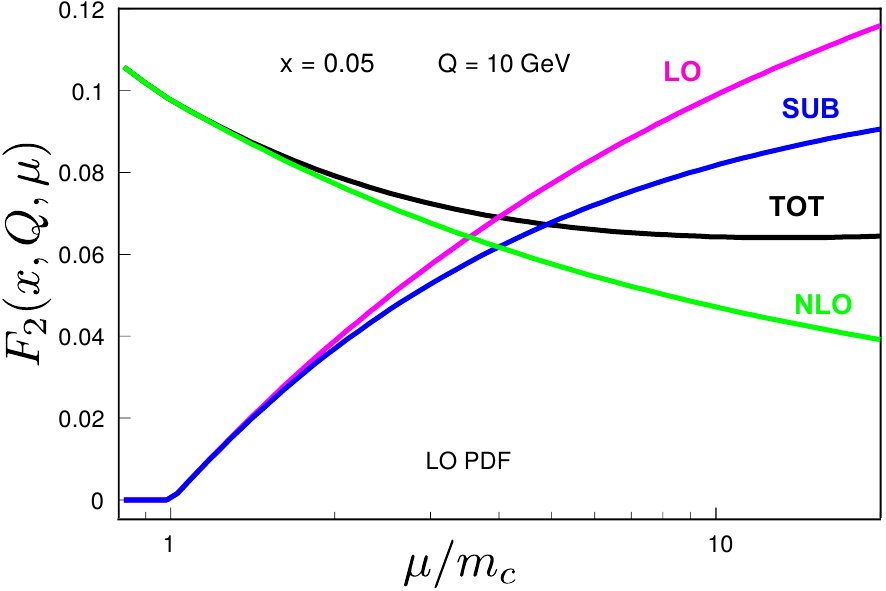}
\caption{Calculation of $F_{2}^{c}$ vs.\ $\mu$ in the VFNS
  illustrating the cancellation of the LO ($\bar{c}W^{+}\to\bar{s}$)
  and the SUB \hbox{$(g\to
    \bar{c})$}$\otimes$\hbox{$(\bar{c}W^+
    \to \bar{s})$}
  contributions in the region $\mu\sim m_{c}$. The $Q$ scale is fixed
  at $10\,$GeV and the charm PDF is matched at $\mu_{c}=m_{c}$ such
  that $f_{c}(x,\mu=m_{c})=0$. \label{fig:acot}}
\end{figure}

\xsec{Explanation \#1: matching of LO and SUB:}
Recall that in the $t$-channel case, the subtraction term SUB removed
the double counting between the LO \mbox{$sW^{+}\to c$} and NLO
\mbox{$gW^{+}\to c\bar{s}$} subprocesses.

The $u$-channel case is analogous in that this subtraction term
removes the double counting between the LO
\mbox{$\bar{c}W^{+}\to\bar{s}$} and NLO \mbox{$gW^{+}\to c\bar{s}$}
subprocesses; both contributions are required to ensure that the resulting
cross section is insensitive to the scale $\mu$.

This is apparent in Fig.~\ref{fig:acot} where we plot the individual
terms versus $\mu$ for fixed values of \xbj and $Q$. In the region
$\mu\sim m_{c}$, the charm PDF $f_{c}(x,\mu)$ (and hence, the LO
contribution) rises very quickly as the DGLAP evolution is driven by
the very large gluon distribution via $g\to c\bar{c}$ splitting, and
combined with a large $\alpha_{s}(\mu)$. The SUB subtraction also
rises quickly as this is driven by the logarithmic term
$\ln(\mu^{2}/m_{c}^{2})$. The difference \mbox{$(LO-SUB)$} is the
physical contribution to the total \mbox{$[TOT=LO+NLO-SUB]$}, and it
is this combination that is smooth across the ``turn on'' of the
charm PDF at the matching scale $\mu_{c}=m_{c}$.  We now see that if
we neglect the LO \mbox{($\bar{c}W^{+}\to\bar{s}$)} contribution, we
lose the cancellation between LO and SUB in the region
$\mu\sim m_{c}$, and our structure function (or cross section) would
have an anomalous shift at the arbitrarily location $(\mu_{c})$ where
we turn on the charm PDF.

As we vary the unphysical scale $\mu$, we are simply shifting
contributions between the separate \mbox{$\{LO,NLO,SUB\}$} terms which
individually exhibit a large $\mu$-dependence. However, the total
combination $(TOT)$, which represents the physical observable, is
relatively insensitive to $\mu$ (up to higher orders), and this
property is evident in Fig.~\ref{fig:acot}.

\xsec{Explanation \#2: removing ``double counting:''}
A~second way to understand why we require the LO process
\mbox{$\bar{c}W^{+}\to\bar{s}$} is to consider the regions of phase
space covered by each of the subprocesses. The singularity of the
$u$-channel NLO \mbox{$gW^{+}\to c\bar{s}$} processes arises from the
phase-space region where the intermediate $\bar{c}$-quark becomes
collinear and close to the mass shell.\footnote{For example, the
  $c$-quark is off-shell by the order of its mass $m_{c}$; this is
  independent of the scale $Q$ and \textit{does not} assume any
  $Q\gg m_{c}$ limit.} This is precisely the phase-space region of the
LO process \mbox{$\bar{c}W^{+}\to\bar{s}$} where the partonic
$\bar{c}$-quark is collinear to the hadron. The SUB term then removes
the ``double counting'' between the LO and NLO contributions; hence,
all three contributions \mbox{$\{LO,NLO,SUB\}$} are necessary to cover
the full phase space.

This is also apparent if we consider the transverse momentum $(p_{T})$
of the final-state charm in the Breit frame. For the LO
\mbox{$\bar{c}W^{+}\to\bar{s}$} process in the Breit frame, the
incoming $W^{+}$ and $\bar{c}$ are collinear, and the produced
$\bar{s}$ must have zero $p_{T}$ in this frame.

For the NLO $gW^{+}\to c\bar{s}$ process, we integrate over the
complete phase space for the exchanged $\bar{c}$ quark, and this will
include the region where the $\bar{c}$-quark is emitted nearly
collinear to the gluon and nearly on-shell; in this region the
$\bar{c}$-quark will have $p_{T}\sim 0$ and we encounter a singularity
from the internal $\bar{c}$-quark propagator. The $p_{T}\sim 0$ region
is precisely that subtracted by the SUB counter
term\footnote{Specifically, the incoming $W^{+}$ and $g$ are
  collinear and the gluon then emits a collinear $c\bar{c}$ pair so
  the final $\bar{s}$ has zero $p_{T}$.} and this ensures that the
combination $(NLO-SUB)$ is free of divergences.

\xsec{Recap:}
To recap, i)~the combination of the LO and SUB terms ensure a minimal
$\mu$ variation at low $\mu$, and ii)~the combination of SUB and NLO
ensures that the mass singularities are cancelled at high $\mu$.

This interplay of terms illustrates some of the intricacies of
QCD, especially since this exchange is across different orders of
$\alpha_{s}$.

Furthermore, note that in the $u$-channel for both the LO and SUB
contributions, the charm quark is collinear to the incoming hadron,
and thus exits in the hadron remnants. While this may be
experimentally difficult to observe, because we are asking for a
\textit{``fully inclusive''} $F_{2}^{c}$, these contributions cannot
be simply ignored. We will discuss this further in the following
section.

\xsec{Defining $F_{2}^{c}$:}
The LO $u$-channel \mbox{$\bar{c}W^{+}\to\bar{s}$} process foreshadows
difficulties that we encounter if we try and extend the concept of
``fully inclusive'' $F_{2}^{c}$ to higher orders. We note that in
Ref.~\cite{Collins:1998rz} Collins extended the proof of factorization
to include heavy quarks such as charm and bottom for an inclusive
structure function $F_{2}$;
analysis of a  \textit{``fully inclusive''} $F_{2}^{c}$
is more complex for a number of reasons. 
Whereas $F_2$ only requires measurement of the outgoing lepton energy
and angle, $F_2^c$ also requires information on the hadronic final state.  At
the parton level, this introduces complications including when the
charm is in the hadronic remnants and brings in both fragmentation and
fracture functions.

To characterize the theoretical issues involved in constructing
$F_{2}^{c}$, we can imagine starting from the (well-defined) inclusive
$F_{2}$, and then dividing the contributions into two sets: one for
$F_{2}^{c}$ for the ``heavy'' charm quark, and the rest into
$F_{2}^{u,d,s}$ for the ``light'' quarks. We will show that this
theoretical procedure encounters ambiguities.

The LO $u$-channel $\bar{c}W^{+}\to\bar{s}$ process does not have any
``apparent'' charm quark in the final state, but this contribution is
essential to balance with the SUB process
\mbox{$f_{g}\otimes\widetilde{f}_{g\to\bar{c}}\otimes\sigma_{\bar{c}W^{+}\to\bar{s}}$.}
Note that for the SUB process the charm quark arises from a gluon
splitting into a collinear $c\bar{c}$ pair which is then part of the
hadron remnants. For the LO process, presumably our $\bar{c}$ quark
also came from a gluon splitting into a collinear $c\bar{c}$
pair. Thus, our $F_{2}^{c}$ must include those cases where the charm
is contained in the hadron remnants.

This issues touches on the fact that, because the charm parton
ultimately fragments into a charmed hadron (typically a $D$ meson), we
must introduce a set of fragmentation functions (FFs) which are
scale-dependent and will factorize final-state singularities in a
similar manner as the PDFs factor the initial-state
singularities.\footnote{For the NLO quark-initiated contributions (not
  shown) we will have final state singularities from processes such as
  $c\to cg$ which will be factorized into the FFs.} Specifically, we may
also allow for the possibility that a gluon or a light quark fragments
into a charmed hadron.

\begin{figure}[t]
\centering \includegraphics[width=0.45\textwidth]{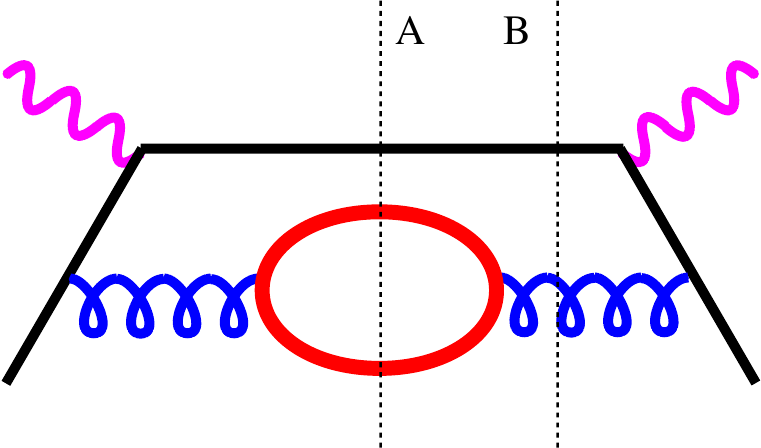}
\caption{A higher order Feynman graph illustrating the complications
  in defining a ``fully inclusive'' $F_{2}^{charm}$.  A light quark
  ($q$) scatters from a vector boson ($V$) with a $c\bar{c}$ in the
  internal loop.  If we cut the amplitude at ``A'' we have charm in
  the final state and this must be included in $F_{2}^{charm}$.  If we
  cut the amplitude with cut ``B'' there is no charm in the final
  state.  Additionally, since this diagram contributes to the beta
  function, this highlights the complications of using an $\alpha_{S}$
  and hard scattering $\hat{\sigma}$ with differing $N_{\rm
    eff}$. \label{fig:bubble}}
\end{figure}

\xsec{The bubble  diagram:}
Some of the theoretical intricacies of defining a \textit{``fully
  inclusive''} $F_{2}^{c}$ are illustrated in Fig.~\ref{fig:bubble}
which shows a higher-order DIS process with a quark-antiquark loop.

Let us compute this diagram in the $n_f=3$ FFNS where the internal
loop is a massive $c\bar{c}$-pair and the external quark is a light
quark \mbox{$\{u,d,s\}$}.  If the final state is represented by Cut-A,
then we have charm quarks in the final state, and this should be
included in $F_{2}^{c}$.

However, if we instead use Cut-B as a final state, there is no charm
in the final state, so this should not be included in $F_{2}^{c}$.
[More precisely, when we renormalize the charm loop with zero-momentum
subtraction, this contribution effectively decouples.] Thus, the
contribution from Cut-A will be included in $F_{2}^{c}$, but the
contribution from Cut-B will not.

This diagram generates additional complications in that multiple quark
flavors are involved. For example, the bubble diagram involves quarks
of both $q=\{u,d,s\}$ and $c$ flavors, so this contribution cannot be
uniquely assigned to $F_{2}^{q}$ or $F_{2}^{c}$. We can introduce
theoretical definitions to make the choice, but then we have to be
careful about double-counting contributions and introducing
uncancelled singularities.
For example, the bubble diagram of Fig.~\ref{fig:bubble} is
encountered in the $F_{2}^{c}$ heavy-quark calculations of
Refs.~\cite{Chuvakin:1999nx,Chuvakin:2000jm}; here, an additional
scale $\Delta$ is introduced to subdivide the contributions.

\xsec{The running of $\alpha_{s}$ in the FFNS:}
The bubble diagram of Fig.~\ref{fig:bubble} also highlights the
difficulty of using a $n_f=3$ FFNS with a VFNS running of
$\alpha_{s}$. In a $n_f=3$ FFNS, internal $c\bar{c}$ loops decouple
from the theory and are not included in the calculation;\footnote{More
  precisely, the heavy quarks are renormalized with zero-momentum
  subtraction and their contributions decouple; this is why we can
  neglect loops from the the top quark and any other heavy particle.
} however, the $\beta$-function with $n_f=4$ requires precisely these
$c\bar{c}$ loop contributions. This deficiency can be
patched order by order by expanding the $\beta$-function and inserting
the required terms at each
order~\cite{Napoletano:2014thesis,Bierenbaum:2009zt,Cascioli:2013era}.
Once again, we cannot unambiguously divide the inclusive $F_{2}$ into
separate ``light'' and ``heavy'' quantities.

\xsec{Extensions to bottom and top:}
While we have used the charm quark to illustrate these features, the
same properties can, in principle, be applied to both the bottom and
top quark.\footnote{Additionally, Collins definitively addressed the
  case of multiple heavy quarks which can allow for both charm and
  bottom in a unified framework; in contrast to some incorrect claims
  in the literature, there is no difficulty in including multiple
  heavy quarks. (\textit{cf.} Ref.~\cite{Collins:1998rz}, Sec.~IX.)}
For the case of the bottom quark, the larger mass $m_{b}$ yields a
smaller $\alpha_{s}(\mu)$ for $\mu\sim m_{b}$ and the evolution of
$f_{b}(x,\mu)$ is thus reduced. Nevertheless, for large-scale
processes (such as at the LHC) we often find it convenient to make use
of $f_{b}(x,\mu)$ and treat the bottom on an equal footing as the
other light quarks.
For the case of the top quark, the very large mass $m_{t}$ yields a
much smaller $\alpha_{s}(\mu)$ for $\mu\sim m_{t}$ and the evolution
of $f_{t}(x,\mu)$ is comparatively reduced.

\xsec{Summary}
To properly define $F_{2}^{c}$ at higher orders, we encounter the
theoretical issues discussed above: as the charm quark fragments into
a charmed meson, we must be careful to ensure that the theoretical
quantity matches what is actually measured experimentally.
This is more complex than simply asking for the portion of $F_{2}$
has a charm in the final state, and is an issue for both the FFNS
and VFNS as we move to higher orders.
We can perform the computation in the FFNS but in the large energy
limit we encounter $\ln(Q^{2}/m_{c}^{2})$ divergences and this, in
part, contributes to the observed differences at large $Q$.

The VFNS includes the charm quark as an active parton for $\mu$ scales
above a matching scale $\mu_{c}$. For large $Q$ scales, the mass
singularities of NLO and SUB terms will cancel to yield a result free of
divergences. For scales $\mu\sim m_{c}$, cancellation between the LO
and SUB contributions ensures a minimal $\mu$ dependence; however, as
this can be delicate to implement numerically, we have the option of
displacing the matching scale $\mu_{c}$ to a larger scale where the
cancellation is more stable~\cite{Bertone:2017ehk,Bertone:2018ids}.

\newpage
\printbibliography

\end{document}